\documentclass[journal]{IEEEtran}

\makeatletter
\newcommand{\thickhline}{%
	\noalign {\ifnum 0=`}\fi \hrule height 1pt
	\futurelet \reserved@a \@xhline
}

\usepackage{tabularx}
\usepackage{graphicx}
\usepackage{subcaption}
\usepackage{epstopdf}
\usepackage{tabularx}
\usepackage{array}
\usepackage{authblk}
\usepackage{algpseudocode}
\usepackage{mathtools}
\newtheorem{remark}{Remark}

\newtheorem{definition}{Definition}%[section]

\newtheorem{theorem}{Theorem}
\usepackage{array}
\usepackage{mdwmath}
\usepackage{amsmath,amssymb,amsfonts}
\usepackage{epsfig}
\usepackage{color,soul}

\begin{document}
\title{Wavelet based multivariate signal denoising using Mahalanobis distance and EDF statistics}
\author{\IEEEauthorblockN{Khuram Naveed and Naveed ur Rehman}
\thanks{K. Naveed and N. Rehman are with the Electrical and Computer Engineering Department, COMSATS University Islamabad (CUI), Park Road, Islamabad, 45550 Pakistan.}}

\maketitle

\begin{abstract} 
A multivariate signal denoising method is proposed which employs a novel multivariate goodness of fit (GoF) test that is applied at multiple data scales obtained from discrete wavelet transform (DWT). In the proposed multivariate GoF test, we first utilize squared Mahalanobis distance (MD) measure to transform input multivariate data residing in M-dimensional space $\mathcal{R}^M$ to a single-dimensional space of positive real numbers $\mathcal{R}_+$, i.e., $\mathcal{R}^M \rightarrow \mathcal{R}_+$, where $M > 1$. Owing to the properties of the MD measure, the transformed data in $\mathcal{R}_+$ follows a distinct distribution. That enables us to apply the GoF test using statistic based on empirical distribution function (EDF) on the resulting data in order to define a test for multivariate normality. We further propose to apply the above test locally on multiple input data scales obtained from discrete wavelet transform, resulting in a multivariate signal denoising framework. Within the proposed method, the reference cumulative distribution function (CDF) is defined as a quadratic transformation of multivariate Gaussian random process. Consequently, the proposed method checks whether a set of DWT coefficients belong to multivariate reference distribution or not; the coefficients belonging to the reference distribution are discarded. The effectiveness of our proposed method is demonstrated by performing extensive simulations on both synthetic and real world datasets.
\end{abstract}


\IEEEpeerreviewmaketitle

%\cite{misczynski2005electrode}
%\cite{marro2002frontal}
%\cite{zotev2007multi}
%\cite{brown2004exercise}
%\cite{gobel2010serious}
%\cite{richardson1989estbsn}
\vspace{-4mm}
\section{Introduction}
Multivariate or multichannel signals are routinely encountered in modern engineering and scientific applications owing to rapid advances in computational and multi-sensor data acquisition technologies. Some noteworthy applications of multivariate data include diagnosis and treatment using biomedical signals such as electrocardiogram (ECG), electroencephalogram (EEG), fatigue and health monitoring within automatic exercise machines, exergaming platforms that employ an array of visual and body sensors and prediction of geophysical, meteorological and oceanographic trends. To make multi-sensor technology cost effective, low cost sensors are being increasingly employed in many applications resulting in multivariate measurements with degraded quality due to the addition of noise from different sources. To remove noise from such data, a process termed as \textit{signal denoising}, is an important preprocessing step in many engineering pplications.

Let $\mathbf{x}_i \in\mathcal{R}^M$ denote the acquired multivariate observations, with $M$ number of channels, modeled as the sum of true signal values $\mathbf{s}_i\in\mathcal{R}^{M}$ and additive noise observations $\boldsymbol{\psi}_i\in\mathcal{R}^{M}$
\begin{equation}
\mathbf{x}_i=\mathbf{s}_i+\boldsymbol{\psi}_i, \ \ \ \ i=1, \cdots, N,
\label{Ch5:noisy}
\end{equation}        
\noindent where $N$ denotes the number of observations.

\noindent In this work, noise observations $\boldsymbol{\psi}_i$ are modeled through independent and identically distributed multivariate Gaussian distribution $\mathcal{N}_M(\mathbf{0},\Sigma)$ having zero mean and the covariance matrix $\Sigma$ i.e.,  $\boldsymbol{\psi}_i\sim\mathcal{N}_M(\mathbf{0},\Sigma)$. The probability density function (pdf) $f_{\boldsymbol{\psi}}(\boldsymbol{\psi}_i)$ of the noise process is given below
\begin{equation}
f_{\boldsymbol{\psi}}(\boldsymbol{\psi}_i) = \frac{1}{\sqrt{(2\pi)^M\ |\Sigma|}}  \exp^{-\frac{1}{2}\left( \boldsymbol{\psi}_i^{T}\Sigma^{-1}\boldsymbol{\psi}_i\right)}.
\label{eq:mwgn}
\end{equation}
To estimate true variable $\mathbf{s}_i$ from noisy observations $\mathbf{x}_i$, multivariate signal denoising methods aim to suppress the multivariate noise without loss of signal details. 

%\subsection{Related Work}
Signal denoising methods are mostly designed for single-channel (univariate or 1D) data or 2D images \cite{vaseghi2008advanced}. Among those, multiscale denoising approaches based on wavelet transform have enjoyed great success in a wide range of practical applications. The utility of wavelet based denoising originates from the ability of the wavelet transform to effectively segregate signal and noise coefficients 
%	in such a way that signal coefficients exhibit sparse spread and higher amplitudes in contrast to the noise coefficients 
through sparse signal representation relative to noise \cite{mallat1999wavelet}. By utilizing that property, simple nonlinear thresholding functions based on, for instance, universal threshold $T_u = \sqrt{2\sigma log(N)}$ as a function of noise variance $\sigma^2$ \cite{donoho1995Visu} or statistical tools such as Bayesian local false discovery rate (BLFDR) \cite{lavrik2008BLFDR} and goodness of fit (GoF) test \cite{ur2017DWTGoF}, may be used to detect noise coefficients which can then either be completely or partially removed to yield denoised signal.

Given the success of wavelet based method in single-channel signal denoising, several multivariate extensions of those methods have been proposed over the years. These extensions are used as important preprocessing step in many applications involving multichannel or multivariate data \cite{jha2010MvDenApplication1,mostacci2010MWDapplication}. Multivariate data typically exhibit interchannel correlations, modeled through the covariance matrix $\Sigma$, as highlighted in \eqref{eq:mwgn} for multivariate wGn distribution. The goal of multivariate denoising is to incorporate interchannel correlations within the denoising process. Note that applying single-channel data denoising methods channel-wise on multivariate data is suboptimal as it completely ignores interchannel correlations, i.e., it considers $\Sigma$ to be a diagonal matrix.           

To this end, the first multivariate extension of wavelet denoising, termed as MWD, was introduced in \cite{aminghafari2006MWD}. Using that method, firstly, multiscale decomposition was obtained by applying DWT separately across all channels of the noisy multivariate signal. Then, universal thresholds were computed
for each channel separately by using $T_m =\sqrt{2\lambda_m\log{N}}$
where $\lambda_m$ denotes eigenvalues of the noise covariance matrix $\Sigma$
and $m = 1,\ldots,M$ denotes channel index. Subsequently, univariate thresholding is performed separately on each channel to detect noise coefficients which are subsequently discarded. 
         
A multivariate signal denoising method using time-frequency (T-F) reassignment based on synchrosqueezing transform (SST) has been presented in \cite{ahrabian2015MWSD}. The method can be seen as a generalization of a similar denoising method that was proposed to denoise single-channel (univariate) data \cite{meignen2012UWSD}. This method, referred to as MWSD in the sequel, exploits high resolution T-F representation to extract signal components using a thresholding technique; the partitioning of the T-F plane is performed in order to select a common set of \textit{multivariate modulated oscillations} within a multivariate data. 
%A simple universal thresholding function, given by $T = CT_u$, is used to reject noise and extracts multivariate oscillations using a scaled universal threshold $T_u$. 

Another method accomplishing the same goal employs multivariate empirical mode decomposition (MEMD) algorithm \cite{rehman2009MEMD} to obtain multiple data-driven scales/ modes of the noisy multivariate signal
%This technique directly employs the univariate interval thresholding  separately on each channel
% Again, the motivation behind this method was a 
that is followed by the use of single-channel interval thresholding (IT) on multiple scales/ modes corresponding to each channel \cite{hao2017MEMD-IT}. This method, called MEMD-IT in the sequel, is a straight forward multichannel extension of \cite{kopsinis2009EMD-IT} where IT was used to extract oscillatory signal parts from the intrinsic mode functions (IMF) of univariate EMD. 
Similarly, \cite{ur2019MMD} presents a new multivariate signal denoising method that performs interval thresholding on Mahalanobis distances (MDs) corresponding to the multivariate IMFs of MEMD. The main result in that paper is a theorem that underpins the extension of interval thresholding procedure on MD by providing an analytical relation between the stationary points of MD and derivatives of individual input data channels.

The majority of the above mentioned multivariate denoising methods operate by employing univariate thresholding functions \textit{channel-wise} to suppress multivariate noise. That is, the thresholding is performed on each channel separately thereby neglecting the multivariate nature of noise due to cross-channel correlations. That yields suboptimal denoising results and inconsistent performance across different channels. To alleviate this problem, it is important to devise fully multivariate thresholding strategies which fully take into account the cross-channel correlation structure of multivariate noise by operating in multidimensional space where multivariate signal resides.    

To achieve that goal, we propose a novel multivariate signal denoising method that is based on a fully multivariate thresholding function. We make use of the squared Mahalanobis distance (MD) measure that maps a multivariate data residing in $\mathcal{R}^M$ to a univariate data in $\mathcal{R}_+$, i.e., $\mathcal{R}^M\rightarrow \mathcal{R}_+$. The choice of using squared-MD measure in our work is motivated by the fact that distribution of MD-transformed data follows a one-to-one relation with that of input multivariate data \cite{mcassey2013empiricalMGOF}. Consequently, we define a multivariate goodness of fit (GoF) test for normality based on the MD-transformed data. 
In this regard, a modification of the robust Anderson Darling (AD) test statistic \cite{anderson1954ADtest,stephens1974EDFGoFComparisons} has been proposed based on EDF of squared-MDs to test for multivariate normality. Subsequently, we propose to apply that test on coefficients corresponding to multiple scales obtained from applying DWT to input noisy data. A multivariate thresholding function is then operated across all channels at multiple scales to obtain the denoised coefficients. Finally, the denoised signal is obtained by taking the inverse DWT of the thresholded coefficients.                       

The proposed denoising method can be seen as multivariate extension of the univariate denoising method proposed in \cite{ur2017DWTGoF}. In \cite{ur2017DWTGoF}, a standard GoF test based on EDF statistic was applied to multiple scales obtained from DWT to perform multiscale denoising. Our proposed work is not a straight forward extension of \cite{ur2017DWTGoF} to multivariate data due to the following novel contributions: i) we employ multivariate GoF test based on Squared-MD measure to map multivariate data to univariate distances while preserving important characteristics of the input multivariate data; ii) we employ a unique definition of empirical multivariate CDF based on squared-MD, as a quadratic transformation of multivariate Gaussian random process, which is convenient to compute and process using the proposed multivariate denoising framework based on hypothesis testing; iii) unlike other multivariate denoising methods \cite{aminghafari2006MWD,ahrabian2015MWSD,hao2017MEMD-IT}, cross-channel correlation structures within input noise are fully incorporated in the proposed method owing to the fully multivariate thresholding function. 

The proposed approach is also fundamentally and significantly different from our previous work related to multivariate denoising \cite{ur2019MMD} that utilizes the properties of MD measure to extend interval thresholding operation within single-channel EMD to multivariate EMD. In \cite{ur2019MMD}, an analytical relation between the stationary points of MD measure and derivatives of individual input data channels is given that justifies the extension of interval thresholding at multiple scales (obtained via MEMD) to multichannel data. On the other hand, this work exploits the (EDF) statistics of squared-MD (or quadratic transformation) to define a novel multivariate GoF test that is subsequently applied at multiple scales obtained from DWT to perform multivariate denoising.                  
	
Furthermore, the proposed GoF test develops key theoretical results involving (i) specification of the reference CDF as a quadratic transformation (squared-MD) of multivariate normal random variables and (ii) its use to derive a modified AD statistics to check for multivariate normality. Therefore, proposed test is completely different from the empirical GoF test in \cite{mcassey2013empiricalMGOF} that computes Mahalanobis distances (MDs) of the multivariate data followed by the application of a test statistic that is based on global deviation between the assumed and the observed EDFs of the MDs.

The paper is organized as follows: The current introduction section is followed by the review of multivariate GoF tests and related topics in Section II. The proposed methodology is presented in Section III, while Section IV presents detailed experimental results and related discussion. Finally, conclusions and avenues for future work are presented in Section V.

\section{Goodness of Fit Tests for Normality}
A goodness of fit (GoF) test checks whether a given set of observations originate from a specified reference distribution model. Within GoF tests, a \textit{test statistic} serves to quantify the difference between the observed and assumed (reference) distribution. For univariate data, GoF tests based on EDF are popular in many engineering applications \cite{ur2017DWTGoF,lei2011SS1} owing to their uniqueness and ease of computation. Few examples of popular test statistics based on EDF include 
%Kolmogrov-Simrnov (KS) statistic \cite{smirnov1948KS}, 
Cramer-Von-Mises (CVM) statistic \cite{cramer1928CVMtest} and Anderson Darling (AD) statistic \cite{anderson1954ADtest}.  In these statistics, reference distribution is modeled through cumulative distribution function (CDF). 

Several attempts have been made to extend GoF tests to check for normality in a multivariate data e.g., Pearson chi-square test \cite{moore1981ChiPearson}, Shapiro-Wilk test \cite{royston1983SPWILK}, multivariate skewness and kurtosis tests \cite{malkovich1973SkewKurt}.
%and tests based on Rosenblatt and Box-Cox transformations \cite{velilla1995BoxCox}. 
However, none of these have found a widespread applicability due to the following bottlenecks: i) these are computationally too expensive to compute or even intractable \cite{mcassey2013empiricalMGOF}; ii) their extension to generalized multivariate distribution case are not available.
To address these issues, an empirical GoF test based on MD for multivariate distributions was proposed in \cite{mcassey2013empiricalMGOF}, however, that was highly sensitive to the size of input data and failed for smaller sized data.

\subsection{Multivariate GoF Tests for Normality}

Similar to the univariate case, multivariate GoF tests measure how well an observed multivariate data coincides with the assumed multivariate data model \cite{d2017goodness}. These tests require a measure of fit, known as test statistic, to quantify the difference between the empirical distribution model of the observed data and the assumed reference model. Finally, hypothesis testing framework is employed to check (statistically) whether the observed data belongs to the assumed model or not.
Just like the univariate case where EDF can be defined uniquely, the multivariate GoF tests also require a unique definition of an empirical distribution model for multivariate observations. That is challenging though since multivariate EDF has multiple possible definitions \cite{hanebeck2008LCD1}, as stated below:
\begin{definition}[Multivariate EDF] 
	Let $\mathbf{x}_i=[x_{i}^{(1)}, x_{i}^{(2)}, \ \ldots, \ \\x_i^{(M)}]\in\mathcal{R}^{M}$ denote the $i$th observation of a multivariate signal with $M$ channels and size $N$, a multivariate EDF $\mathcal{F}\left({\color{blue}\mathrm{t}_1, \ \mathrm{t}_2, \ldots,\mathrm{t}_M} \right) : \mathcal{R}^M \rightarrow \mathcal{R}^M$ may be defined as follows: 
	\begin{equation}
	\begin{split}
	\mathcal{F} & \left(\mathrm{t}_1, \ \mathrm{t}_2, \ldots,\mathrm{t}_M\right) =\\ 
	& \frac{1}{N}\sum_{i=1}^N \ \mathbf{1}.\left(x_{i}^{(1)} \leq \mathrm{t}_1, x_{i}^{(2)} \leq \mathrm{t}_2 \ \ldots \ x_{i}^{(M)} \leq \mathrm{t}_M\right)
	\end{split}
	\label{medf}
	\end{equation}
	%\vspace{-2mm}
	%	\noindent where .
	However, \eqref{medf} is one of the several ways to define a multivariate EDF. This is because an EDF is a discrete approximation of the CDF which can be defined in $2^M$ ways for a given dataset with $M$ dimensions, consequently, an EDF is also defined in $2^M$ ways \cite{hanebeck2009LCD2}. For instance, four possible EDFs for bivariate case are specified as follows
	\begin{equation}
	\mathcal{F}_1 (\mathrm{t}_1, \ \mathrm{t}_2) = \frac{1}{N}\sum_{i=1}^N \ \mathbf{1}.(x_{i}^{(1)} \leq \mathrm{t}_1, x_{i}^{(2)}\leq \mathrm{t}_2),
	\end{equation}
	\begin{equation}
	\mathcal{F}_2(\mathrm{t}_1, \ \mathrm{t}_2) =\frac{1}{N}\sum_{i=1}^N \ \mathbf{1}.(x_{i}^{(1)} > \mathrm{t}_1, x_{i}^{(2)} > \mathrm{t}_2),
	\end{equation}
	\begin{equation}
	\mathcal{F}_3 (\mathrm{t}_1, \ \mathrm{t}_2) =  \frac{1}{N}\sum_{i=1}^N \ \mathbf{1}.(x_{i}^{(1)} \leq \mathrm{t}_1, x_{i}^{(2)} > \mathrm{t}_2),
	\end{equation}
	\begin{equation}
	\mathcal{F}_4 (\mathrm{t}_1, \ \mathrm{t}_2) = \frac{1}{N}\sum_{i=1}^N \ \mathbf{1}.(x_{i}^{(1)} > \mathrm{x}_1, x_{i}^{(2)} \leq \mathrm{x}_2).
	\label{biedf}
	\end{equation}
\end{definition}
%\vspace{-2mm}

Since multivariate EDF has multiple definitions, testing for all possible EDFs is essential for accurate and robust decision, yet that is cumbersome and impractical. Therefore, an alternate representation of multivariate pdf was proposed in \cite{hanebeck2008LCD1,hanebeck2009LCD2} whereby multidimensional pdf $g(\mathbf{x})$ was integrated across symmetric kernel functions $\mathcal{K}$ to yield a unique localised cumulative distribution (LCD) $\mathcal{L}(\mathbf{m},\mathbf{a})$
\begin{equation}
\mathcal{L}(\mathbf{m},\mathbf{a}) = \int_{\mathcal{R}^N} g(\mathbf{x}_i) \mathcal{K}(\mathbf{x}_i-\mathbf{m},\mathbf{a})d\mathbf{x}_i,
\label{eq:lcd}
\end{equation}
\noindent where  $\mathcal{K}(.,.) \in \mathcal{R}_+ \rightarrow [0,1]$ is a symmetric and integrable kernel located at position $\mathbf{m} \in \mathcal{R}^M$ having width characterized by the vector $\mathbf{a}$. Subsequently, the following multivariate test statistic $\gamma$ based on modified Cramer Von Mises (CVM) measure was proposed: 
\begin{equation}
\gamma = \int_{\mathcal{R}^N} h(\mathbf{a}) \int_{\mathcal{R}^N} (\tilde{\mathcal{L}}(\mathbf{m},\mathbf{a}))  - \mathcal{L}(\mathbf{m},\mathbf{a})) \ d\mathbf{m} \ d\mathbf{a},
\label{eq:cvm}
\end{equation}
\noindent where $\tilde{\mathcal{L}}(\mathbf{m},\mathbf{a})$ is reference LCD and $h(\mathbf{a})$ is a suitable weighting function.

Having defined a unique estimate of the multivariate CDF (i.e., LCD) and test statistic for multivariate data via \eqref{eq:lcd} and \eqref{eq:cvm} respectively, a binary hypothesis testing framework could be used to check how well given multivariate observations fit a reference model. However, a downside of LCD is its enormous computational cost which prohibits its use in practical applications, e.g., even for a reference multivariate Gaussian distribution, an approximation of \eqref{eq:cvm}, given in \cite{hanebeck2009LCD2}, requires quite large computational resources.
%\vspace{-5mm}
\subsection{Mahalanobis Distance }
In this section, we recall the definition and some important properties of Mahalanobis distance (MD) measure. In our proposed work, we are interested in distribution of squared MD values obtained from observations originating from reference multivariate normal distribution. To this end, we illustrate the quadratic transformation of multivariate random variables that paves the way for an important Theorem 1 in the next section.        
\begin{definition}[Mahalanobis distance]
Let $\mathbf{x}_i=[x_{i}^{(1)},\cdots,x_{i}^{(M)}]\in\mathcal{R}^M$ denote a random multivariate observation from a set of multivariate observations of size $N$. The Mahalanobis distance (MD) $\Delta_i$ between the $i$th observation $\mathbf{x}_i$ and the mean of the multivariate observations $\boldsymbol{\mu}=E[\mathbf{x}_i;\ \forall \ i=1,\cdots, N]$ is defined as %\in\mathcal{R}^M$ is defined as 
\begin{equation}
\Delta_i =\sqrt{(\mathbf{x}_i-\boldsymbol{\mu})^{T}\Sigma^{-1}(\mathbf{x}_i-\boldsymbol{\mu})}.
\label{eq:MD}
\end{equation}
\noindent where $\Sigma$ denotes the covariance matrix which characterizes the inter-channel dependencies within the random variables.
\end{definition}

\begin{remark}
MDs corresponding to data from a multivariate probability distribution function follow a distinct probability distribution \cite{mcassey2013empiricalMGOF}.
\end{remark}

\begin{remark}
	Squared MD is a quadratic transformation of multivariate random observations $\mathbf{x}_i$ through the covariance matrix $\Sigma$.
\end{remark}

In order to illustrate this remark, we first define a quadratic transformation of random vector $x$ as follows: 
\begin{definition}[Quadratic Transformation of Random Variables]
	Let $\mathbf{x}$ denote a \textit{real} vector of $P$ random observations $\{x_1,\cdots,x_P\}$ then the quadratic transformation of random variables $Q(\mathbf{x})$ is defined as 
	\begin{equation}
	%	\begin{split}
	Q(\mathbf{x}) =\mathbf{x}^T A \mathbf{x}
	%	&= \sum_{m=1}^{P} \sum_{n=1}^{P} a_{m,n} \ x_m \ x_n
	%	\end{split}
	\label{QuadForm1}
	\end{equation}
	where $A\in \mathcal{R}^{P\times P}$ is a real, symmetric and positive definite matrix such that $A=A^T>0$.
\end{definition}

From \eqref{eq:MD}, the squared MD for random variables $\mathbf{x}_i$ is given by
\begin{equation}
\Delta_i^2=(\mathbf{x}_i-\boldsymbol{\mu})^{T}\Sigma^{-1}(\mathbf{x}_i-\boldsymbol{\mu})
\label{eq:MSD}
\end{equation}
where $\Sigma$ is a symmetric and positive definite matrix i.e., $\Sigma=\Sigma^T\succ0$ which further implies that ${\Sigma^{-1}=\Sigma^{-1}}^T\succ0$.
We next recall that spectral decomposition of $\Sigma^{-1}$ can be performed using its eigenvalues and corresponding eigenvectors 

\begin{definition}[Spectral decomposition of $\Sigma^{-1}$]
	%\begin{definition}[Symmetric]
	Given a symmetric and positive definite matrix $\Sigma^{-1}$, there exists an orthogonal matrix $B$ such that $B^TB=I$. The spectral decomposition of $\Sigma^{-1}$ can then be achieved, as given below
	\begin{equation}
	B^T \Sigma^{-1} B = diag(\lambda_1,\lambda_2, \cdots. \lambda_P)=
	\begin{bmatrix}
	\lambda_1 & 0 & 0 & 0\\ 
	0& \lambda_2 & 0 & \vdots\\ 
	\vdots &  & \ddots & \vdots \\
	0 & \cdots & \cdots & \lambda_M\\ 
	\end{bmatrix}
	\label{EigenDec}
	\end{equation}
	where $\boldsymbol{\lambda}=[\lambda_1, \cdots, \lambda_M]^T$ are the $M$ distinct eigenvalues of $\Sigma^{-1}$ obtained by solving $|\Sigma^{-1}-\boldsymbol{\lambda}B|=0$. The columns of matrix $B$ contain the $M$ eigenvectors of $\Sigma^{-1}$.
\end{definition}

\section{Proposed Multivariate Denoising Framework}
%\vspace{-2mm}
In this section, we present a novel signal denoising method for multivariate data; a new multivariate data normality test based on quadratic transformation MD and AD statistic which underpins the proposed denoising method is explained first.     
%\vspace{-2mm}
\subsection{Multivariate GOF test based on MD and AD statistic} 
%\vspace{-1mm}
In a multivariate GoF test, statistically significant difference between the EDFs of observed multivariate data and reference distribution function is sought. To perform this task effectively, unique definition of multivariate EDF is required. That is challenging since typical multivariate EDF representations, described in \eqref{medf}-\eqref{biedf}, are not unique. While some kind of averaging over all possible EDFs is an option \cite{hanebeck2008LCD1}, the resulting representation is cumbersome and computationally expensive. 

To address that problem, we propose to first transform multivariate data in $R^M$ to univariate space $R_+$ through Mahalanobis distance measure. Since there is a one-to-one relation between the empirical distribution functions of a given multivariate dataset and its corresponding MDs \cite{mcassey2013empiricalMGOF}, it suffices to utilize the EDF of MDs rather than the cumbersome EDFs of corresponding input multivariate data to perform the multivariate GoF test. {A graphical explanation of the rationale behind the use of squared-MD for definition of a valid multivariate GoF test for normality is given in Fig. \ref{MD-Rationale}. Specifically, it is highlighted in the figure that had we used standard approaches for computing multivariate EDFs in our work, no unique multivariate EDF would have been available for us to compare against the reference EDF for multivariate GoF testing. That is because $2^M$ number of CDFs or EDFs may be defined for an M-variate signal thus making it difficult to decide which definition of EDF should be used. Hence, the GoF test for normality - an integral part of our denoising method - could not have been defined properly had we used the standard EDF definition.}

In our case, the \textit{reference EDF} for multivariate wGn can be given by a quadratic transformation of Gaussian random vector $\mathbf{x}$ via squared-MD, as given in \eqref{eq:MSD}. That is followed by the estimation of local EDF of observed multivariate noisy samples of input data, which we term as \textit{test EDF} in the sequel.
We then propose a modified Anderson Darling (AD) test statistic to quantify the statistical similarity between the \textit{test EDF} and the \textit{reference EDF}. Finally, hypothesis testing is performed by selecting a threshold over the AD test statistic to test the observed data for multivariate normality. We illustrate each of these steps in detail in the following subsections: 

\begin{figure}[t!]
	%	\begin{minipage}[b]{1\linewidth}
		\centering
		
	\includegraphics[width=9cm]{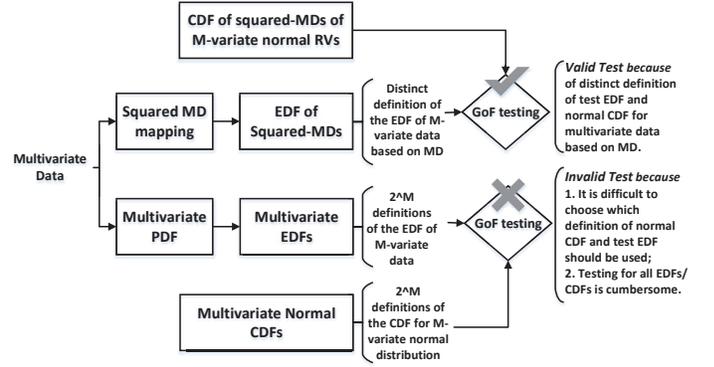}
	%  \vspace{2.0cm}
	%\centerline{(a) correlated bivariate data}\medskip
	%	\end{minipage}
	%	\vspace{-4mm}
	\caption{Rationale of the proposed GoF test based on MD.}
	\label{MD-Rationale}
	\vspace{-3mm}
\end{figure}
\subsubsection{Specification of the Reference CDF}

Our proposed test checks whether observed multivariate samples come from multivariate Gaussian (normal) random vectors. As mentioned before, the crux of our method is the transformation of input multivariate data samples to univariate signal, i.e., $\mathcal{R}^M\rightarrow \mathcal{R}_+$, through a quadratic transformation via squared-MD measure. Therefore, we first specify the reference distribution model $\mathcal{F}_0(\cdot)$ based on the quadratic transformation of multivariate Gaussian random vectors.
\begin{theorem}
	\textit{Let $\boldsymbol{\psi}_i$ denote a vector-valued Gaussian random variable, i.e., $\boldsymbol{\psi}_i\sim\mathcal{N}_M(\mathbf{0},\Sigma)$, where $\Sigma$ is symmetric and positive definite (and therefore $\Sigma^{-1}$ is also symmetric and positive definite). Given that eigenvalues $\boldsymbol{\lambda}$ of $\Sigma^{-1}$ are distinct, the pdf $f_{\mathbf{y}}(y)$ and CDF $\mathcal{F}_0(t)$ of the quadratic transformation $y = \boldsymbol{\psi}^T \Sigma^{-1} \boldsymbol{\psi}$ are given from \cite{mathai1992quadraticTx}, as follows}
	\begin{equation}
	f_{\mathbf{y}}(y) = \sum_{n=0}^{\infty} (-1)^n \ c_n \frac{y^{\frac{M}{2}+n-1}}{\Gamma(\frac{M}{2}+n)}, \ \ \ 0 < y < \infty
	\label{nullpdf}
	\end{equation}
	\begin{equation}
	\mathcal{F}_0(t) = \sum_{n=0}^{\infty} (-1)^n \ c_n \frac{t^{\frac{M}{2}+n}}{\Gamma(\frac{M}{2}+n+1)}, \ \ \ y\leq t < \infty
	\label{nullCDF}
	\end{equation}
	\textit{where $\Gamma(\cdot)$ is Gamma function and the coefficients $c_n$ for $n=0$ and $n\geq 1$ are respectively given below}
	\begin{equation}
	c_0 = \prod_{m=0}^{M}(2\lambda_m)^{-\frac{1}{2}},
	\end{equation}
	%	and the coefficients $c_n $ for  are given as follows
	\begin{equation}
	c_n = \frac{1}{n}\sum_{r=0}^{n-1} h_{n-r}\ c_r, \ \ \ \ n \geq 1,
	\end{equation}
	\textit{where $h_m$ are given as follows}
	\begin{equation}
	h_m = \frac{1}{2}\sum_{m=0}^{M} (2\lambda_m)^{-m}, \ \ \ \ n \geq 1.
	\end{equation}
\end{theorem}

The derivation of \eqref{nullpdf} and \eqref{nullCDF} is given in \cite{mathai1992quadraticTx} where the relations for the parameters $c_n$ are also given.

\begin{figure}[t!]
%	\begin{minipage}[b]{1\linewidth}
	\centering
	\includegraphics[width=8.5cm]{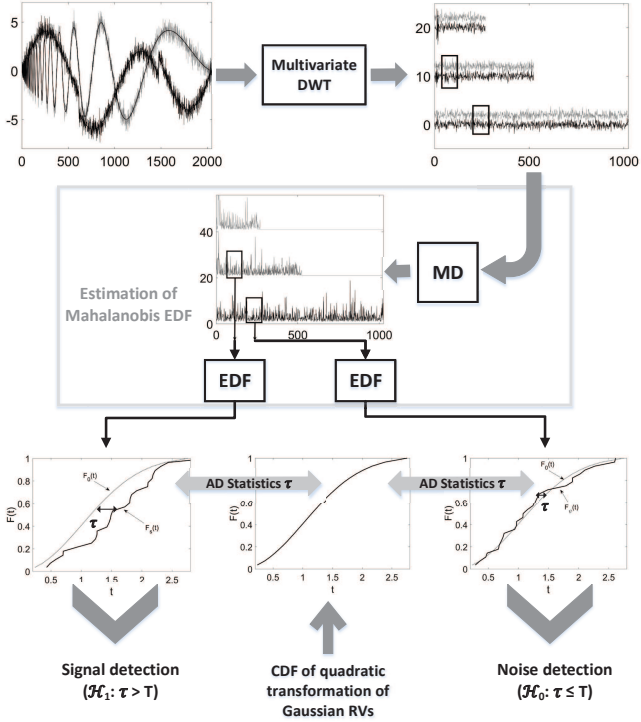}
	\caption{Graphical illustration of the proposed method. Firstly, multiscale decomposition of bivariate signal is computed via DWT (top row). That is followed by estimation of \textit{test EDF}  locally (observe the windows in the figure) using the Mahalanobis distance (MD) measure (middle row) at multiple data scales. In the bottom row, AD statistic is computed which is a function of the \textit{test EDF} estimated above and the \textit{reference EDF} that is given by the relation \eqref{nullCDF}. In case $\tau\leq T$, noise is detected at the given scale and the corresponding coefficients are discarded (set to zero), otherwise we deduce that the coefficients belong to the signal and are therefore retained.}
	\label{Ch5:BlockDaig}
		\vspace{-3mm}
\end{figure}

\subsubsection{Mahalanobis EDF}
We define a unique EDF of multivariate data, termed as \textit{Mahalanobis EDF} or test EDF, based on the quadratic transformation of multivariate observations.
%\vspace{-1mm}
\begin{definition}[Mahalanobis EDF] 
	Let $\mathbf{x}_i \in \mathcal{R}^M \ \forall \ i=1,\cdots,N$ denote a set of $N$ \textit{zero-mean} multivariate measurements, then test EDF $\mathcal{F}(t)$ based on squared-MD is defined as
	\begin{equation}
	\mathcal{F}(t) = \frac{1}{N}\sum_{i=1}^N \ \mathbf{1}.\left(\mathbf{x}_i^{T} \ \Sigma^{-1}\ \mathbf{x}_i \leq t\right),
	\label{EDFMD}
	\end{equation}
	\noindent where $\Sigma$ is the covariance matrix of the multivariate data $\mathbf{x}_i \ \forall \ i=1,\cdots,N$ and $\mathcal{F}(t) : \mathcal{R}^M \rightarrow \mathcal{R}_+$ denotes the EDF of multivariate data defined over support $t$. 
\end{definition}

Uniqueness of the EDF of MD measure, $\mathcal{F}(t)$, for a given multivariate signal (or probability distribution) enables us to define a multivariate GoF test utilizing $\mathcal{F}(t)$ given in \eqref{EDFMD}, instead of the cumbersome multidimensional EDF in \eqref{medf}. %We next quantify the similarity between the \textit{test EDF} \eqref{EDFMD} and the \textit{reference EDF} \eqref{nullCDF} through a robust AD test statistic which is defined next.

\subsubsection{A Modified AD Statistic}

We next quantify the similarity between the \textit{test EDF} \eqref{EDFMD} and the \textit{reference EDF} \eqref{nullCDF} through a robust Anderson Darling (AD) test statistic. AD statistic belongs to a class of EDF statistics that are based on the comparison between a reference CDF $\mathcal{F}_0(t)$ with the test EDF \cite{anderson1954ADtest}. Other notable test statistic within this class include KS and CVM statistic, though the AD statistic has been proven to be relatively more robust \cite{stephens1974EDFGoFComparisons,d2017goodness}.

%In our case, we employ AD statistic to quantify the statistical difference between the reference CDF $\mathcal{F}_0(t)$ \eqref{nullCDF} and the test (Mahalanobis) EDF as defined in \eqref{EDFMD}. 

In our case, the AD statistic $\tau$ between the \textit{test EDF} \eqref{EDFMD} and the \textit{reference EDF} \eqref{nullCDF} is computed via
\begin{equation}
\label{ADmv}
\tau = \int_{-\infty}^{\infty} \left(\mathcal{F}_0(t)-\mathcal{F}(t)\right)^2 \mho(\mathcal{F}_0(t))\ d\mathcal{F}_0(t),
\end{equation}
\noindent  where $\mho(\mathcal{F}_0(t))=\left(\mathcal{F}_0(t)(1-\mathcal{F}_0(t))\right)^{-1}$ is the weighting function employed to give more weight to the tail of the distribution function rendering flexibility to the statistic \cite{anderson1954ADtest}.

A computationally convenient numeric representation of \eqref{ADmv} is given from \cite{d2017goodness}, which in our case becomes  
\begin{equation}
\begin{split}
\label{ADmv1}
\tau = L-\sum_{l=1}^{\mathit{L+1}}& \frac{(2l-1)}{\mathit{L}}(ln(\mathcal{F}_0(\mathbf{x}_l^T\Sigma^{-1}\mathbf{x}_l)\\&-ln(\mathcal{F}_0(\mathbf{x}_{L+1-l}^T \Sigma^{-1} \mathbf{x}_{L+1-l}))), 
\end{split}
\end{equation}
\noindent By substituting \eqref{nullCDF} in \eqref{ADmv1}, numeric form of the proposed modified AD statistic to test for multivariate normality is obtained
\begin{equation}
\small
\begin{split}
\label{ADmv2}
\tau = L&-\sum_{l=1}^{\mathit{L+1}} \frac{(2l-1)}{\mathit{L}}\bigg[\bigg. ln\bigg(\sum_{n=0}^{\infty} (-1)^n \ c_n \frac{(\mathbf{x}_l^T\Sigma^{-1}\mathbf{x}_l)^{\frac{M}{2}+n}}{\Gamma(\frac{M}{2}+n+1)}\bigg)\\&-ln \bigg(\sum_{n=0}^{\infty} (-1)^n \ c_n \frac{(\mathbf{x}_{L+1-l}^T \Sigma^{-1} \mathbf{x}_{L+1-l})^{\frac{M}{2}+n}}{\Gamma(\frac{M}{2}+n+1)}\bigg)\bigg. \bigg], 
\end{split}
\end{equation}
\noindent where $L$ denotes the length of the input data.
\subsubsection{Hypothesis testing}
Having quantified the difference $\tau$ between estimated test EDF $\mathcal{F}(t)$ \eqref{EDFMD} and the reference CDF $\mathcal{F}_0(t)$ \eqref{nullCDF}, we use the following \textit{hypothesis testing} framework based on GoF test to check for the normality of given multivariate data observations
\begin{eqnarray}
\begin{split}
&\mathcal{H}_0: \ \ \mathcal{F}(t) \cong \mathcal{F}_0(t) \ \ \Leftrightarrow \ \ \  \tau<T, \nonumber \\
%\text{                       vs.} \nonumber \\
&\mathcal{H}_1: \ \ \mathcal{F}(t) \ncong \mathcal{F}_0(t) \ \ \Leftrightarrow \ \ \ \tau\geq T,
\end{split}
\label{hypotestmulti}
\end{eqnarray} 
\noindent where $\mathcal{H}_0$ denotes the null hypothesis, i.e., the case where observations fit the reference model of multivariate normal distribution and $\mathcal{H}_1$ denotes the alternate hypothesis; $T$ denotes the threshold value.

The selection of the threshold value $T$ is vital in the above framework. It is typically selected as a function of a given \textit{probability of false alarm}, $P_{fa}$, which corresponds to the probability of falsely choosing the alternate hypothesis $\mathcal{H}_1$ whereas observations actually belonged to the null hypothesis $\mathcal{H}_0$, i.e., data originated from reference distribution but the method decided otherwise. Mathematically, the relation for $P_{fa}$ is given below
% The threshold $T$ is selected as a function of very small probability of false alarm $P_{fa}$ given as follows
\begin{equation}
\begin{split}
P_{fa} = Prob(\mathcal{H}_1 \ | \mathcal{H}_0)=& \ Prob(\tau>T \ | \mathcal{H}_0) \\ &=\int_{\{y \ | \ \tau>T\}} p(y|\mathcal{H}_0)\mathtt{d}y,
\end{split}
\label{PfaVT}
\end{equation}
where $y_i=\boldsymbol{\psi}_i^T\Sigma^{-1}\boldsymbol{\psi}_i$ correspond to quadratic transformation of Gaussian random variables and therefore pdf $p(y|\mathcal{H}_0)=f_{\mathbf{y}}(y)$, as given in \eqref{nullpdf}. Consequently, $P_{fa}$ can be written as
\begin{equation}
P_{fa} = \int_{\{y; \ \tau>T \ | \ \mathcal{H}_0\}} f_{\mathbf{y}}(y) \ \mathtt{d}y,
\label{PfaMv}
\end{equation}
where ${\{y; \ \tau>T \ | \ \mathcal{H}_0\}}$ are the MDs corresponding to Gaussian or normally distributed multivariate observations for which the null hypothesis $\mathcal{H}_0$ has been falsely rejected. 
%Simply stated, threshold $T$ yielding least false rejections of $\mathcal{H}_0$ is selected.

Naturally, we are interested in selecting the threshold $T$ which yields the minimum value of $P_{fa}$. Typically, in hypothesis testing, the value of $P_{fa}$ is specified a priori depending on the requirements of a given application. In our case, we select $P_{fa}=\alpha$ in the range of $\alpha=10^{-3}-10^{-5}$.% and obtain the threshold $T$ accordingly. 

\subsection{Multivariate signal denoising based on squared-MD}
%\vspace{-1mm}

Traditionally, GoF tests have been utilized in detection problems, e.g., spectrum sensing in cognitive radio \cite{lei2011SS1,wang2009SS2}, since the framework facilitates the detection of noise ($\mathcal{H}_0$) and `signal plus noise' ($\mathcal{H}_1$). In \cite{ur2017DWTGoF}, that framework was modified to make it applicable to univariate signal denoising. That was achieved by performing the GoF based hypothesis testing at multiple scales obtained from DWT. Using that approach, observations detected as noise at multiple scales, i.e., corresponding to $\mathcal{H}_0$, were discarded while those associated with signal were retained to reconstruct the denoised signal.

Here, we present a multivariate extension of the denoising algorithm \cite{ur2017DWTGoF} that is based on the multivariate GoF test presented in the previous section. We specifically consider the multivariate denoising problem given in \eqref{Ch5:noisy} where noise $\boldsymbol{\psi}_i$ is modeled as independent and identically distributed by zero mean multivariate Gaussian distribution $\mathcal{N}(\mathbf{0},\Sigma)$, with the covariance matrix $\Sigma$. We propose a multiscale approach for data denoising based on DWT, similar to \cite{ur2017DWTGoF}, owing to the following properties of DWT: i) the distribution of multivariate Gaussian noise samples in the transform domain is preserved; ii) DWT yields sparse signal representation thus enabling suitable segregation between noise and signal coefficients in the transform domain. The multivariate GoF hypothesis test, given in the previous section, can therefore be used to test for the presence of noise at multiple scales provided the noise distribution at multiple scales is known a priori. 

Let $\mathcal{T}$ denotes a multivariate DWT transform, then multiscale decomposition of the noisy signal $\mathbf{X} = \{\mathbf{x}_{i^{'}} \ \forall \ {i^{'}}=1,\ldots,N\}$ is obtained as follows
\begin{equation}
\mathbf{d}_i^k = \mathcal{T}(\mathbf{X}), \ \ \forall \ i = 1,\ldots,N/2^k,
\label{muldec}
\end{equation}
\noindent where $\mathbf{d}_i^k = [d_{i^k}^{(1)}, d_{i^k}^{(2)}, \cdots, d_{i^k}^{(M)}]$ is a multivariate wavelet coefficient at the scale $k$ and index $i$, which is obtained using DWT on a noisy multivariate signal, depicted in Fig. \ref{Ch5:BlockDaig} (top).

Owing to the properties of the DWT, $\mathbf{d}_i^k$ must either correspond to noise or signal coefficient. Moreover, the linearity of the DWT operation ensures that the coefficients corresponding to noise still exhibit multivariate Gaussian or Normal distribution at multiple data scales, see \textit{Theorem 2}. Consequently, the GoF test for multivariate normality (described in the previous section) is employed on multiple scales to detect whether the coefficients $\mathbf{d}_i^k$ originated from the reference multivariate Gaussian distribution $\mathcal{N}_M(\mathbf{0},\Sigma)$ or not. The coefficients for which the null hypothesis $\mathcal{H}_0: \mathbf{d}_i^k\sim\mathcal{N}_M(\mathbf{0},\Sigma)$ is satisfied are set to zero while the remaining coefficients are retained.%, see Fig. \ref{Ch5:BlockDaig} (lower center).

To implement the multivariate GoF test based on MD measure, reference EDF given by $\mathcal{F}_0(t)$, in \eqref{nullCDF}, is utilized. To compute $\mathcal{F}_0(t)$, estimation of noise covariance matrix $\Sigma$ is first performed from noisy observations. In this regard, we use the minimum covariance determinant (MCD) estimator \cite{rousseeuw1999MCD} on the first-scale DWT coefficients $\{\mathbf{d}_i^1;\forall \ i=1,\cdots,N/2\}$ which are mostly composed of noise \cite{aminghafari2006MWD}. Owing to its robustness to outliers, MCD is able to disregard the presence of signal traces within the first scale coefficients to estimate a robust covariance matrix as follows
\begin{equation}
\Sigma = MCD(\{\mathbf{d}_i^1; \ \forall \ i = 1 \ ... \ N/2\}). 
\label{cov}
\end{equation}
Moreover, the orthogonality of DWT retains the original covariances of noisy data in each scale \cite{mallat1999wavelet}, hence, the covariance of noise at each multichannel scale remains $\Sigma$.

To compute the \textit{test EDF} locally at multiple scales, spatial windows of multivariate wavelet coefficients $\mathbf{W}_i^k=[\mathbf{d}_{i-L/2}^k, \cdots, \mathbf{d}_i^k, \cdots, \mathbf{d}_{i+L/2}^k]^T$ of size $(L+1) \times M$ are chosen around each coefficient $\mathbf{d}_i^k$, as depicted in Fig. \ref{Ch5:BlockDaig} (top right). That is followed by the computation of Mahalanobis EDF $\mathcal{F}_i^k(t)$ for each window $\mathbf{W}_i^k$ using \eqref{EDFMD}. Next, the AD statistic $\tau_i^k$ is computed between reference CDF $\mathcal{F}_0(t)$ and test EDF $\mathcal{F}_i^k(t)$ for each window $\mathbf{W}_i^k$ at scale $k$, using \eqref{ADmv2}. 

A scale adaptive threshold value $T_k$ is estimated using \eqref{PfaMv} where computation of the integral requires estimation of $\{y_i; \ \tau_i^k>T_k\ |\ \mathcal{H}_0\}$ which are quadratic transformations $y_i={\mathbf{d}_{i}^{'k}}^T\Sigma^{-1}\mathbf{d}_{i}^{'k}$ of multiscale noise coefficients  $\mathbf{d}_{i}^{'k}=\mathcal{T}(\boldsymbol{\psi}_i)$ that are falsely detected as signal, i.e., false detection of $\mathcal{H}_1$ given $\mathcal{H}_0$. To this end, noise samples $\boldsymbol{\psi}_i$ are sampled from multivariate normal distribution parameterized by the covariance matrix $\Sigma$ estimated in \eqref{cov} that helps to imitate the actual noise samples in the noisy signal and subsequently estimate a threshold for very small false alarm probability $P_{fa}$.%The threshold $T_k$ yielding least false alarms is estimated using \eqref{PfaMv} which defines the significance level of the test statistics to reject the null hypothesis.
\begin{figure}[t]
	\centering
	\begin{subfigure}{.24\textwidth}
		\centering
		\includegraphics[width=\linewidth]{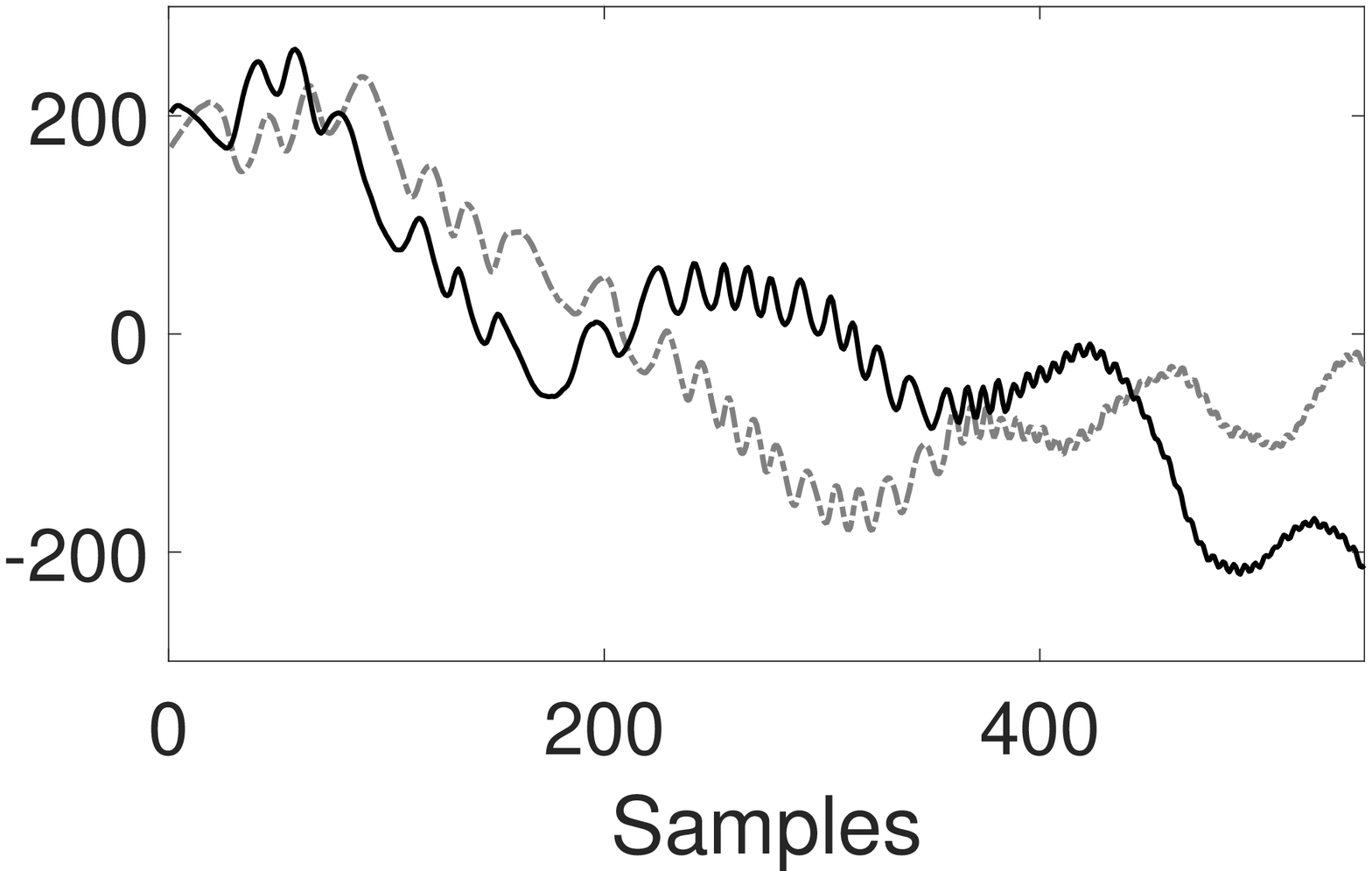}
		\caption{}
%		\label{Sofar}
	\end{subfigure}
	%
	%	\hspace{-8mm}
	\begin{subfigure}{.24\textwidth}
		\centering
		\includegraphics[width=\linewidth]{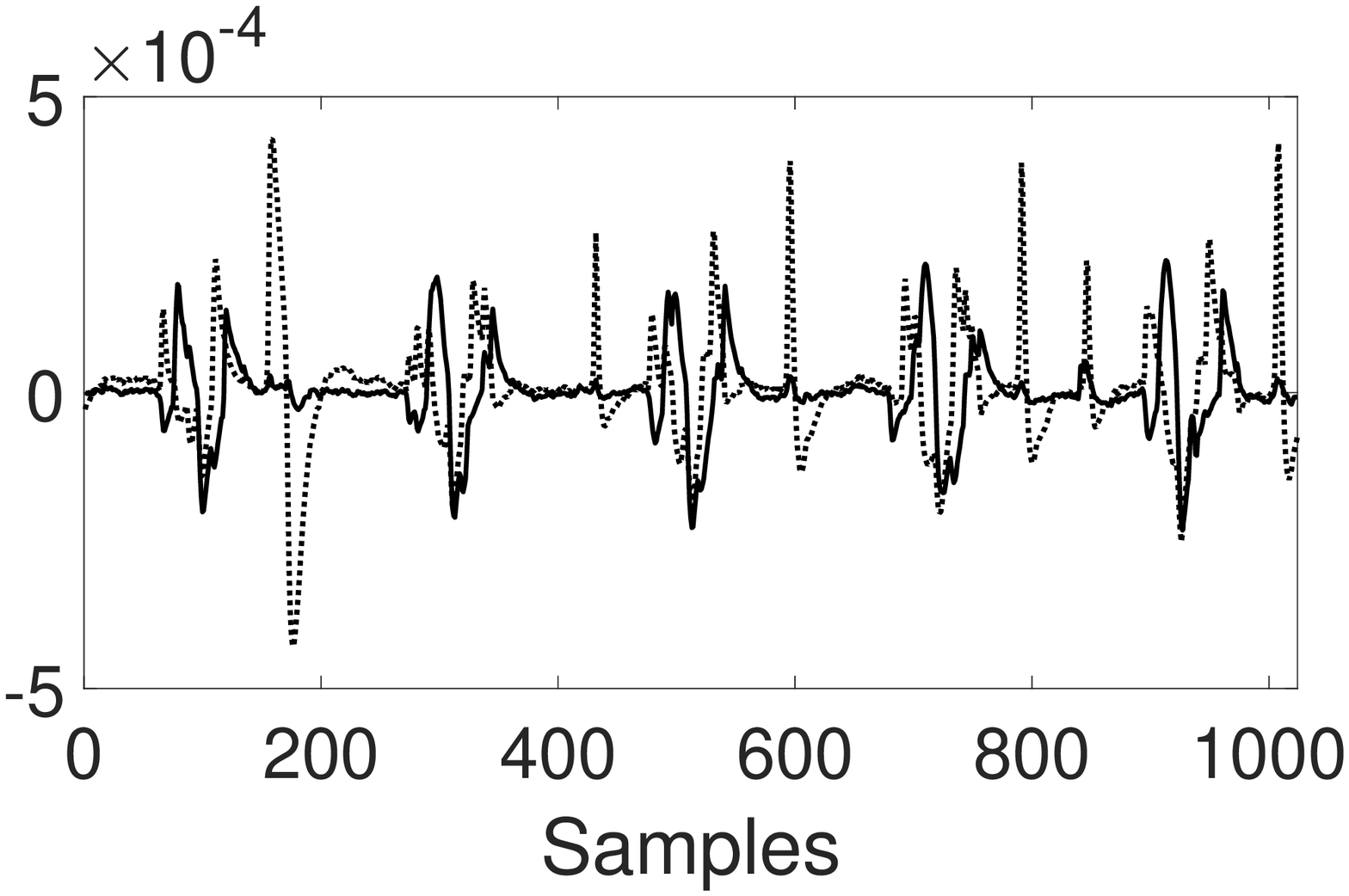}
		\caption{}
		\label{EOG}
	\end{subfigure}
	
	\begin{subfigure}{.24\textwidth}
		\centering
		\includegraphics[width=\linewidth]{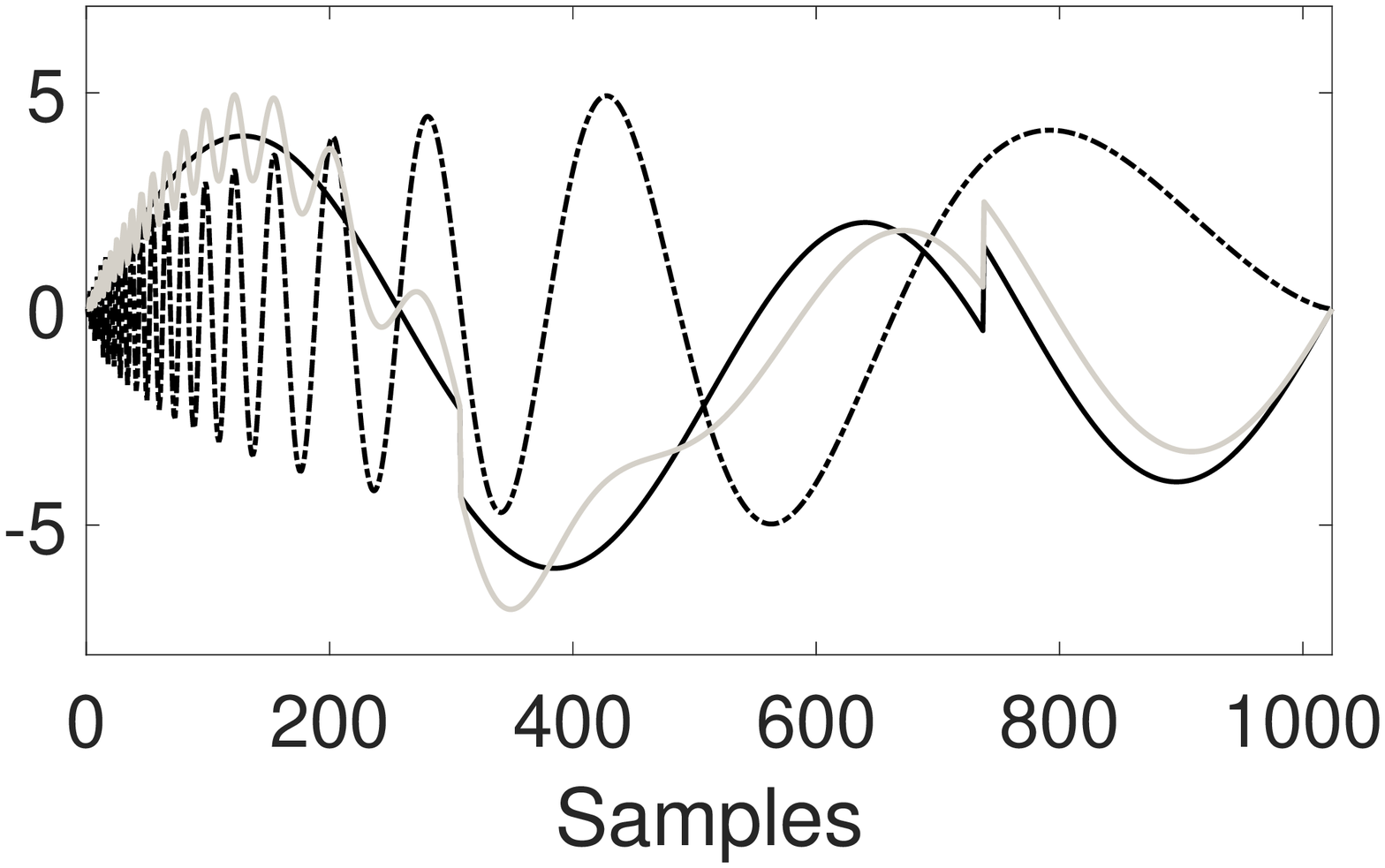}
		\caption{}
		\label{HvyDopp}
	\end{subfigure}
	%	\hspace{-8mm}
	\begin{subfigure}{.24\textwidth}
		\centering
		\includegraphics[width=\linewidth]{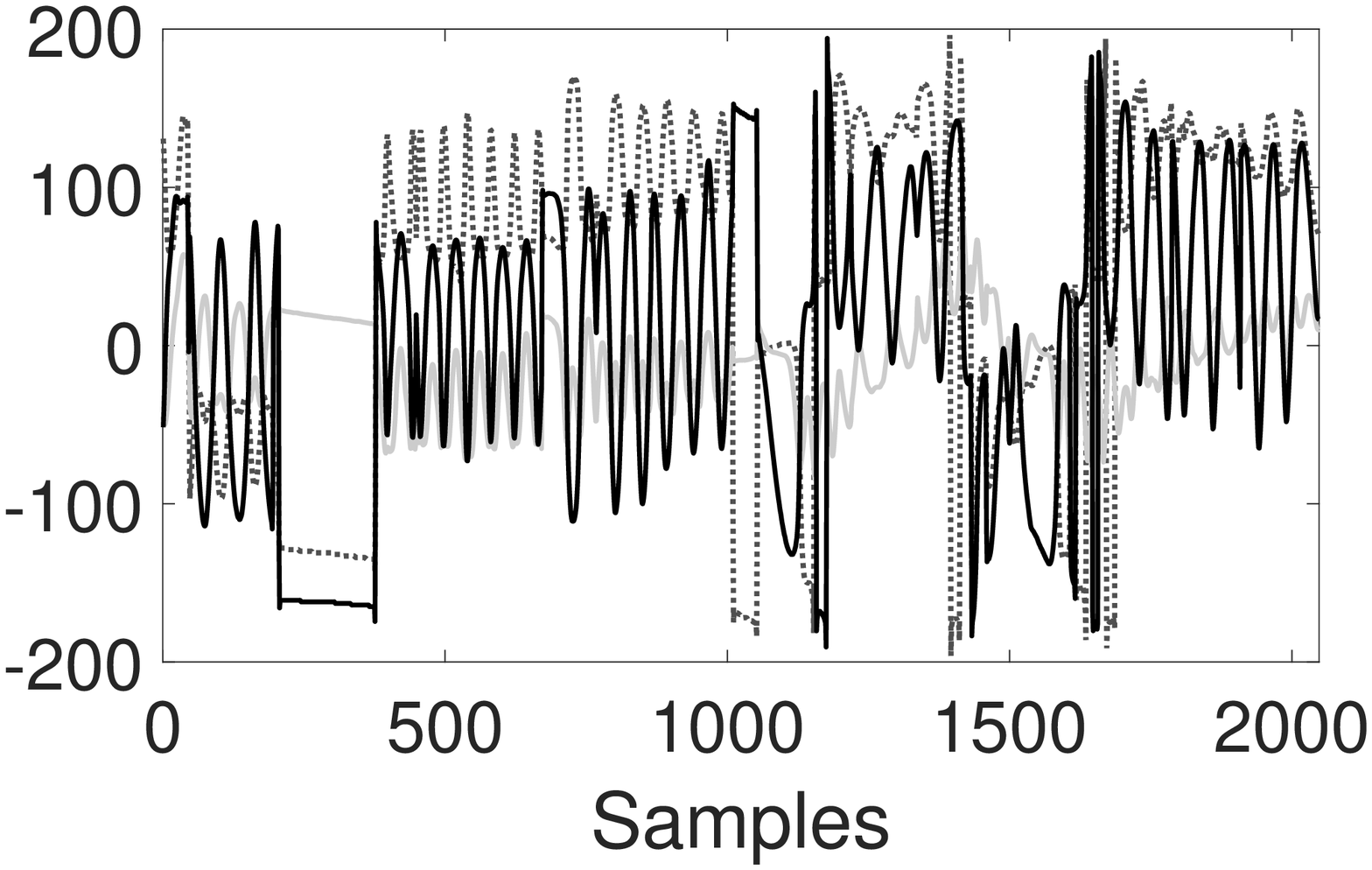}
		\caption{}
		\label{WLS}
	\end{subfigure}
	
	\begin{subfigure}{.24\textwidth}
		\centering
		\includegraphics[width=\linewidth]{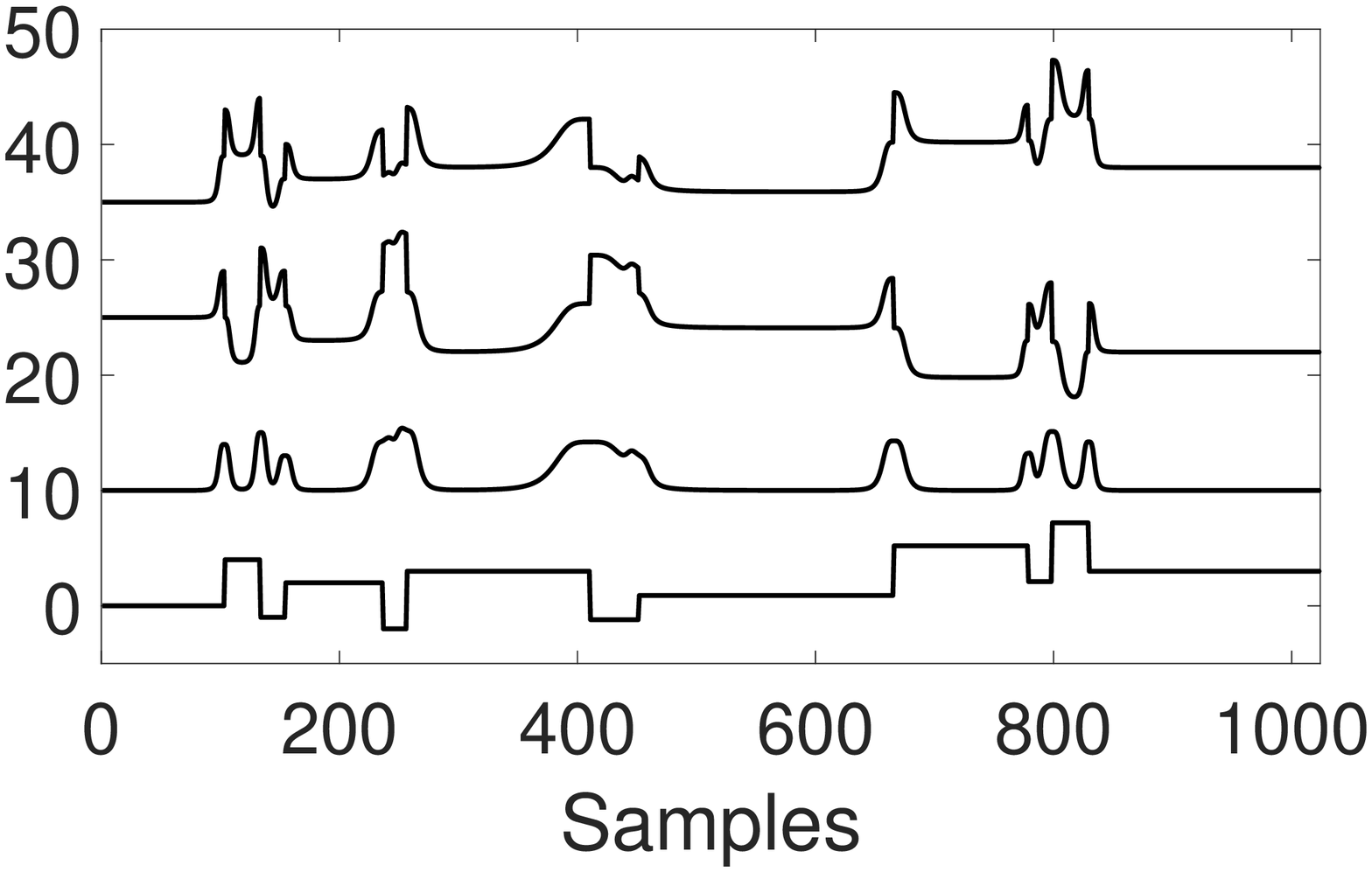}
		\caption{}
		\label{HvyDopp}
	\end{subfigure}
	%	\hspace{-8mm}
	\begin{subfigure}{.24\textwidth}
		\centering
		\includegraphics[width=\linewidth]{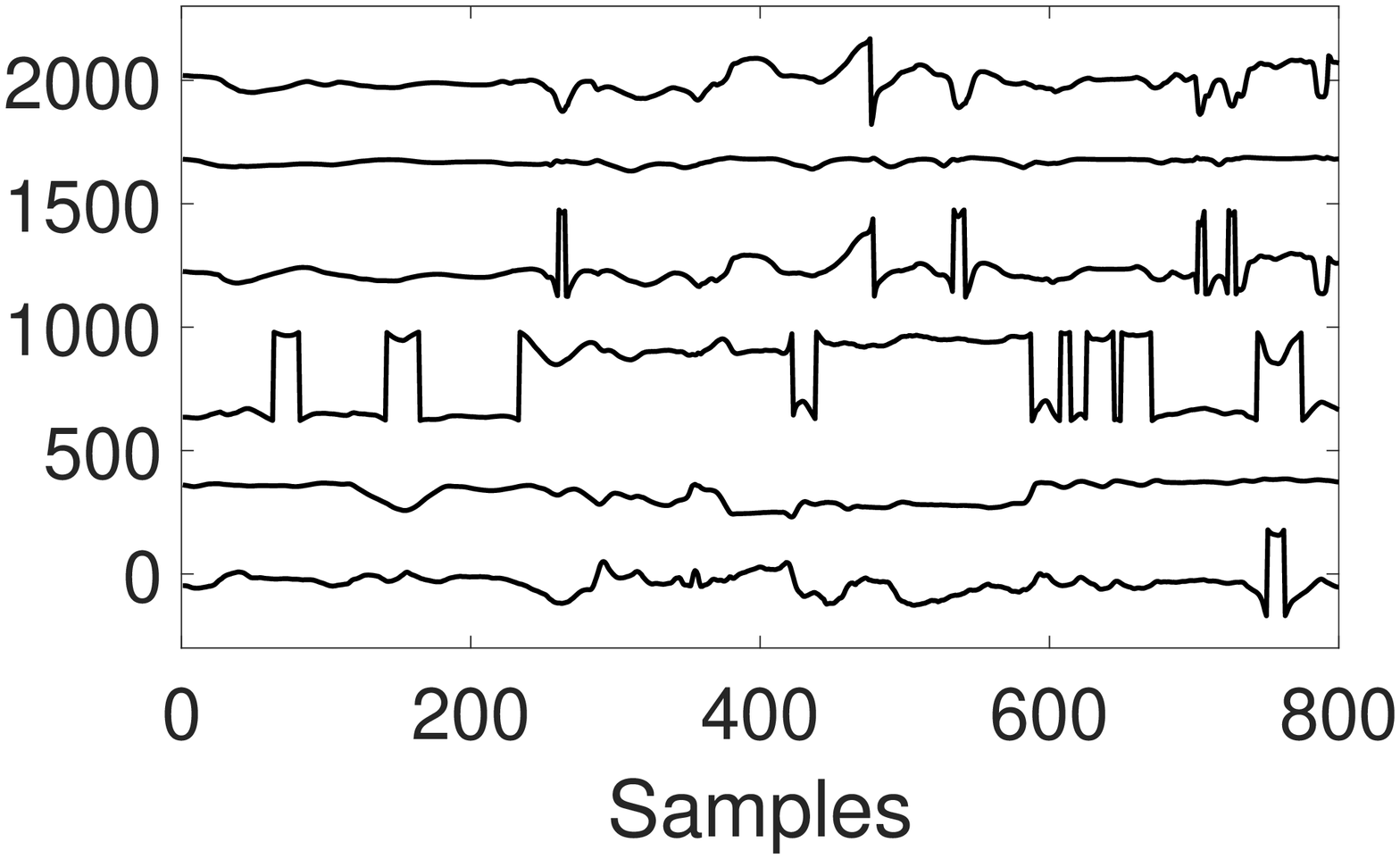}
		\caption{}
		\label{WLS}
	\end{subfigure}
	\caption{Multivariate test signals used in our experiments; (a) Biviariate \textit{Sofar} signal; (b) Bivariate \textit{EOG} signal; (c) Trivariate Synthetic \textit{HeavyDoppler} signal; (d) Trivariate \textit{Health-Moitoring} signal; (e) Quadrivariate Synthetic \textit{BumpsBlocks} signal; (f) Hexavariat Tai-Chi signal.}
	\label{InpTestSig}
		\vspace{-3mm}
\end{figure}

Finally, the hypothesis test given in \eqref{hypotestmulti} is conducted by comparing the estimated $\tau_i^k$ against the scale dependent threshold $T_k$. In case $\tau_i^k < T_k$, the corresponding coefficient is detected as a noise coefficient and discarded whereas if $\tau_i^k\geq T_k$, the corresponding coefficient belongs to signal is retained. Mathematically, the proposed hypothesis testing framework can be specified as the following multivariate thresholding function
\begin{equation}
\hat{\mathbf{d}}_i^k =
\begin{cases}
\mathbf{0} & \quad \tau_i^k < T_k,\\
{\mathbf{d}}_i^k & \quad \tau_i^k \geq T_k.
\end{cases}
\label{thrmv}
\end{equation}

Graphical illustration of how the proposed method operates is shown in Fig. \ref{Ch5:BlockDaig}. Note from the lower row of the figure where two test EDFs, $\mathcal{F}_s(t)$ and $\mathcal{F}_\psi(t)$, have been plotted along with the reference EDF $\mathcal{F}_0(t)$. The EDF $\mathcal{F}_\psi(t)$ is close to $\mathcal{F}_0(t)$ and therefore yields a smaller value of $\tau$ resulting in $\tau < T$, hence, suggesting noise detection. Contrarily, the EDF $\mathcal{F}_s(t)$ is different from $\mathcal{F}_0(t)$ resulting in $\tau\geq T$ corresponding to the detection of signal coefficient.
\begin{figure*}[t]
	\centering
	
	\begin{subfigure}{.32\textwidth}
		\centering
		\includegraphics[width=\linewidth]{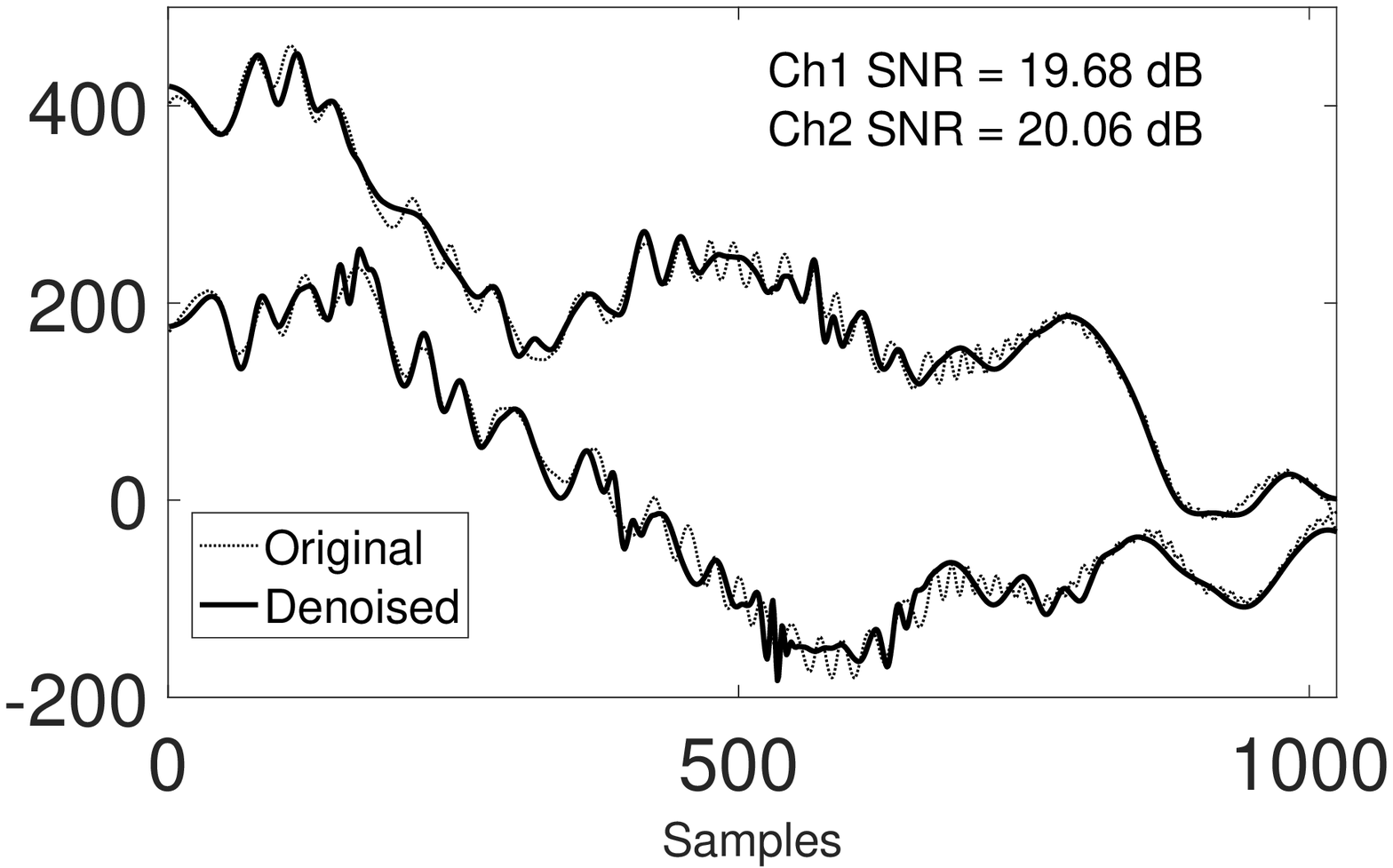}
		\caption{MWD; $\rho=0$}
		\label{WLS}
	\end{subfigure}
	\hspace{-6mm}
	\begin{subfigure}{.32\textwidth}
		\centering
		\includegraphics[width=\linewidth]{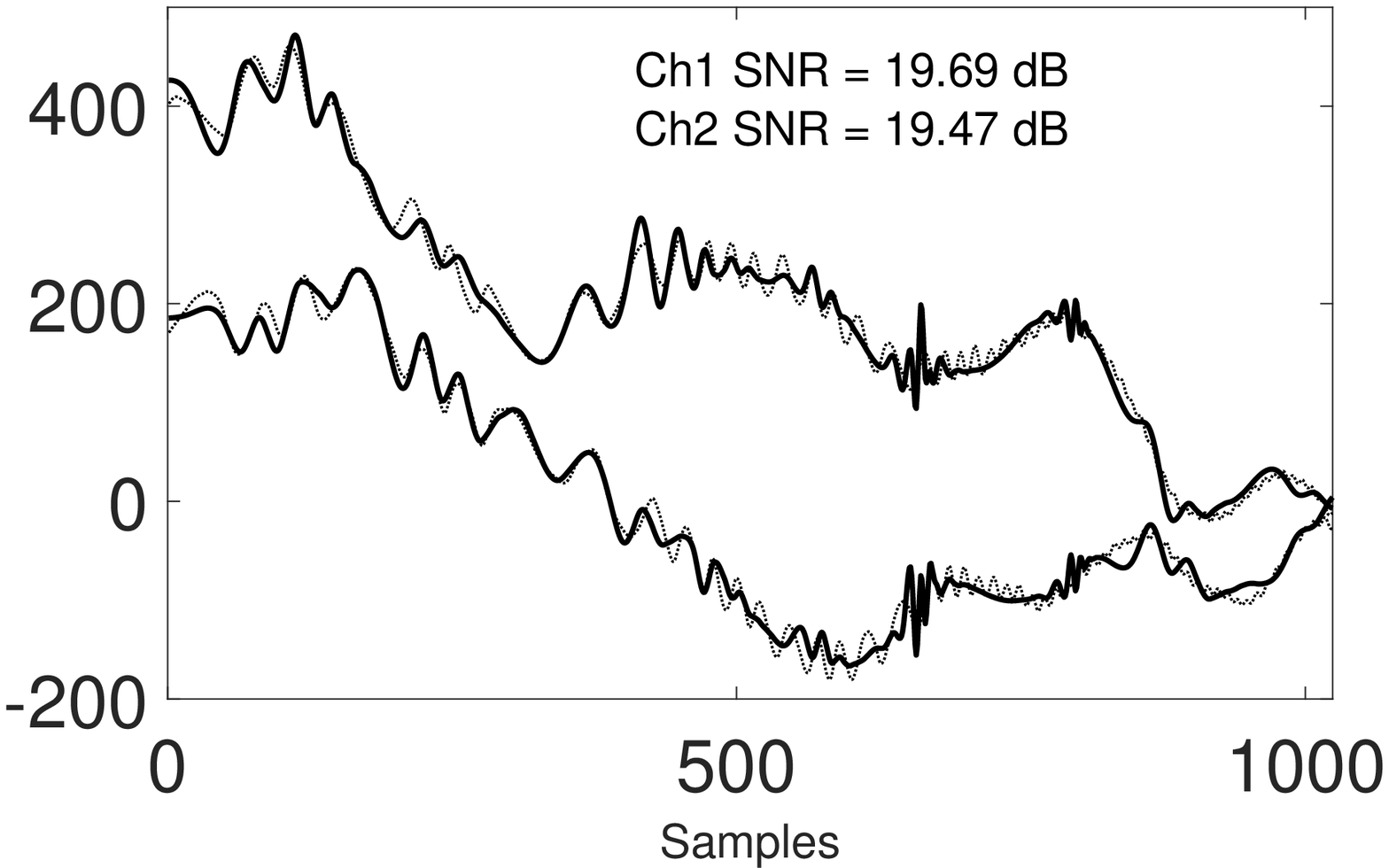}
		\caption{MWD; $\rho=0.25$}
		\label{WLS}
	\end{subfigure}
	\hspace{-6mm}
	\begin{subfigure}{.32\textwidth}
		\centering
		\includegraphics[width=\linewidth]{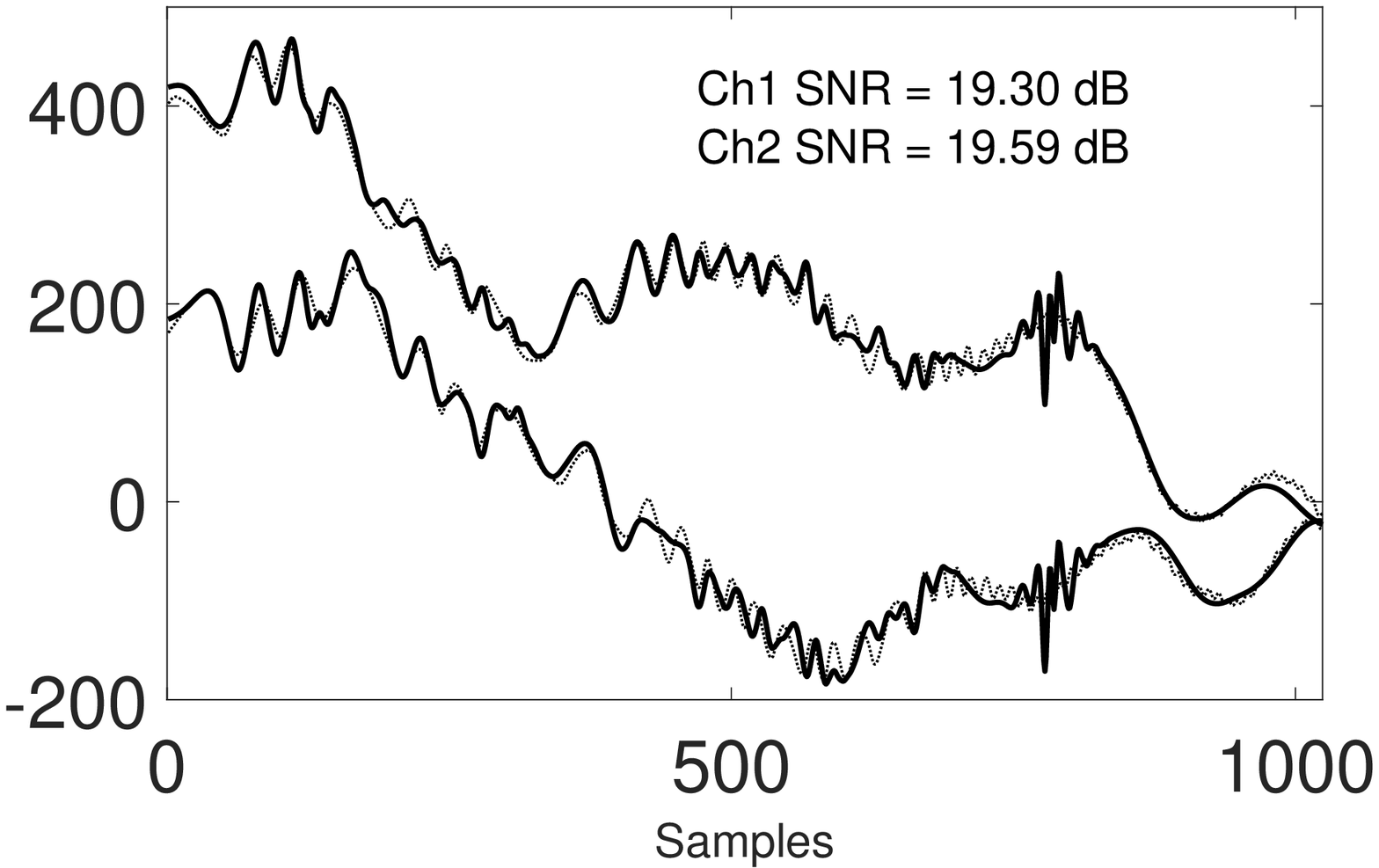}
		\caption{MWD; $\rho=0.75$}
		\label{HvyDopp}
	\end{subfigure}
	
	\begin{subfigure}{.32\textwidth}
		\centering
		\includegraphics[width=\linewidth]{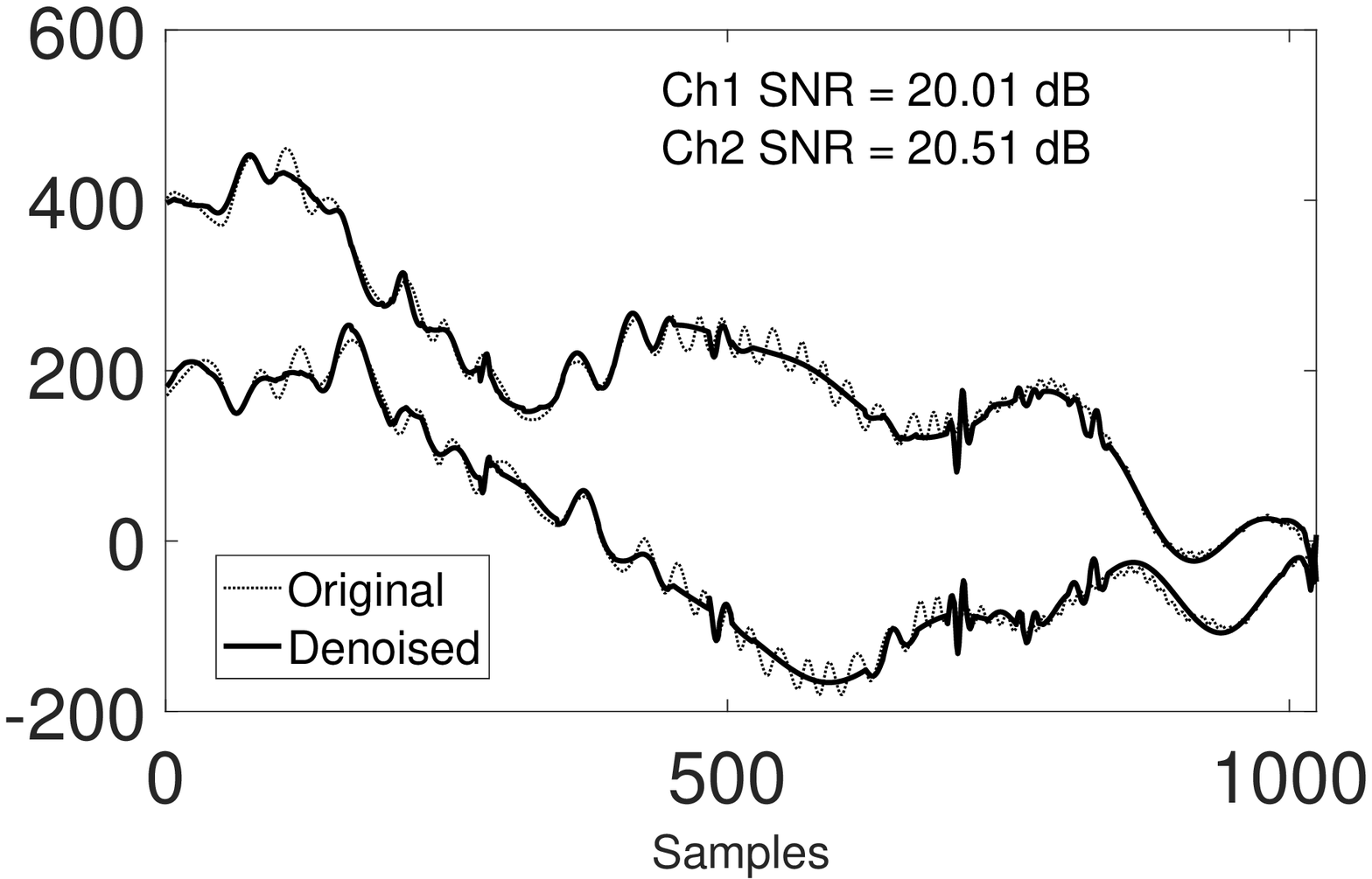}
		\caption{MMD; $\rho=0$}
		\label{WLS}
	\end{subfigure}
	\hspace{-6mm}
	\begin{subfigure}{.32\textwidth}
		\centering
		\includegraphics[width=\linewidth]{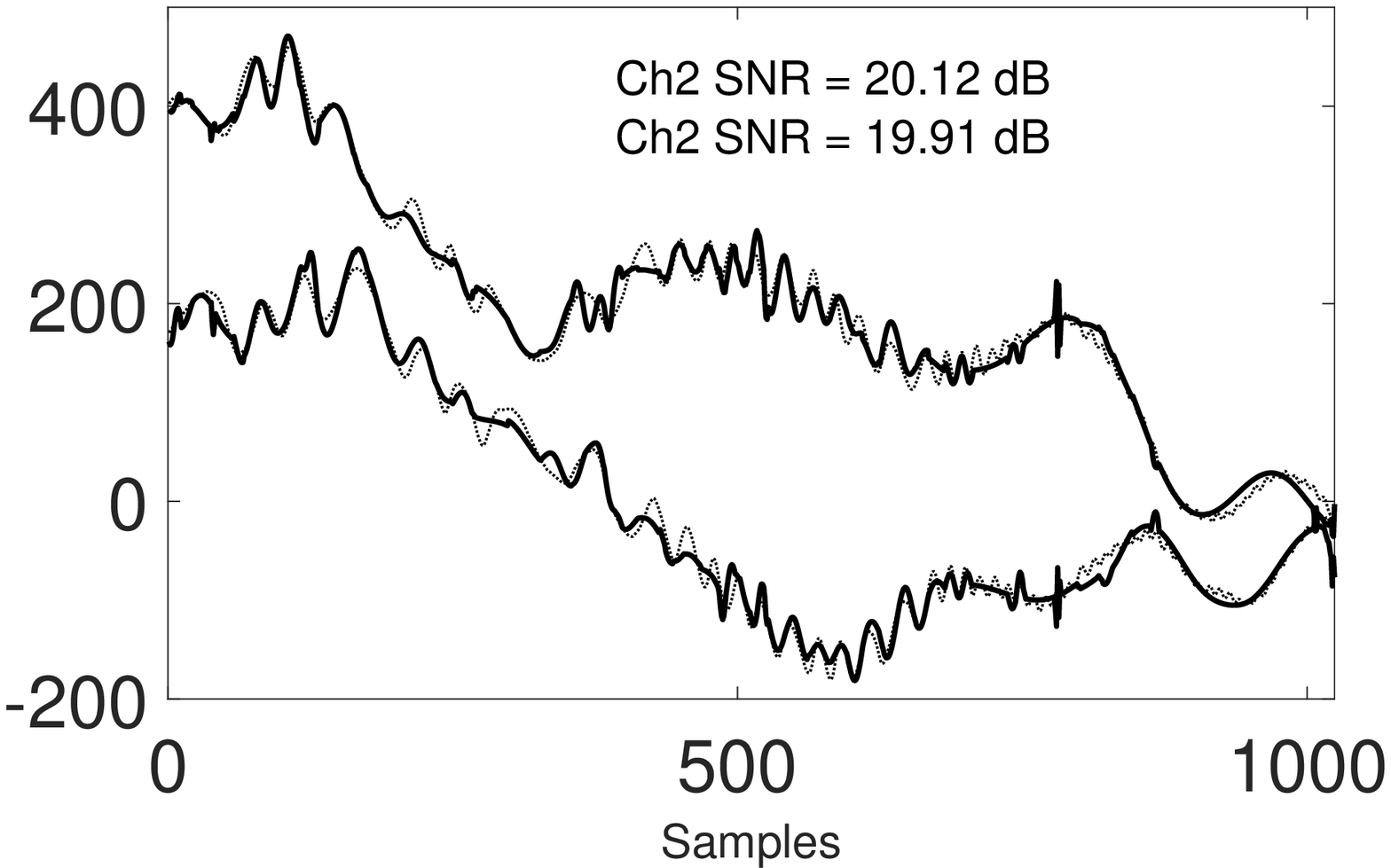}
		\caption{MMD; $\rho=0.25$}
		\label{HvyDopp}
	\end{subfigure}
	\hspace{-6mm}
	\begin{subfigure}{.32\textwidth}
		\centering
		\includegraphics[width=\linewidth]{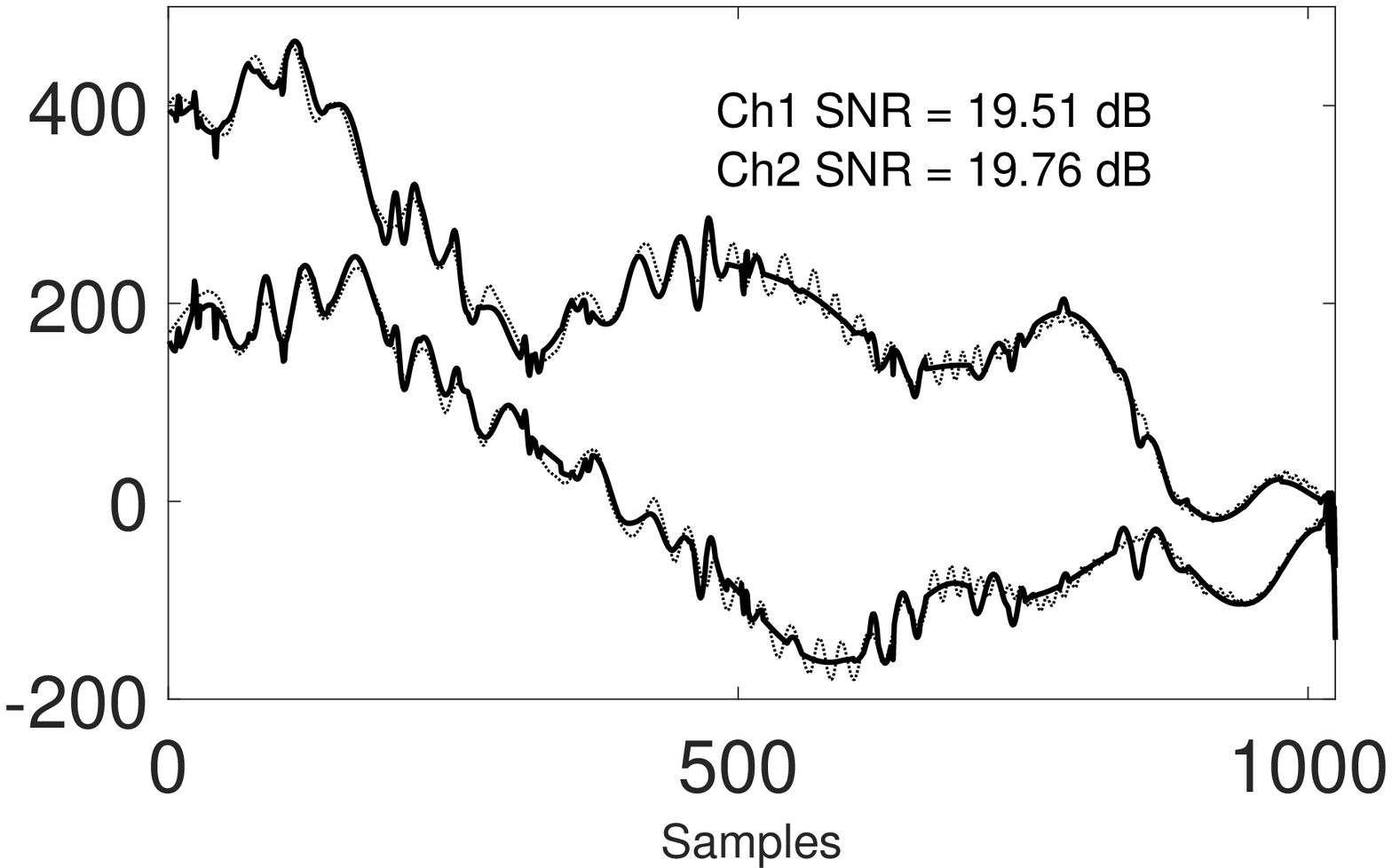}
		\caption{MWD; $\rho=0.75$}
		\label{WLS}
	\end{subfigure}
	
	\begin{subfigure}{.32\textwidth}
		\centering
		\includegraphics[width=\linewidth]{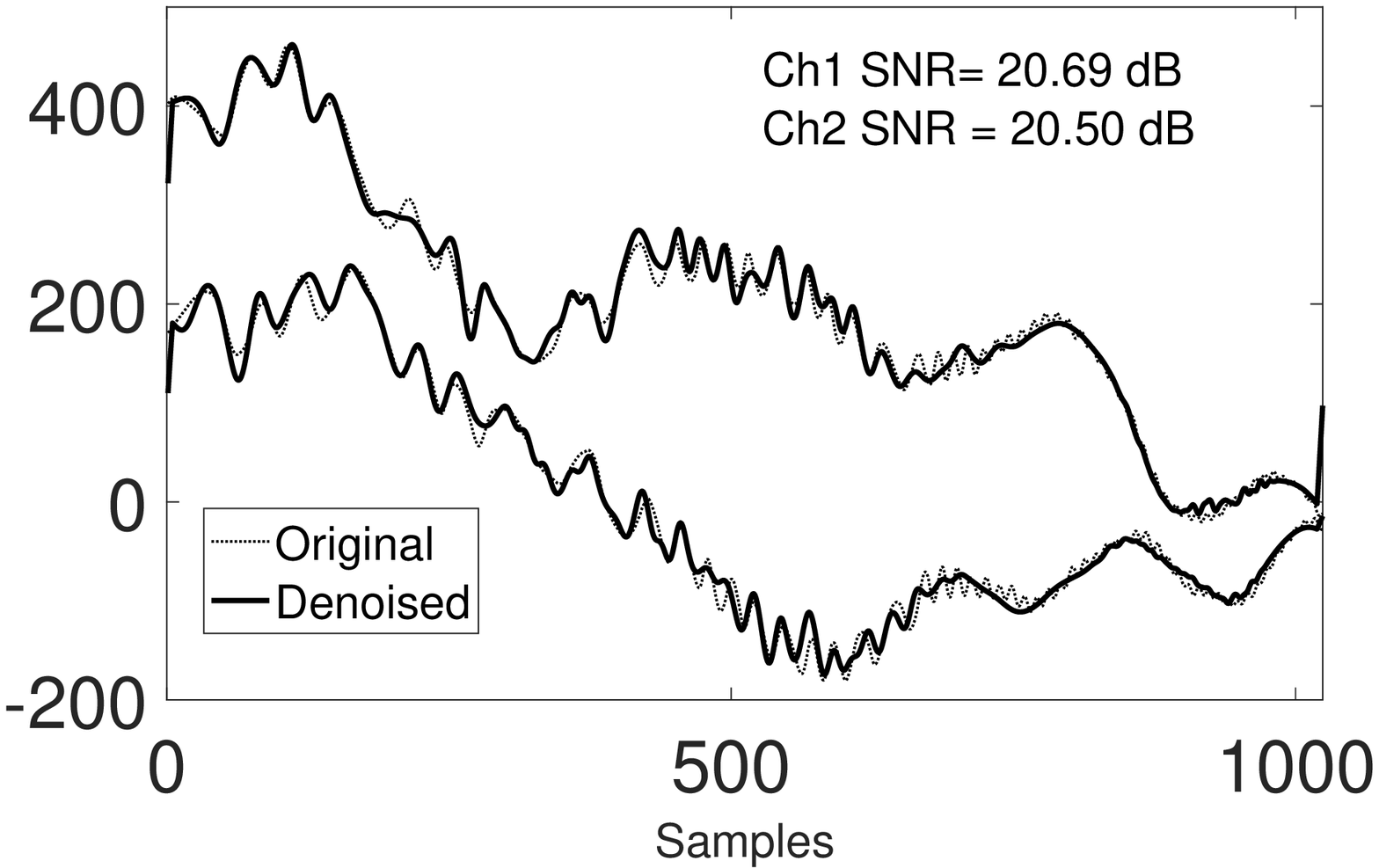}
		\caption{MGWD;$\rho=0$}
%		\label{Sofar}
	\end{subfigure}
	\hspace{-6mm}
	\begin{subfigure}{.32\textwidth}
		\centering
		\includegraphics[width=\linewidth]{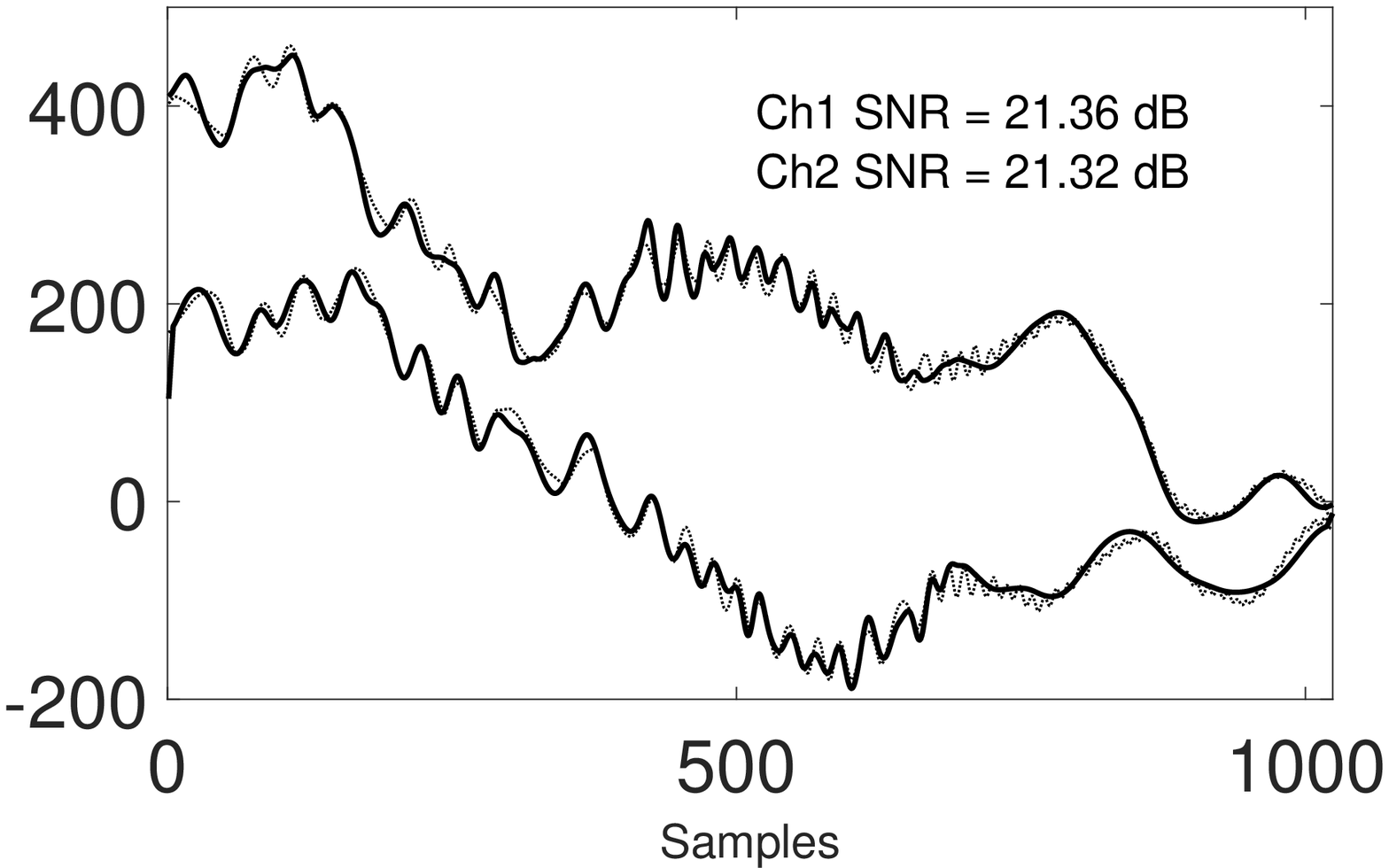}
		\caption{MGWD; $\rho=0.25$}
		\label{EOG}
	\end{subfigure}
	\hspace{-6mm}
	\begin{subfigure}{.32\textwidth}
		\centering
		\includegraphics[width=\linewidth]{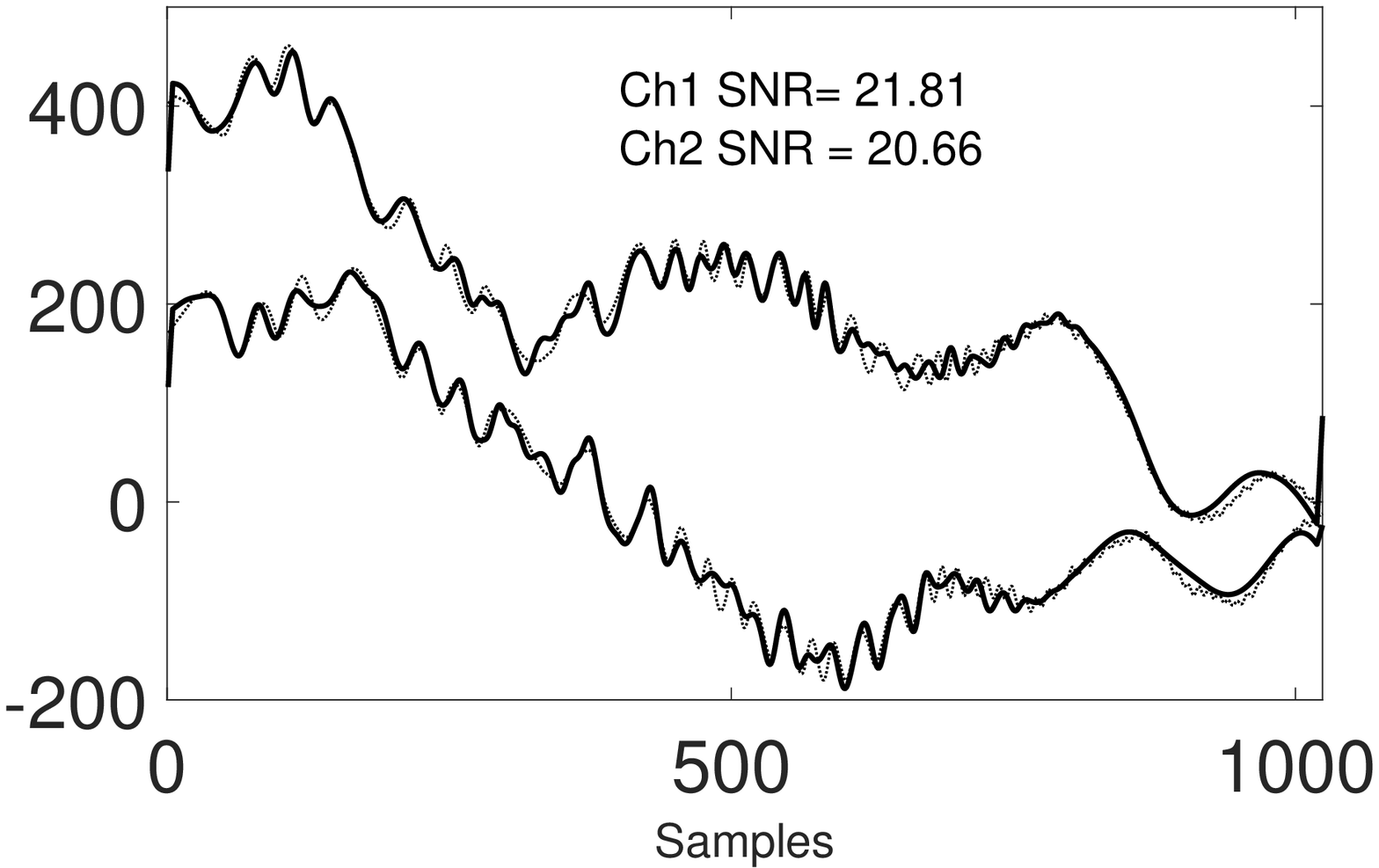}
		\caption{MGWD; $\rho=0.75$}
		\label{HvyDopp}
	\end{subfigure}
	\caption{A toy example demonstrating the effect of correlation among noise channels on the denoising results from the proposed MGWD method (displayed in left column) and the best performing state of the art method MWD (displayed in left column). }
	\label{ToyEx}
	\vspace{-3mm}
\end{figure*}

\begin{table*}[t]
	\centering
	\caption{Input and output SNR values obtained from various comparative multivariate signal denoising methods on synthetic signals used in this work. Channel-wise results as well as averaged SNR values obtained across all channels are reported.}
	\small
	\centering
	\scalebox{0.8}{
		\resizebox{1.25\textwidth}{!}{
			\setlength\extrarowheight{3pt}
			\begin{tabular}{|c||ccccc||ccccc||ccccc||ccccc||}\thickhline
				\textbf{Avg. Input SNR}
				%				&\multicolumn{4}{c}{\textbf{-10}} 
				&\multicolumn{4}{c}{\textbf{-5}} & &\multicolumn{4}{c}{\textbf{0}} & &\multicolumn{4}{c}{\textbf{5}} & &\multicolumn{4}{c}{\textbf{10}} &\\
				\thickhline
				\textbf{Test Signal}
				&\multicolumn{19}{c}{\textbf{HeavySine \& Doppler Trivariate Signal}} &\\
				\hline 
				\textbf{Channels}
				%				&\textbf{C1} &\textbf{C2} & \textbf{C3} & & \textbf{Avg}
				&\textbf{C1} &\textbf{C2} & \textbf{C3} & &\textbf{Avg} &\textbf{C1} &\textbf{C2} & \textbf{C3} & &\textbf{Avg} &\textbf{C1} &\textbf{C2} & \textbf{C3} & &\textbf{Avg} &\textbf{C1} &\textbf{C2} & \textbf{C3} & &\textbf{Avg} \\
				\textbf{Inp. SNR (Balanced)}
				%				&\textbf{-10} & \textbf{-10} & \textbf{-10} & & \textbf{-10} 
				&\textbf{-5} &\textbf{-5} & \textbf{-5} & &\textbf{-5} & \textbf{0} &\textbf{0} & \textbf{0} & & \textbf{0} & \textbf{5} & \textbf{5} & \textbf{5} & & \textbf{5} &\textbf{10} &\textbf{10} & \textbf{10} & &\textbf{10} \\
				\textbf{Inp. SNR (Unbalanced)}
				%				&\textbf{-12} & \textbf{-10} & \textbf{-8} & & \textbf{-10} 
				&\textbf{-3} &\textbf{-5} & \textbf{-7} & &\textbf{-5} & \textbf{-2} &\textbf{0} & \textbf{-2} & &\textbf{0} & \textbf{3} & \textbf{5} & \textbf{7} & &\textbf{5} &\textbf{12} &\textbf{10} & \textbf{-8} & & \textbf{10} \\
				\hline
				\textbf{MWD}
				%				&3.83 & 0.31 & 4.08 && 2.74 
				& 8.93 & 6.69 & 7.48 && 7.70 & 13.26 & \textbf{10.59} & 11.45 && 11.77 & 16.54 & 14.19 & 14.54 && 15.16 & 19.75 & 17.99 & 17.65 && 18.46 \\
				%				&\textbf{3.20} & -0.22 & 4.82 && 2.60 
				& 10.28 & 6.93 & 7.35 && 8.19 & \textbf{12.87} & \textbf{11.02} & 12.03 && 11.97 & \textbf{16.41} & \textbf{15.05} & 14.75 && 15.40 & 19.20 & 18.28 & 17.63 && 18.47 \\
				\hline
				\textbf{MWSD}
				%				&-6.33 & -4.98 & -6.16 && -5.82 
				& 3.67 & 3.97 & 3.76 && 3.80 & 7.00 & 7.92 & 7.29 && 7.40 & 8.56 & 9.90 & 9.01 && 9.16 & 9.42 & 11.17 & 10.03 && 10.21 \\
				%				&-6.91 & -4.92 & -6.01 && -5.95 
				& 5.23 & 4.17 & 2.24 && 3.88 &  5.78 & 7.98 & 8.19 && 7.32 & 8.22 & 10.20 & 9.69 && 9.37 & 9.23 & 11.27 & 10.24 && 10.25 \\
				\hline
				\textbf{MEMD-IT}
				%				&  5.16 & 2.26 & 4.75 && 4.06 
				& \textbf{10.95} & 5.37 & 9.05 && 8.59 & 14.22 & 5.83 & 12.52 && 10.89 & 16.73 & 7.08 & 14.01 && 12.69 & 18.14 & 9.15 & 15.89 && 14.47 \\
				%				& 3.30 & 3.45 & 4.62 && 3.79 
				& 9.44 & 5.79 & \textbf{10.57} && 8.60 & 12.39 & 6.84 & 11.60 &&  10.28 & 14.61 & 5.62 & 12.15 && 10.79 & 19.37 & 5.89 &  13.88 && 13.11 \\
				\hline
				\textbf{MMD}
				%				&  5.16 & 2.26 & 4.75 && 4.06 
				& 8.26 & 6.51 & 7.71 && 7.49 & 12.28 & \textbf{10.87} & 11.91  && 11.69 & 16.66 & \textbf{15.13} & 16.13 && 15.97 & 20.72 & \textbf{19.21} & 20.37  && 20.10 \\
				%				& 3.30 & 3.45 & 4.62 && 3.79 
				& 9.81 & 6.42 & 5.59 && 7.28 & 10.29 & \textbf{10.81} & 13.56 && 11.55 & 15.14 & 14.95 & 18.00 && 16.04 & 18.97 & \textbf{19.23} & 21.67 && 19.95 \\
				\hline
				\textbf{MGWD}
				%				&\textbf{4.84} & \textbf{3.89} & \textbf{4.89} & & \textbf{ 4.54} 
				& 9.75 & \textbf{7.09} & \textbf{9.19} && \textbf{8.68} & \textbf{14.46} & 9.60 & \textbf{13.66} && \textbf{12.58} & \textbf{17.37} & 14.33 & \textbf{16.69} && \textbf{16.07} & \textbf{21.49} & 18.54 & \textbf{20.53} && \textbf{20.19} \\
				%				&2.21 & \textbf{3.43} & \textbf{6.09} & & \textbf{3.91} 
				& \textbf{12.27} & \textbf{7.22} & 7.90 && \textbf{9.13} & 12.42 & 9.49 & \textbf{14.95} && \textbf{12.28} & 15.62 & 13.12 & \textbf{18.01} && \textbf{15.58} & \textbf{19.49} & 18.42 & \textbf{21.71} && \textbf{19.87} \\
				\thickhline
				\textbf{Test Signal}
				&\multicolumn{19}{c}{\textbf{Bumps-Blocks Quadrivariate Signal}} &\\
				\hline 
				\textbf{Channels}
				%				&\textbf{C1} &\textbf{C2} & \textbf{C3} & \textbf{C4} & \textbf{Avg} 
				&\textbf{C1} &\textbf{C2} & \textbf{C3} & \textbf{C4} & \textbf{Avg} &\textbf{C1} &\textbf{C2} & \textbf{C3} & \textbf{C4} & \textbf{Avg} &\textbf{C1} &\textbf{C2} & \textbf{C3} & \textbf{C4} & \textbf{Avg} &\textbf{C1} &\textbf{C2} & \textbf{C3} & \textbf{C4} & \textbf{Avg} \\
				\textbf{Inp. SNR (Balanced)}
				%				&\textbf{-10} & \textbf{-10} & \textbf{-10} & \textbf{-10} & \textbf{-10} 
				&\textbf{-5} &\textbf{-5} & \textbf{-5} & \textbf{-5} & \textbf{-5} & \textbf{0} &\textbf{0} & \textbf{0} & \textbf{0} & \textbf{0} & \textbf{5} & \textbf{5} & \textbf{5} & \textbf{5} & \textbf{5} &\textbf{10} &\textbf{10} & \textbf{10} & \textbf{10} & \textbf{10} \\
				\textbf{Inp. SNR (Unbalanced)}
				%				&\textbf{-12} & \textbf{-11} & \textbf{-9} & \textbf{-8} & \textbf{-10} 
				&\textbf{-3} &\textbf{-4} & \textbf{-6} & \textbf{-7} & \textbf{-5} & \textbf{-2} &\textbf{-1} & \textbf{1} &\textbf{2} & \textbf{0} & \textbf{3} & \textbf{4} & \textbf{6} &\textbf{7} & \textbf{5} &\textbf{12} &\textbf{11} & \textbf{9} & \textbf{8}&  \textbf{10} \\
				\hline
				\textbf{MWD}
				%				&\textbf{4.30} & 1.30 & 1.84 & \textbf{4.97} & 3.10 
				& 7.02 & 6.31 & \textbf{7.96} & 6.21 & 6.88 & 10.17 & 9.96 & 11.58 & 10.01 & 10.10 & 13.21 & 13.93 & 15.17 & 13.71 & 14.05 & \textbf{16.04} & 17.04 & 17.51 & 17.58 & 17.04 \\
				%				&\textbf{2.37} & 1.24 & 1.88 & \textbf{5.58} & 2.77 
				& 0.30 & 0.91 & 4.33 & 5.12 & 2.67 & \textbf{8.75} & \textbf{7.59} & \textbf{7.19} & \textbf{5.21} & \textbf{7.19} &
				\textbf{9.32} & 9.14 & \textbf{13.06} & 11.33 & 10.64 & 12.13 & 13.01 & 16.22 & 15.18 & 14.04 \\
				\hline
				\textbf{MWSD}
				%				&0.7442 & -0.5106 & -5.52 & 2.28 & -0.75 
				& 0.66 & 2.07 & 3.90 & 4.50 & 2.78 & 1.78 & 3.83 & 6.82 & 7.96 & 5.10 & 2.34 & 4.74 & 8.64 & 10.02 & 6.44 & 2.47 & 4.94 & 9.40 & 11.16 & 6.99 \\
				%				&-0.84 &  -1.66 & -5.43 & 3.36 & -1.15 
				& -3.85 & -1.73 & 0.76 & 3.12 & -0.42 & 1.29 & 2.59 & 3.36 & 2.81 & 2.51 &
				1.28 & 3.44 & 7.24 & 8.58 & 5.14 & 2.19 & 4.62 & 8.77 & 10.46 & 6.51 \\
				\hline
				\textbf{MEMD-IT}
				%				& 3.49 & 2.59 & 2.53 & 3.98 & 3.15 
				& 5.09 & 3.60 & 3.40 & 5.58 & 4.42 & 2.27  & 2.84 & 2.95 & 2.27 & 2.58 & 0.22 & 4.51 & 1.14 & 0.83 & 1.67 & 1.31 & 3.03 & 2.37 & 1.28 & 2.00 \\
				%				& 4.19 & 2.77 & 3.49 & 6.66 & 4.28 
				& \textbf{3.29} & \textbf{2.87} & 3.37 & \textbf{6.54} & \textbf{4.02} & 1.21 & 2.29 & 1.72 & 1.24 & 1.61 & 0.86 & 2.15 & 1.25 & 0.78 & 1.26 & 0.72 & 2.15 & 1.10 & 0.55 & 1.13 \\
				\hline
				\textbf{MMD}
				%				&  5.16 & 2.26 & 4.75 && 4.06 
				& 6.00 & 5.43 & 7.06 & \textbf{6.87} & 6.34 & 9.69 & 9.65 & 10.98 & \textbf{10.64} & 10.24 & 12.86 & 14.06 & 15.54 & \textbf{14.57} & \textbf{14.25} & 14.90 & 17.24 & 19.63 & \textbf{18.95} & \textbf{17.71} \\
				%				& 3.30 & 3.45 & 4.62 && 3.79 
				& 0.74 & 1.12 & 4.62 & 5.72 & 3.05 & 7.57 & 6.39 & 6.25 & \textbf{5.17} & 6.34 & 7.68 & 8.39 & 12.15 & \textbf{12.27} & 10.13 & 12.14 & 13.15 & 16.32 & \textbf{16.39} & \textbf{14.49} \\
				\hline
				\textbf{MGWD}
				%				&3.75 & \textbf{3.04} & \textbf{3.44} & 3.66 & \textbf{3.47} 
				& \textbf{7.07} & \textbf{6.56} & 7.85 & 6.64 & \textbf{6.95} & \textbf{10.42} & \textbf{10.03} & \textbf{11.49} & \textbf{10.16} & \textbf{10.40} & \textbf{13.42} & \textbf{14.10} & \textbf{15.66} & 13.90 & 14.17 & 15.82 & \textbf{17.32} & \textbf{20.23} & 16.11 & 17.37 \\
				%				&0.58 & \textbf{2.53} & \textbf{4.40} & 1.70 & \textbf{6.00} 
				& 1.56 & 2.37 & \textbf{5.98} & 5.90 & 3.95 & 8.23 &  6.93 & 6.41 & 4.83 & 6.60 & 8.57 & \textbf{9.15} & 12.23 & 11.52 & \textbf{10.72} & \textbf{12.30} & \textbf{13.29} & \textbf{16.44} & 14.67 & 14.15 \\
				\thickhline
	\end{tabular}}}
	\label{synthetictable}
	\vspace{-3mm}
\end{table*}				

Note that unlike existing multivariate denoising methods that apply thresholding operation separately on each channel, the proposed thresholding function \eqref{thrmv} is purely multivariate since it operates collectively on all channels of multivariate coefficient $\mathbf{d}_i^k = [d_{i^{(1)}}^k, d_{i^{(2)}}^k, \cdots, d_{i^{(M)}}^k]$ by incorporating the cross-channel-correlation $\Sigma$ of noise in the decision process. 

Finally, the denoised multivariate signal $\hat{\mathbf{s}}$ is obtained by applying the inverse DWT transformation on the thresholded coefficients $\hat{\mathbf{d}}_i^k$ 
\begin{equation}
\hat{\mathbf{s}}_i = \mathcal{T}^{-1}(\hat{\mathbf{d}}_i^k),
\end{equation}
\noindent where $\hat{\mathbf{s}}_i$ denotes the estimate of the original signal $\mathbf{s}_i$. In the subsequent discussion, we call our proposed method as \textit{multivariate GoF based wavelet denoising (MGWD)} method.
\section{Simulations and discussion}
\subsection{Materials and methods}
In this section, we compare the performance of the proposed method against the established multivariate denoising methods on a wide range of synthetic and real world multivariate signals. The comparative methods used in our analysis include the MWD \cite{aminghafari2006MWD}, MWSD \cite{ahrabian2015MWSD}, EMD-IT \cite{hao2017MEMD-IT} and the recent MMD method \cite{ur2019MMD}. We also compare performance of the proposed method with the state of the art univariate denoising methods namely \textit{BLFDR} \cite{lavrik2008BLFDR} and \textit{EMD-IT} \cite{kopsinis2009EMD-IT} when used for suppressing multivariate noise.  
\begin{table*}[t]
	\caption{Input versus output SNR values of various comparative multivariate signal denoising methods on real signals. Channel-wise results as well as averaged SNR values obtained across all channels are reported.}
	\small
	\centering
	\scalebox{1}{
		\resizebox{0.85\textwidth}{!}{
			\setlength\extrarowheight{3pt}
			\begin{tabular}{|c||cccc||cccc||cccc||cccc||}\thickhline
				%				\textbf{Channels}
				%				&\textbf{C1} &\textbf{C2} & \textbf{C3} &\textbf{C4} & \textbf{Avg} &\textbf{C1} &\textbf{C2} & \textbf{C3} &\textbf{C4} & \textbf{Avg} &\textbf{C1} &\textbf{C2} & \textbf{C3} &\textbf{C4} & \textbf{Avg} &\textbf{C1} &\textbf{C2} & \textbf{C3} &\textbf{C4} & \textbf{Avg} &\textbf{C1} &\textbf{C2} & \textbf{C3} &\textbf{C4} & \textbf{Avg} \\
				\textbf{Avg. Input SNR}
				%				&\multicolumn{4}{c}{\textbf{-10}} 
				&\multicolumn{3}{c}{\textbf{-5}} & &\multicolumn{3}{c}{\textbf{0}} & &\multicolumn{3}{c}{\textbf{5}} & &\multicolumn{3}{c}{\textbf{10}} &\\
				\thickhline
				\textbf{Test Signal}
				&\multicolumn{15}{c}{\textbf{Sofar Bivariate Signal}} &\\
				\hline 
				\textbf{Channels}
				%				&\textbf{C1} & & \textbf{C2} & & \textbf{Avg} 
				&\textbf{C1} & & \textbf{C2} & \textbf{Avg} &\textbf{C1} & & \textbf{C2} & \textbf{Avg} &\textbf{C1} & & \textbf{C2} & \textbf{Avg} &\textbf{C1} & & \textbf{C2} & \textbf{Avg} \\
				\textbf{Inp. SNR (Balanced)}
				%				&\textbf{-10} & & \textbf{-10} &  & \textbf{-10} 
				& \textbf{-5} & & \textbf{-5} & \textbf{-5} & \textbf{0} & &\textbf{0} & \textbf{0} & \textbf{5} & & \textbf{5} & \textbf{5} &\textbf{10} & &\textbf{10} & \textbf{10} \\
				\textbf{Inp. SNR (Unbalanced)}
				%				&\textbf{-12} & & \textbf{-8} & & \textbf{-10} 
				&\textbf{-3} & &\textbf{-7} & \textbf{-5} &\textbf{-2} & &\textbf{2} & \textbf{0} & \textbf{3} & & \textbf{7} & \textbf{5} &\textbf{12} & &\textbf{8} & \textbf{10} \\
				\thickhline
				\textbf{MWD}
				%				&3.84  &&  3.98  & & 3.91 
				& 8.62 && 9.04 & 8.83 & 13.08 && 13.00 & 13.04 & 16.65 && 16.43 & 16.54 & 19.46 && 19.72 & 19.59 \\
				%				&\textbf{3.24} && 3.95 && 3.60 
				& 6.28 && 10.06 & 8.17 & 11.11 && 13.77 & 12.44 & 15.28 && 18.16 & 16.72 & 18.43 && 21.26 & 19.84 \\
				\hline
				\textbf{MWSD}
				%				&-4.13 && -5.65 &&-4.89 
				& 0.59 && 1.32 & 0.95 & 2.04 && 2.66 & 2.35 & 3.40 && 3.64 & 3.52 & 4.35 && 4.14 & 4.25 \\
				%				&-4.71  && -4.76 && -4.74 
				& -0.43 && 2.18 & 0.87 & 1.40 && 3.26 & 2.33 & 2.90 && 3.98 & 3.44 & 4.07 && 4.20 & 4.14 \\
				\hline
				\textbf{MEMD-IT}
				%				& 1.10 && 0.77 && 0.93 
				& 5.23 && 6.02 & 5.63 & 8.41 && 7.67 & 8.04 & 11.89 && 12.59 & 12.24 &  14.27 && 13.52 & 13.89 \\
				%				& 0.36 && 3.88 && 2.12 
				& 4.22 && 6.59 & 5.40 & 7.72 && 8.16 & 7.94 & 11.86 && 13.19 & 12.53 & 12.65 && 13.39 & 13.024 \\
					\hline
				\textbf{MMD}
				%				& 1.10 && 0.77 && 0.93 
				& 8.92 && 8.62 & 8.77 & 12.70 && 13.17 & 12.93 & 16.65 && 16.50 & 16.58 & 20.21 && \textbf{20.42} & 20.31\\
				%				& 0.36 && 3.88 && 2.12 
				& 7.25 && 10.08 & 8.67 & 11.66 && \textbf{14.88} & 13.27 & 15.18 && 18.16 & 16.67 & 18.80 && 21.78 & 20.29 \\
				\hline
				\textbf{MGWD}
				%				&\textbf{5.05} & & \textbf{5.50} &  & \textbf{5.27} 
				& \textbf{9.83} && \textbf{9.89} & \textbf{9.86} & \textbf{14.17} && \textbf{13.88} & \textbf{14.03} & \textbf{17.12} && \textbf{16.92} & \textbf{17.02} & \textbf{20.71} && 20.02 & \textbf{20.37} \\
				%				&1.78 && \textbf{5.74} && \textbf{3.76} 
				&\textbf{7.32} && \textbf{11.50} & \textbf{9.41} & \textbf{12.16} && 14.71 & \textbf{13.43} & \textbf{15.64} && \textbf{18.27} & \textbf{16.96} & \textbf{19.01} && \textbf{21.81} & \textbf{20.36} \\
				\thickhline
				\textbf{Test Signal}
				&\multicolumn{15}{c}{\textbf{Eye Roll EOG Bivariate Signal}} &\\
				\hline 
				\textbf{Channels}
				%				&\textbf{C1} & & \textbf{C2} & & \textbf{Avg} 
				&\textbf{C1} & & \textbf{C2} & \textbf{Avg} &\textbf{C1} & & \textbf{C2} & \textbf{Avg} &\textbf{C1} & & \textbf{C2} & \textbf{Avg} &\textbf{C1} & & \textbf{C2} & \textbf{Avg} \\
				\textbf{Inp. SNR (Balanced)}
				%				&\textbf{-10} & & \textbf{-10} &  & \textbf{-10} 
				& \textbf{-5} & & \textbf{-5} & \textbf{-5} & \textbf{0} & &\textbf{0} & \textbf{0} & \textbf{5} & & \textbf{5} & \textbf{5} &\textbf{10} & &\textbf{10} & \textbf{10} \\
				\textbf{Inp. SNR (Unbalanced)}
				%				&\textbf{-12} & & \textbf{-8} & & \textbf{-10} 
				&\textbf{-3} & &\textbf{-7} &  \textbf{-5} &\textbf{-2} & &\textbf{2} &  \textbf{0} & \textbf{3} & & \textbf{7} &  \textbf{5} &\textbf{12} & &\textbf{8} & \textbf{10} \\
				\thickhline
				\textbf{MWD}
				%				&-0.123  && -0.561 &&  -0.342 
				& 2.60 && 2.82 & 2.71 & 5.28 && 6.01 & 5.65 & 8.40 && 9.30 & 8.85 & 11.81 && 12.35 & 12.08 \\
				%				&\textbf{-0.86}  && -0.41 && -0.63 
				& 1.01 && 4.53 & 2.77 & 4.07 && 7.93 & 6.00 & 7.29 && 11.15 & 9.22 & 10.59 && 14.35 & 12.47 \\
				\hline
				\textbf{MWSD}
				%				&-5.37 && -5.53 && -5.45 
				& 2.84 && 3.87 & 3.36 & 5.87 && 7.30 & 6.58 & 8.69 && 10.29 & 9.49 & 10.42 && 12.00  & 11.21 \\
				%				&-5.75 && -4.65 && -5.19 
				& 1.41 && 5.38 & 3.39 & 4.80 && 8.43 & 6.62 & 7.78 && 10.97 & 9.37 & 9.95 && 12.28 & 11.12 \\
				\hline
				\textbf{MEMD-IT}
				%				& -4.61 && -3.99 && -4.30 
				& -4.17 && -4.58 & -4.38 & -3.52 && -3.42 & -3.47 & 0.98 && 0.88 & 0.93 & 1.40 && 1.37 & 1.38 \\
				%				&-6.22 && -3.51 && -4.87 
				& -2.66 && -1.66 & -2.16 & -1.47 && 0.05 & -0.71 & 0.76 && 2.86 & 1.81 & 0.58 && 2.82 & 1.70 \\
					\hline
				\textbf{MMD}
				%				& 1.10 && 0.77 && 0.93 
				& -0.64 && -0.40 & -0.52 & 3.74 && 3.87 & 3.81 & 7.87 && 8.10 & 7.99 & 11.85 && 12.07 & 11.96 \\
				%				& 0.36 && 3.88 && 2.12 
				&-2.42 && 1.20 & -0.60 & 2.02 && 5.48 & 3.75 & 5.84 && 9.75 & 7.79 & 10.32 && 13.92 & 12.12 \\
				\hline
				\textbf{MGWD}
				%				&\textbf{0.022} && \textbf{0.279} && \textbf{0.150} 
				& \textbf{3.55} && \textbf{4.49} & \textbf{4.02} & \textbf{6.31} && \textbf{7.90} & \textbf{7.11} & \textbf{9.28} && \textbf{10.51} & \textbf{9.90} & \textbf{12.49} && \textbf{13.82} & \textbf{13.15} \\
				%				&-2.37 && \textbf{1.17} && \textbf{-0.60} 
				& \textbf{2.42} && \textbf{5.76} & \textbf{4.09} & \textbf{5.47} && \textbf{8.88} & \textbf{7.17} & \textbf{7.91} && \textbf{11.88} & \textbf{9.89} & \textbf{11.00} && \textbf{15.53} & \textbf{13.26} \\
				\thickhline
				\textbf{Test Signal}
				&\multicolumn{15}{c}{\textbf{Health Monitoring Trivariate Signal}} &\\
				\hline 
				\textbf{Channels}
				%				&\textbf{C1} &\textbf{C2} & \textbf{C3} & & \textbf{Avg} 
				&\textbf{C1} &\textbf{C2} & \textbf{C3} &\textbf{Avg} &\textbf{C1} &\textbf{C2} & \textbf{C3} & \textbf{Avg} &\textbf{C1} &\textbf{C2} & \textbf{C3} & \textbf{Avg} &\textbf{C1} &\textbf{C2} & \textbf{C3} & \textbf{Avg} \\
				\textbf{Inp. SNR (Balanced)}
				%				&\textbf{-10} & \textbf{-10} & \textbf{-10} & & \textbf{-10} 
				&\textbf{-5} &\textbf{-5} & \textbf{-5} & \textbf{-5} & \textbf{0} &\textbf{0} & \textbf{0} & \textbf{0} & \textbf{5} & \textbf{5} & \textbf{5} & \textbf{5} &\textbf{10} &\textbf{10} & \textbf{10} & \textbf{10} \\
				\textbf{Inp. SNR (Unbalanced)}
				%				&\textbf{-12} & \textbf{-10} & \textbf{-8} & & \textbf{-10} 
				&\textbf{-3} &\textbf{-5} & \textbf{-7} & \textbf{-5} & \textbf{-2} &\textbf{0} & \textbf{-2} & \textbf{0} & \textbf{3} & \textbf{5} & \textbf{7} & \textbf{5} &\textbf{12} &\textbf{10} & \textbf{-8} & \textbf{10} \\
				\hline
				\textbf{MWD}
				%				&\textbf{2.49} & -0.09 & \textbf{0.91} && \textbf{1.10} 
				& \textbf{6.48} & 0.82 & 5.26 & 4.26 & 9.27 & 1.92 & 8.51 & 6.57 & 11.76 & 2.54 & 11.24 & 8.51 & \textbf{14.90} & 0.47 & 14.33 & 9.93 \\
				%				&1.58 & -0.15 & 0.84 && 0.75 
				& \textbf{7.81} & 1.13 & 3.31 & 4.18 & 7.77 & 3.82 & 10.04 & 7.28 & 10.47 & 1.51 & 13.26 & 8.41 & 13.35 & 0.46 & \textbf{16.52} & 10.18 \\
				\hline
				\textbf{MWSD}
				%				&-3.42 & -5.13 & -4.65 && -4.40 
				& 3.59 &  3.47 & 4.17 & 3.74 & 5.88 & 5.50 & 7.23 & 6.20 & 6.98 & 6.71 & 9.08 & 7.59 & 7.65 & 7.44 & 10.15 & 8.41 \\
				%				&-3.9192 & -5.48 & -4.29 && -4.56 
				& 4.50 & 3.32 & 2.49 & 3.44 & 4.96 & 5.57 & 7.91 & 6.15 & 6.66 & 6.76 & 9.46 & 7.6317 & 7.43 & 7.41 & 10.26 & 8.36 \\
				\hline
				\textbf{MEMD-IT}
				%				& 2.87 & 2.00 & 1.44 && 2.10 
				& 4.76 & 3.54 & 2.21 & 3.50 & 5.11 & 3.67 & 2.41 & 3.737 & 5.24 & 3.76 & 2.40 & 3.80 & 5.39 & 3.81 & 2.61 & 3.94 \\
				%				& 3.44 & 2.79 & 2.01 && 2.75 
				& 4.16 & 3.39 & 2.20 & 3.25 & 4.45 & 3.39 & 2.30 & 3.38 & 4.95 & 3.73 & 2.57 & 3.75 & 4.64 & 3.46 & 2.94 & 3.68 \\
					\hline
				\textbf{MMD}
				%				& 1.10 && 0.77 && 0.93 
				& 6.23 & 5.28 & 4.67 & 5.39 & \textbf{9.48} & 8.96 & 8.72 & 9.05 & \textbf{12.40} & 12.55 & \textbf{11.94} & 12.30 & 15.06 & 15.77 & \textbf{14.56} & 15.13 \\
				%				& 0.36 && 3.88 && 2.12 
				& 6.89 & 4.90 & 3.02 & 4.93 & \textbf{8.25} & \textbf{9.14} & 9.92 & 9.14 & 10.83 & 12.64 & \textbf{13.25} & 12.24 & \textbf{13.75} & 15.64 & 15.86 & 15.08 \\
				\hline
				\textbf{MGWD}
				%				&0.89 & \textbf{0.78} & 0.82 & & 0.83 
				& 5.32 & \textbf{5.42} & \textbf{5.36} & \textbf{5.30} & 9.21 & \textbf{9.69} & \textbf{8.91} & \textbf{9.27} & 12.18 & \textbf{13.17} & 11.63 & \textbf{12.33} & 14.52 & \textbf{15.77} & 14.39  & \textbf{14.90} \\
				%				&\textbf{-2.96} & \textbf{-1.15} & \textbf{0.78} && \textbf{-1.11} 
				& 7.01 & \textbf{5.28} & \textbf{3.39} & \textbf{5.20} & 7.54 & \textbf{9.57} & \textbf{10.08} & 9.06 & \textbf{11.07} & \textbf{13.27} & 12.54 & \textbf{12.29} & 13.47 & \textbf{15.73} & 16.22 & \textbf{15.14} \\
				\thickhline
	\end{tabular}}}
	\label{realtable}
	\vspace{-3mm}
\end{table*}				

Test signals employed in our experiments include a mix of synthetic and real signals which are shown in Fig. \ref{InpTestSig}. The synthetic signals are formed by combining Donoho's `Bumps', `Blocks', `Heavy Sine' and `Doppler' signals while the real signals include biomedical, health monitoring and oceanographic float drift recordings. Specifically, the bivariate \textit{Sofar} signal, shown in Fig. \ref{InpTestSig} (a), is composed of oceanographic float drift recordings of latitude and longitude drifts of the water flowing through eastern Mediterranean sea, which was acquired as a part of an `Eastern Basin' experiment \cite{richardson1989estbsn}. The bivariate electrooculography (EOG) signal  contains the electrical activity of brain during eye-rolling movement and was taken from the BCI Competition-IV Dataset-I\footnote{Available online at http://www.bbci.de/competition/iv/.}; it is shown in Fig. \ref{InpTestSig} (b). Fig. \ref{InpTestSig} (c) shows the trivariate \textit{HeavyDoppler} signal which is constructed by appending the `Heavy Sine' signal with the scaled version of `Doppler' signal in the first two channels respectively while the third channel is obtained by adding the `Heavy Sine' and scaled `Doppler' signals. 

The trivariate Health Monitoring dataset, shown in Fig. \ref{InpTestSig} (d), was obtained from UCI Machine Learning Repository\footnote{Available online at https://archive.ics.uci.edu/ml/datasets/.}
%	/Weight+Lifting+Exercises+monitored+with+Inertial+Measurement+Units.}, 
where signal in each channel was recorded through wearable sensors and contained measurements of roll, pitch and yaw of a person's arm during a weight lifting exercise. The quadrivariate \textit{BumpsBlocks} signal was obtained by appending `Blocks', `Bumps', difference of `Blocks' \& `Bumps' and sum  of `Blocks' \& `Bumps' in first to the fourth channel respectively, and plotted in Fig. \ref{InpTestSig} (e); offset was added to channels for better visualization. The Hexavariate \textit{Tai-Chi} signal contained multichannel recordings from two 3D sensors attached to left and right ankles of a human body moving in a Tai-Chi sequence \cite{rehman2009MEMD} and shown in Fig. \ref{InpTestSig} (f) where selected channel were given an offset for better visualization. 

We performed several experiments on the input signals by introducing both balanced and unbalanced multivariate noise. In the balanced case, noise power in all channels was equal while noise with different variance (power) was added across channels in the unbalanced case. Noisy versions of the input signals were obtained by adding multivariate additive wGn to the signal. The denoising performance is assessed through signal-to-noise ratio (SNR) of the denoised signal; we report output SNR values for each channel individually as well as the average reconstructed SNRs over all channels. 

For a fair comparison, we used Daubechies wavelet filters with eight vanishing moments (db-$8$) along with $k=5$ decomposition levels for all wavelet based approaches. Rest of the parameters for comparative methods were set according to specifications provided by the authors in their publications. The parameters specific to the proposed MGWD were chosen to be $L=M\times 28$ and $P_{fa}=0.005$ in subsequent experiments.
\subsection{Effect of correlation of noise on denoising performance}
This experiment is specifically designed to show how the consideration of noise correlation within the proposed \textit{MGWD} method enables it to perform better in case of correlated noise when compared to the established \textit{MMD} and \textit{MWD} methods. In this regard, the input `Sofar'  bivariate signal, known for its strongly correlated subtle variations, is corrupted by artificially adding multichannel noise with and without correlations. 
%The experiment is designed by  of `Sofar' signal by adding the correlated and uncorrelated bivariate noise at input SNR $=10$ dB. 
Subsequently, comparative methods are applied on the noisy signals and the resulting denoised signals are displayed in Fig. \ref{ToyEx}. Here, the original signal is also plotted in the background to demonstrate how closely the denoised signal resembles the original one. In addition, the output SNRs for denoised signal from each method are also printed for a more meaningful analysis of the recovered signal.
In Fig. \ref{ToyEx}, the case of uncorrelated noise (i.e., correlation coefficient of noise $\rho=0$) is shown in the first column while the denoised signals for correlated noise at $\rho=0.25$ \& $0.75$ are respectively displayed in second and third columns. 
 
Observe from Fig. \ref{ToyEx} that the denoised signals by the proposed \textit{MGWD} method (shown in bottom row) obtain much better estimate of the original signal for all correlated and uncorrelated noise cases compared to the denoised signals from \textit{MMD} and \textit{MWD} methods (shown in top and middle row respectively). Furthermore, the efficacy of the use of noise correlations within the proposed framework is also verified by the ability of our method to extract exaggerated signal details with increase in noise correlations that is also evident from the improved output SNRs. On the contrary, the \textit{MWD} method observes loss of signal details and artifacts when moving from uncorrelated to correlated noise cases, see top row of Fig. \ref{ToyEx}. 

Furthermore, the \textit{MWD} method yields significant artifacts which seem to strengthen with increase in noise correlations. The \textit{MMD} method, owing to the use of channel dependencies within its framework, recovers increased amount of signal details in case of correlated noises though with exaggerated amplitudes that results in comparatively lower SNRs.

\subsection{Output SNR versus input SNR}
Table. \ref{synthetictable} and Table. \ref{realtable} give the output SNR values obtained by applying the \textit{MMD}, \textit{MWD}, \textit{MWSD}, \textit{MEMD-IT} and the proposed \textit{MGWD} methods on synthetic and real signals corrupted with multivariate wGn noise of averaged input SNR of $-5$ dB, $0$ dB, $5$ dB and $10$ dB. We report results from both the balanced and unbalanced cases. For each denoising method, the SNR values in the upper row correspond to the balanced noise while the lower row reports output SNR values for the unbalanced case. Both tables report output SNR values of all channels separately along with the average reconstructed joint SNRs to demonstrate how well the comparative methods performed across different channels. The sub-columns labeled as `C$m$' list the output SNRs for the $m$th channel while average output SNRs are listed under the label `Avg' in the Table I. Each reported output SNR value was obtained by averaging $K = 50$ iterations for each input noise realizations.
\begin{figure*}[t]
	\centering
	\begin{subfigure}{.22\textwidth}
		\centering
		\includegraphics[width=\linewidth]{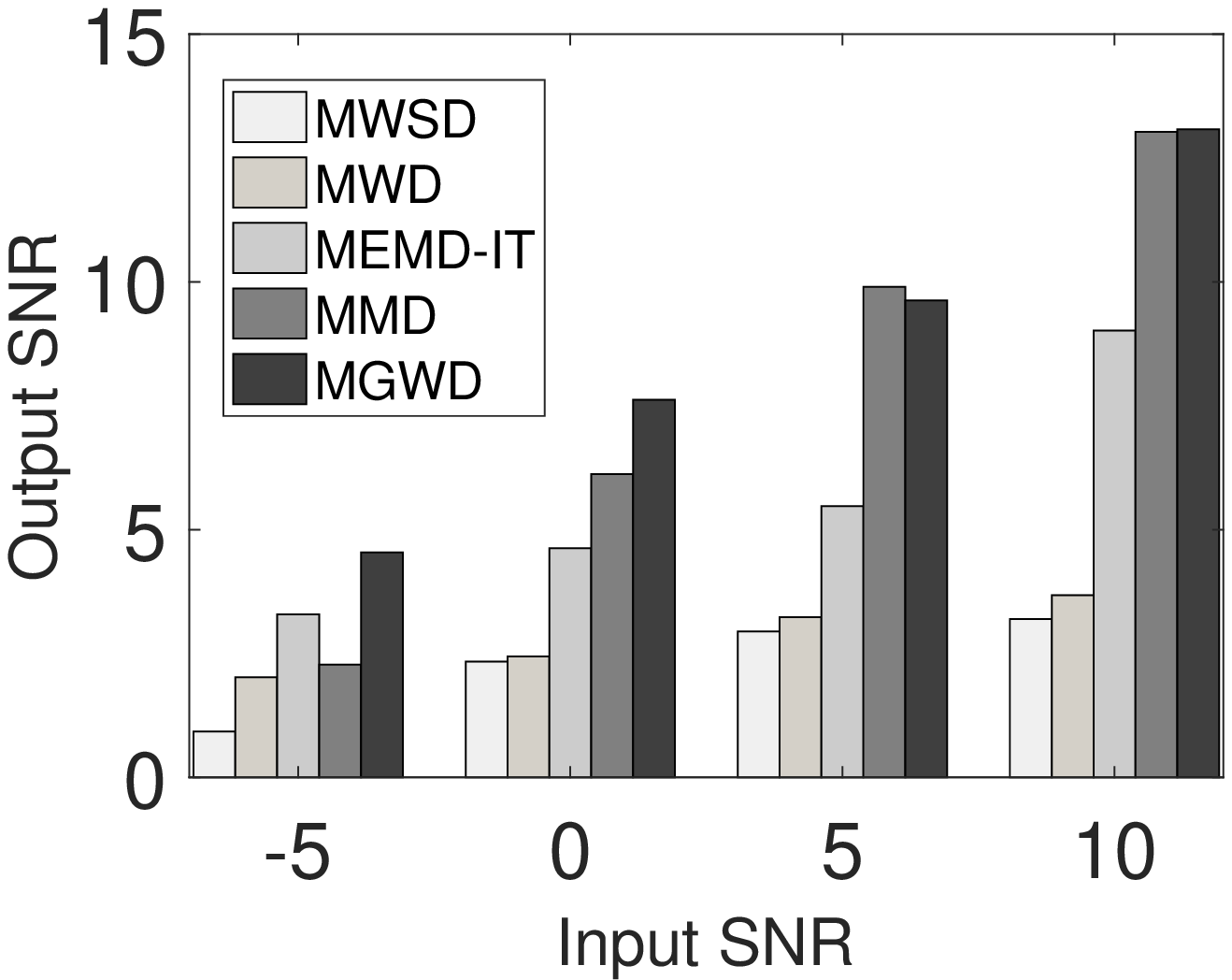}
		\caption{}
		\label{HvyDopp}
	\end{subfigure}
	%	\vspace{-6}
	\begin{subfigure}{.22\textwidth}
		\centering
		\includegraphics[width=\linewidth]{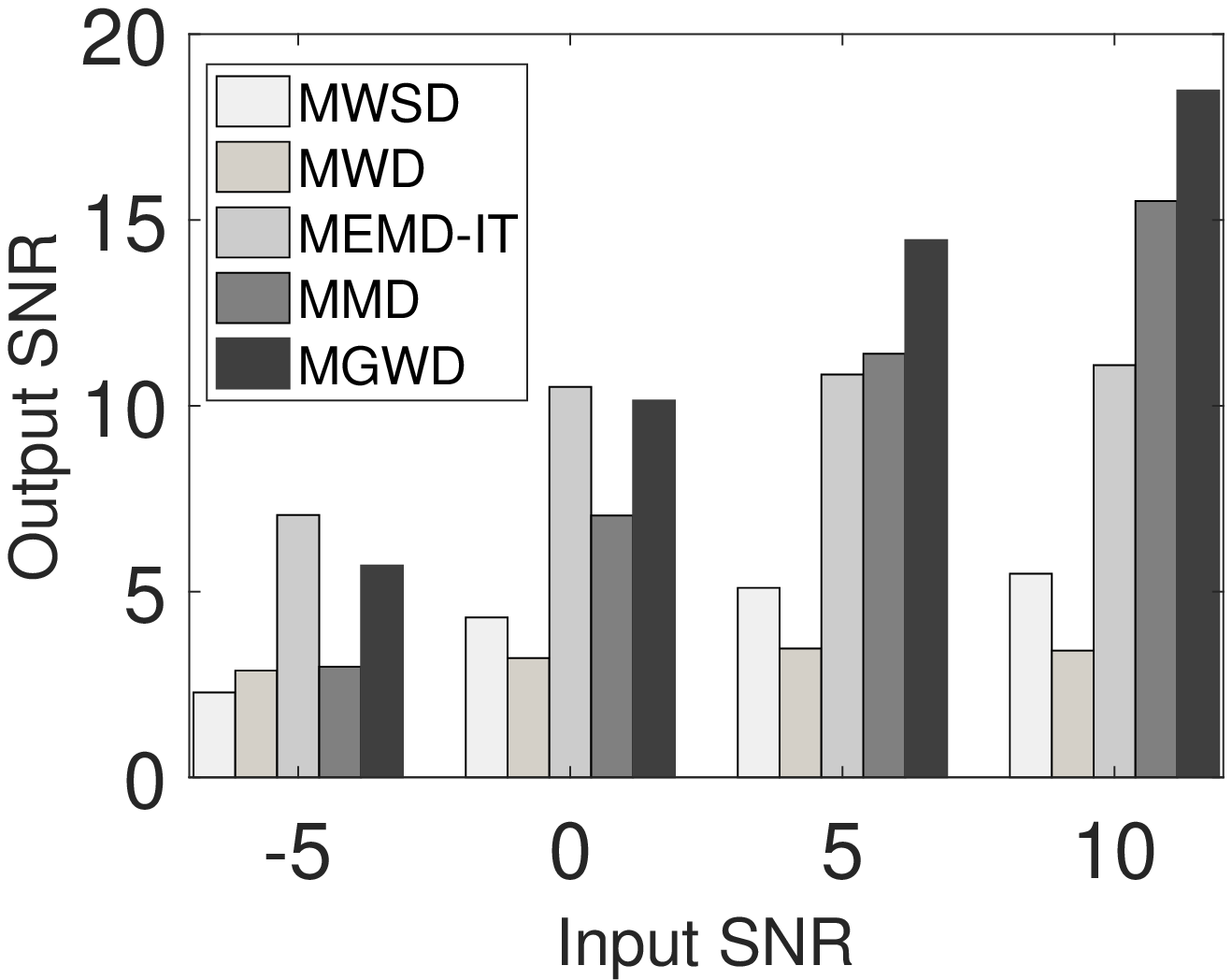}
		\caption{}
		\label{WLS}
	\end{subfigure}
	%	\vspace{-6}
	\begin{subfigure}{.22\textwidth}
		\centering
		\includegraphics[width=\linewidth]{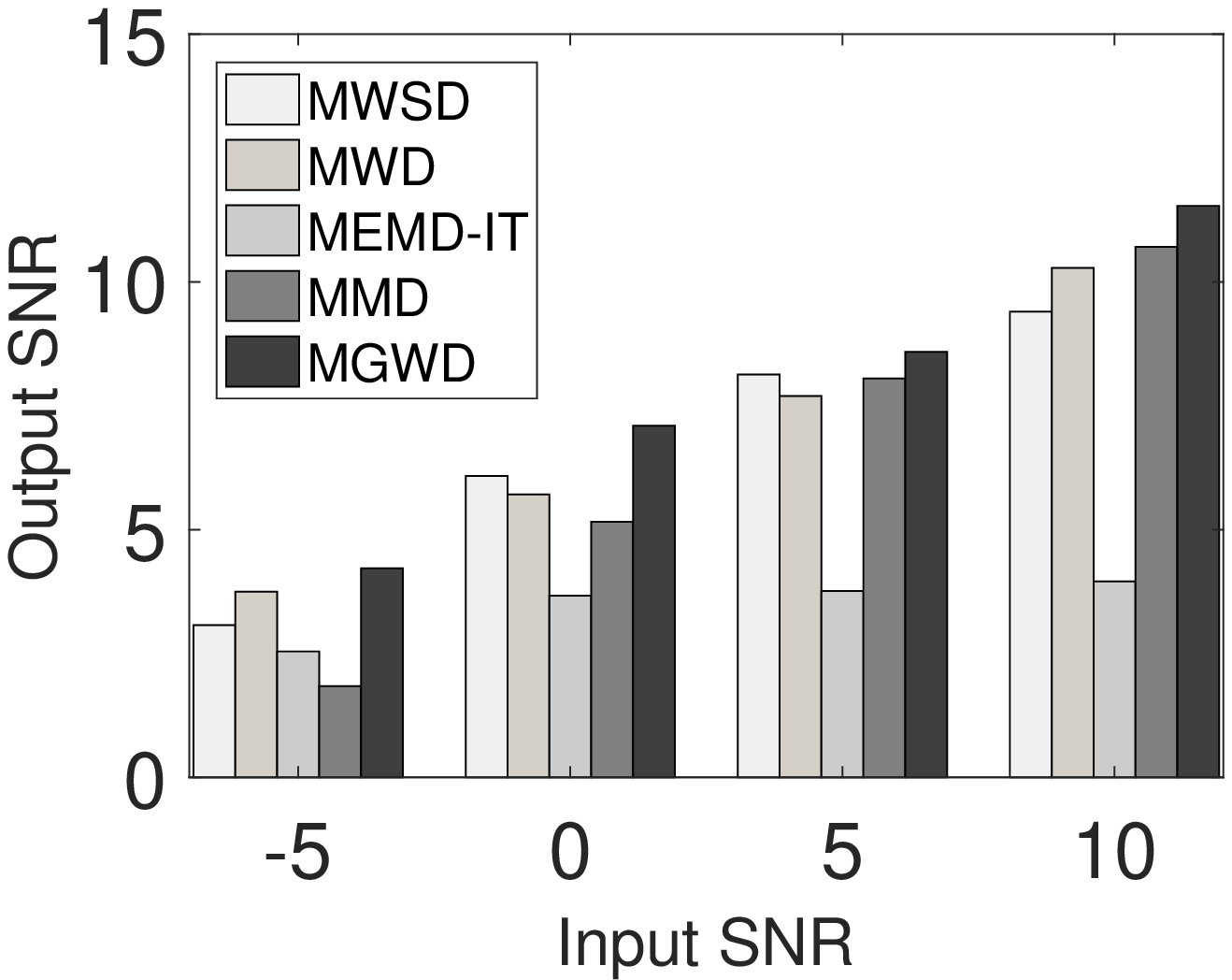}
		\caption{}
		\label{HvyDopp}
	\end{subfigure}
	%	\vspace{-6}
	\begin{subfigure}{.22\textwidth}
		\centering
		\includegraphics[width=\linewidth]{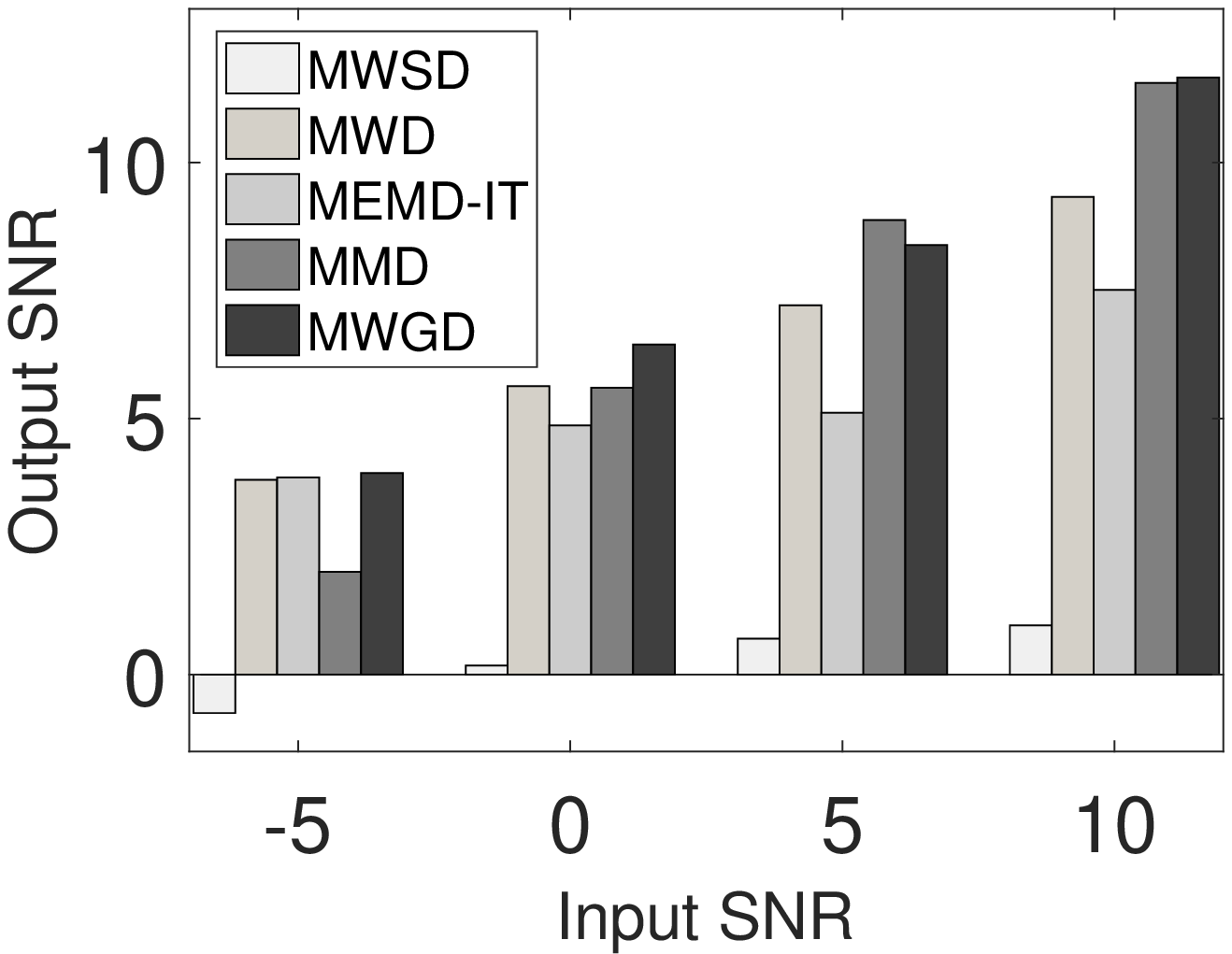}
		\caption{}
		\label{WLS}
	\end{subfigure}
	%	\vspace{-6}
	
	\begin{subfigure}{.22\textwidth}
		\centering
		\includegraphics[width=\linewidth]{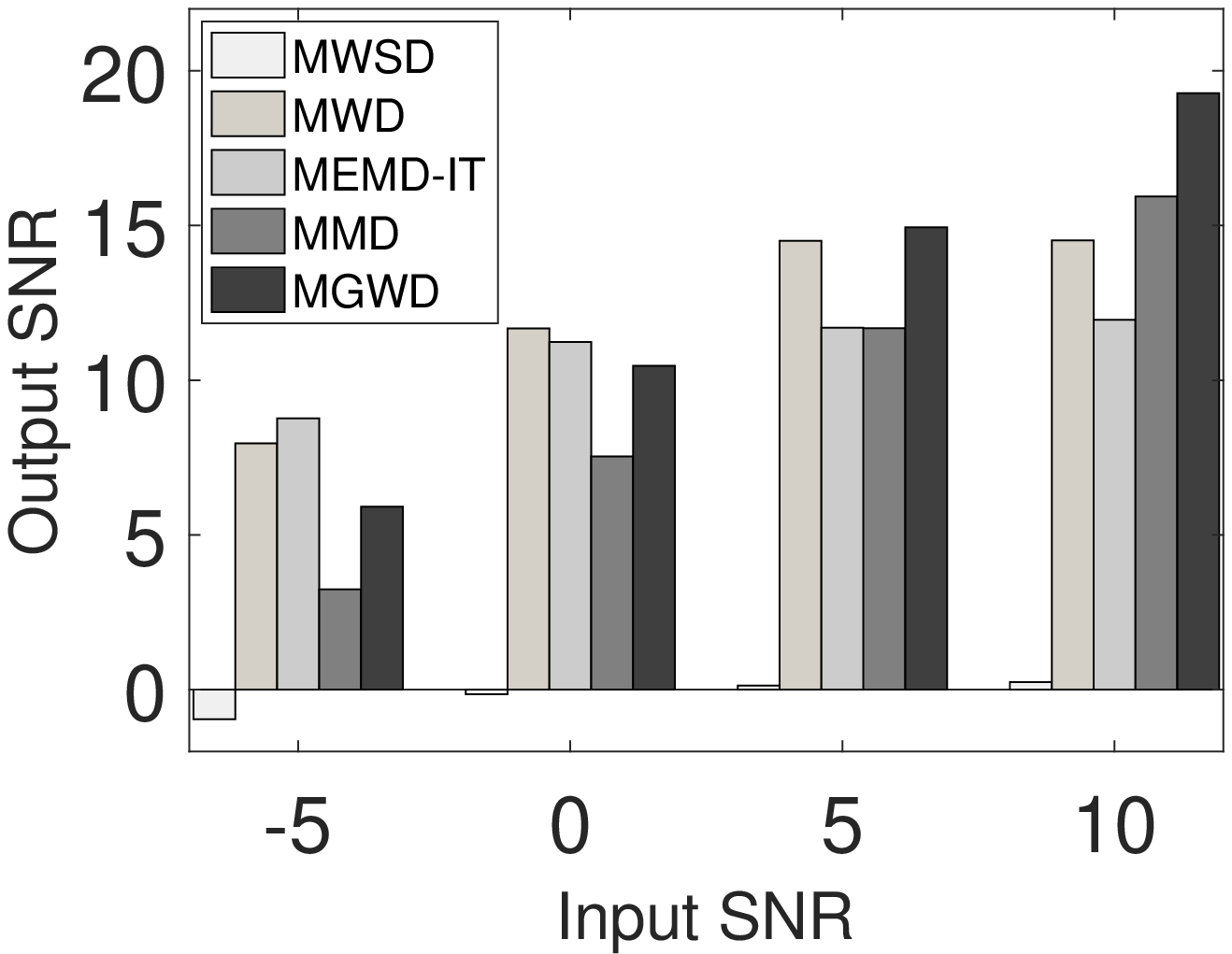}
		\caption{}
		\label{HvyDopp}
	\end{subfigure}
	%	\vspace{-6}
	\begin{subfigure}{.22\textwidth}
		\centering
		\includegraphics[width=\linewidth]{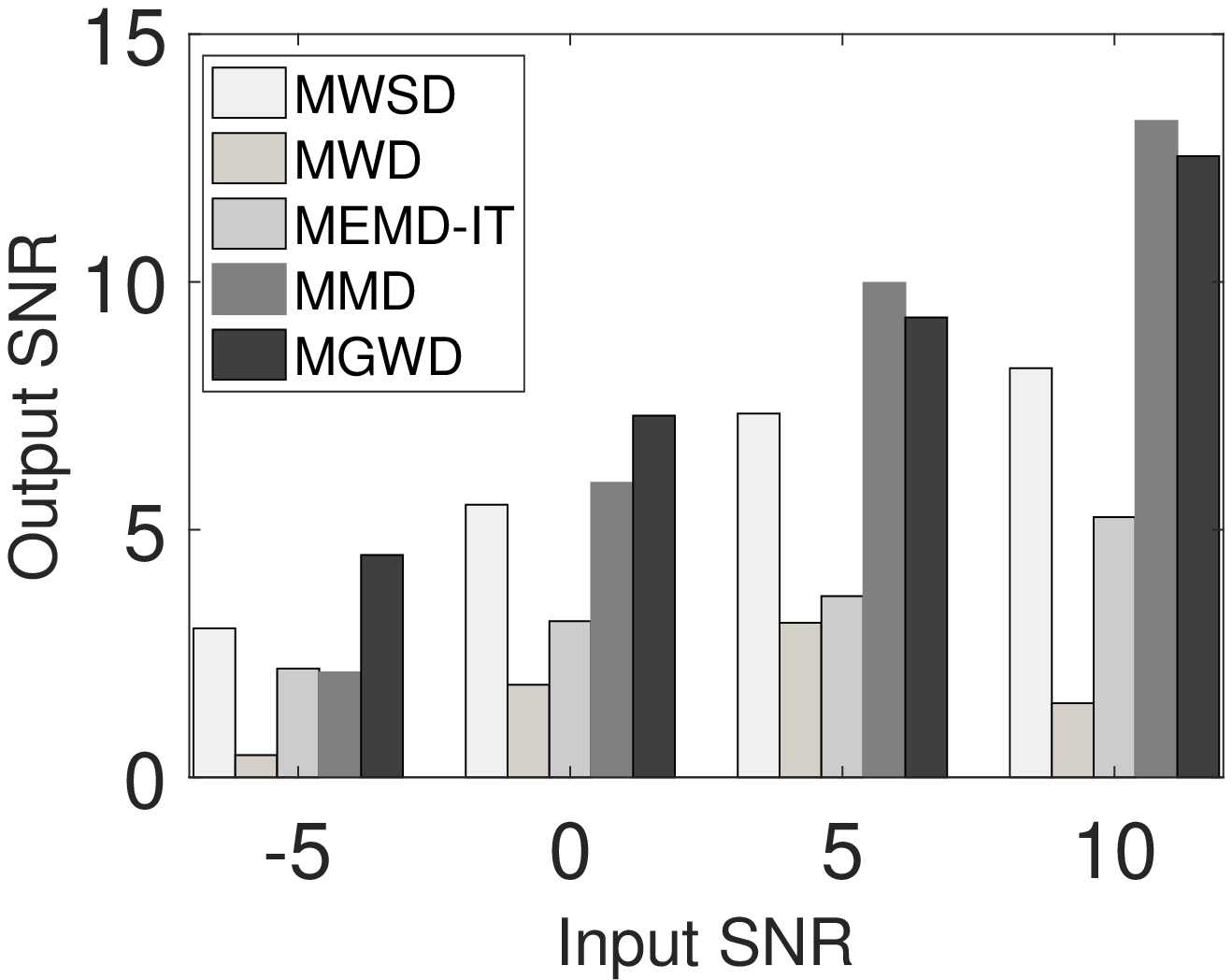}
		\caption{}
		\label{WLS}
	\end{subfigure}
	\begin{subfigure}{.22\textwidth}
		\centering
		\includegraphics[width=\linewidth]{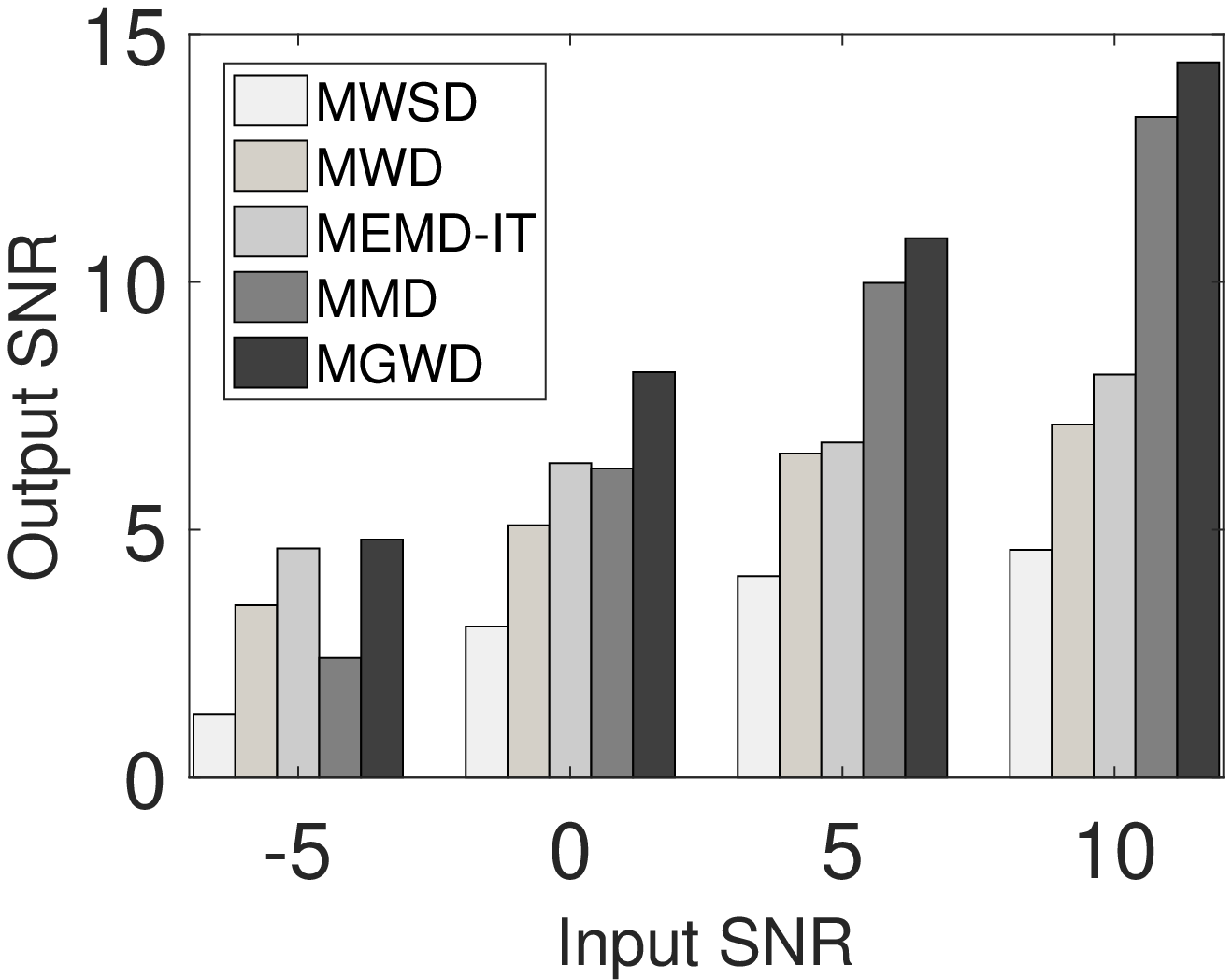}
		\caption{}
		\label{HvyDopp}
	\end{subfigure}
	\caption{Denoising performance comparison of various multivariate signal denoising methods at different noise levels on the hexavariate `Tai Chi signal'. The bar graphs corresponding to the output SNRs by various methods for channel $1\rightarrow 6$ are plotted in (a)-(h) respectively. The average reconstructed SNR values across all six channels are plotted in (g).}
	\label{BarTaichi}
	\vspace{-3mm}
\end{figure*}

Observe from Table. \ref{synthetictable} that for synthetic \textit{Heavy Sine \& Doppler} signal, the proposed \textit{MGWD} method outperformed the comparative methods in most cases while \textit{MWD} and \textit{MMD} remained competitive and manage to beat our method at times. This superior performance by \textit{MGWD} may be attributed to the multichannel dependencies of noise used within it. 

For synthetic `Bumps-Blocks' quadrivariate signal, the \textit{MGWD} mostly yields highest output SNRs apart from a few instances of unbalanced noise. Moreover, the proposed method shows consistent performance across all channels when compared to the \textit{MMD} and \textit{MWD} which also remain competitive. Observe that both \textit{MEMD-IT} and \textit{MMD} yield comparatively lower SNRs for the first channel containing 'Blocks' signal may be due the fact that MEMD is more suited to oscillatory signals and less for piece-wise constant.

Moving on to the assessment of quantitative results for real world signals where the cases `Sofar' and `Eye Roll EOG' bivariate signals are of particular importance because of the strong correlation within two channels of both signals. The subtle variations in the `Sofar' signal and sharp peaks within the `EOG' signal further add to the complexity of these signals. Here, the proposed \textit{MGWD} method showed superior results against the comparative methods for both these signals where margin of differences between the proposed and the second best method are very high. Although, for higher input SNR values, the margin reduces gradually but still in majority of cases it remains statistically significant. Once again, the \textit{MMD} method yields second best results owing to its ability to incorporate channel dependencies. 
Here, \textit{MWD} also showed competitive performance that may be due to the use of PCA.

Another real signal with strong correlations and highly varying dynamic structure is the `Health Monitoring' signal containing \textit{roll}, \textit{pitch} and \textit{yaw} movements of a person's arms during a weight lifting exercise. Here, it is evident from Table \ref{realtable} that the \textit{MGWD} and \textit{MMD} jointly yield highest output SNR values which may be attributed to their ability to consider inter-dependencies across channels. Observe that the \textit{MWD} shows inconsistent performance across channels that may be due to the exclusive univariate thresholding functions for each channel. On the contrary, \textit{MWSD} and \textit{MEMD-IT} show below par performance and lacked consistency across noise levels.

Fig. \ref{BarTaichi} displays denoising results for the hexavariate \textit{Taichi} signal using bar graphs of averaged output SNRs by various denoising methods. These results were obtained for balanced multivariate noise case for a range of input SNRs $=-5,\ 0,\ 5, \ 10$ dB. Each of the Fig. \ref{BarTaichi} (a-f) respectively plots the input versus output SNR values for channel $1$ to $6$ while the  Fig. \ref{BarTaichi} (g) plots the average reconstructed SNRs for the multivariate signal. The challenge posed by this input signal to the denoising methods involves its diverse signal content across channels where higher number of channels add to its complexity. Here, the proposed \textit{MGWD} method yields highest output SNRs in most cases where on a few instances in channel $2$, $4$ and $6$, \textit{MMD} and \textit{MWD} yield better performance. 

\begin{figure*}[h!]
	\centering
	\begin{subfigure}{.4\textwidth}
		\centering
		\includegraphics[width=\linewidth]{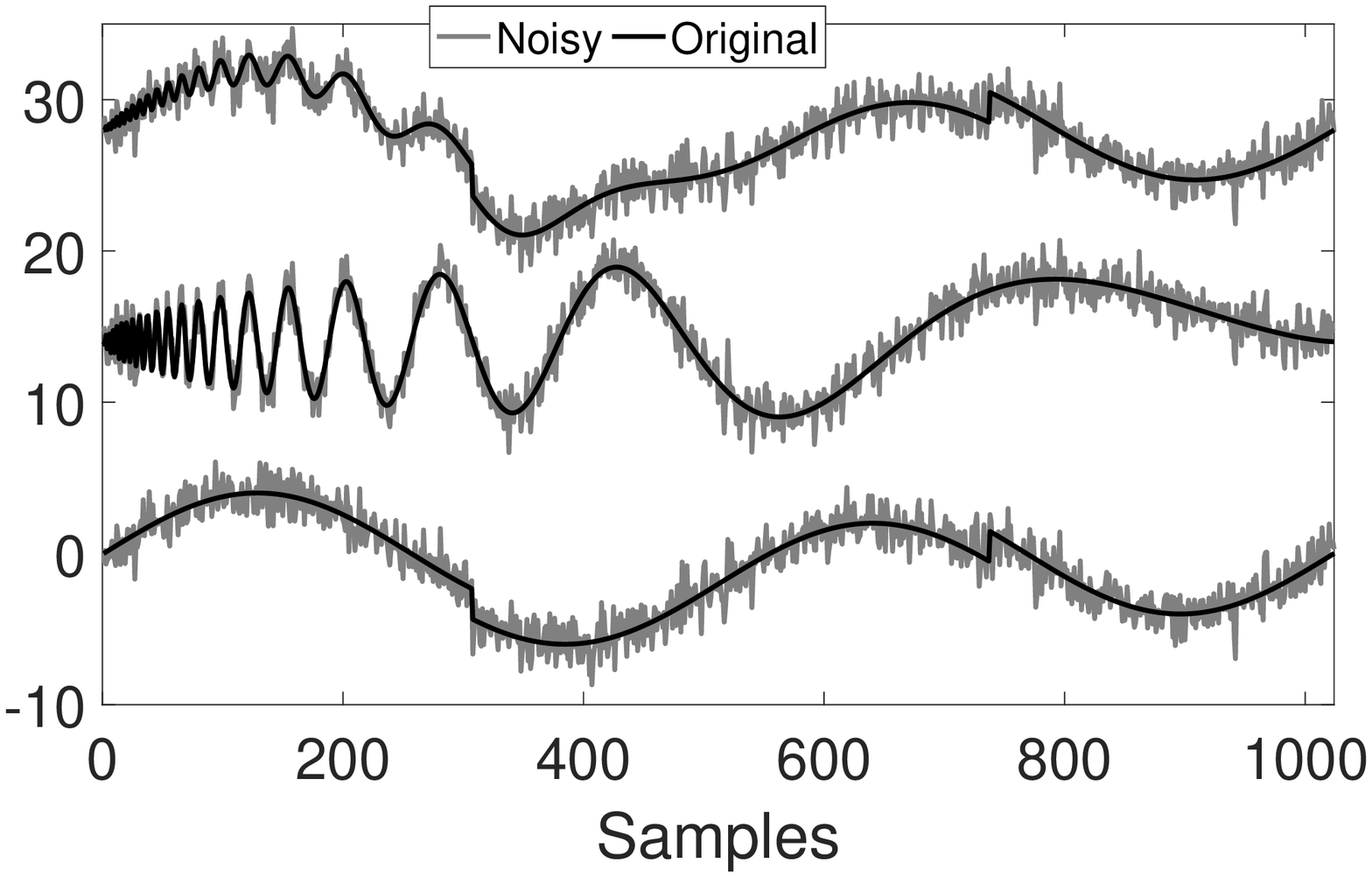}
		\caption{Noisy}
		%		\label{HvyDopp}
	\end{subfigure}
	\hspace{-8mm}
	\begin{subfigure}{.4\textwidth}
		\centering
		\includegraphics[width=\linewidth]{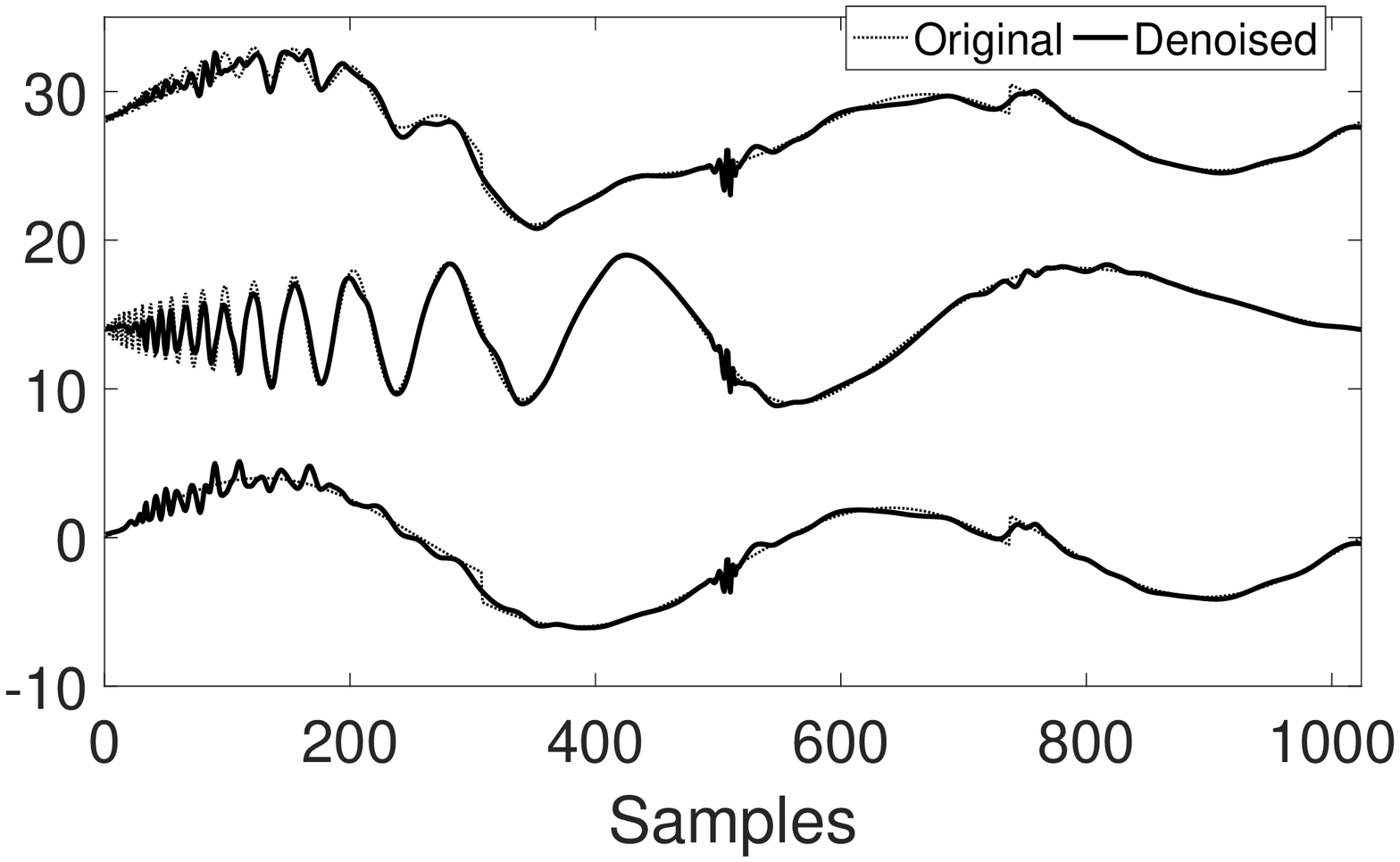}
		\caption{MWD}
		%		\label{WLS}
	\end{subfigure}
	%	\vspace{-6}
	\begin{subfigure}{.4\textwidth}
		\centering
		\includegraphics[width=\linewidth]{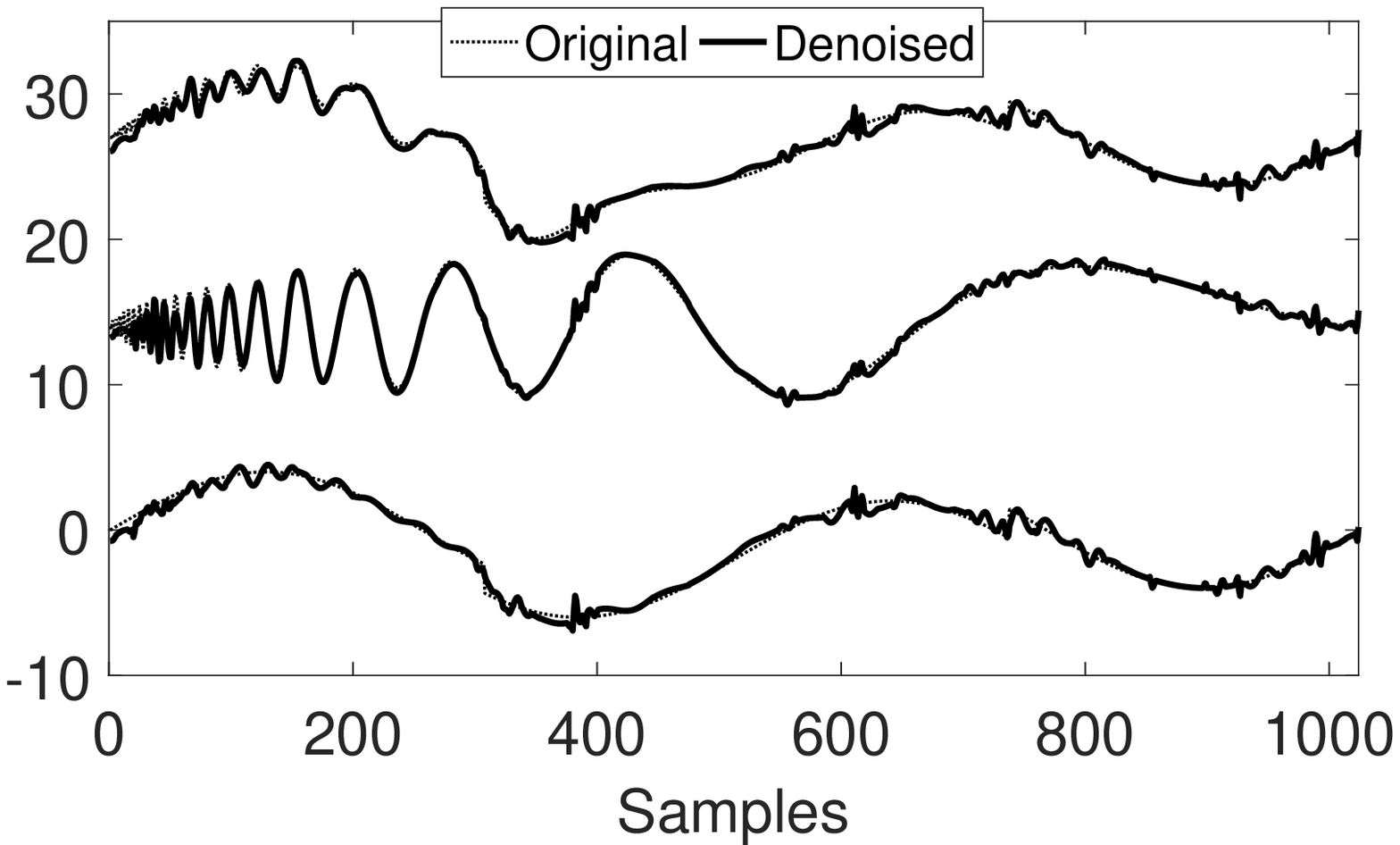}
		\caption{MMD}
		%		\label{HvyDopp}
	\end{subfigure}
	\hspace{-8mm}
	\begin{subfigure}{.4\textwidth}
		\centering
		\includegraphics[width=\linewidth]{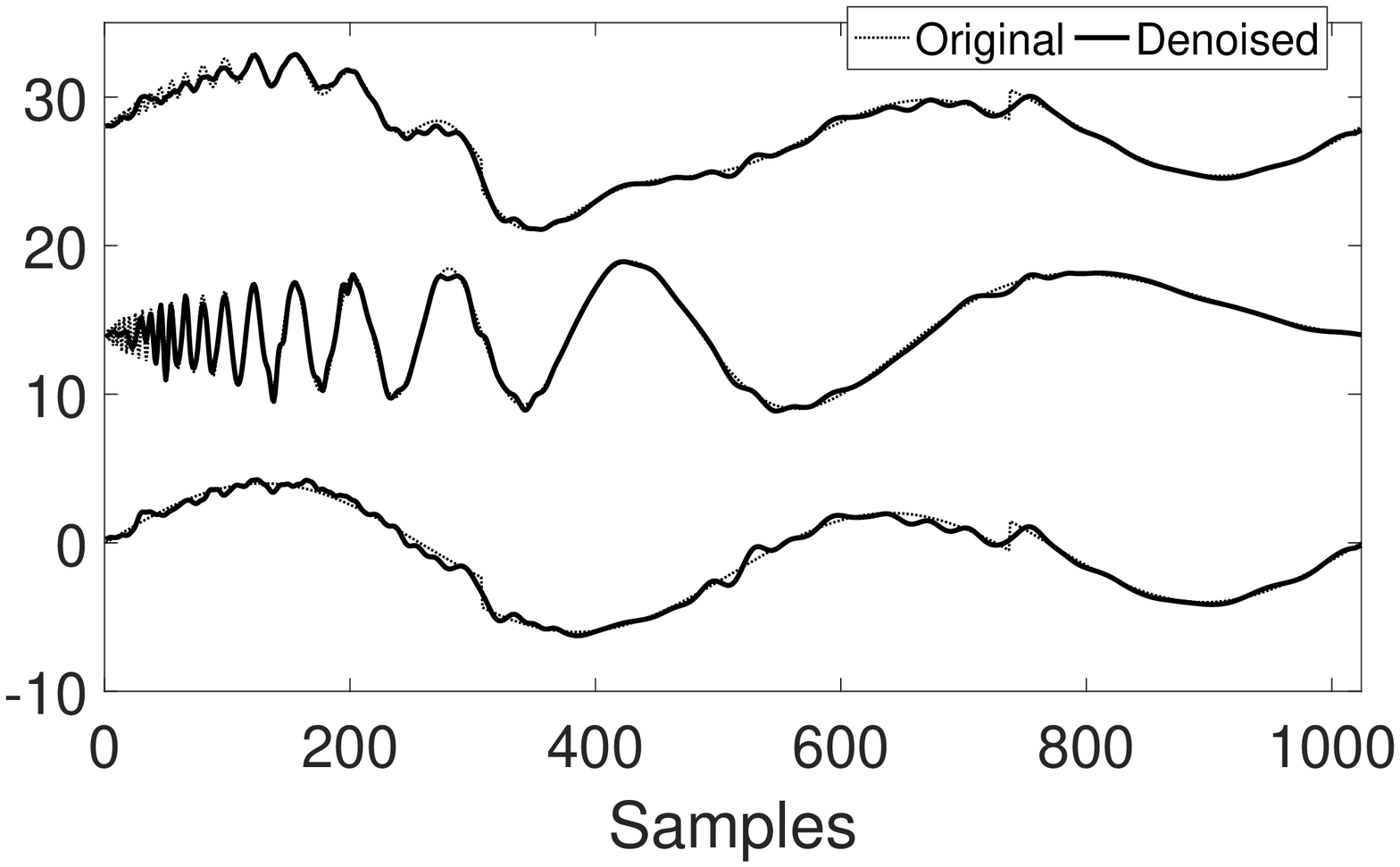}
		\caption{MGWD}
		%		\label{WLS}
	\end{subfigure}
	%	\vspace{-6}
	\caption{Noisy `Heavy Sine \& Doppler' signal and its denoised versions by different methods at the input SNR level $=10$ dB.}
%		\vspace{-4mm}
	\label{HvyDopp}
\end{figure*}

Furthermore, the proposed \textit{MGWD} along with the \textit{MMD} yielded consistent denoising results despite the diversity of the spectrum of \textit{Tai Chi} signal across channels. A possible reason could be that these methods perform purely multivariate thresholding where \textit{MGWD} detects and rejects multichannel noise by explicitly considering the noise correlations. On the contrary, \textit{MMD} detects signal oscillations by incorporating correlations of noisy IMFs. Moreover, the significantly higher SNRs by the \textit{MGWD} method indicate its robustness to signal with diverse structure, i.e., lack of correlation in the signal content across channels. The rest of the methods including \textit{MWD}, \textit{MWSD} and \textit{MEMD-IT} perform extremely inconsistently which may due to the use of the univariate thresholding functions for removing correlated multichannel noise.

\subsection{Qualitative analysis of denoising performance}
\begin{figure*}[t!]
	\centering
	\begin{subfigure}{.4\textwidth}
		\centering
		\includegraphics[width=\linewidth]{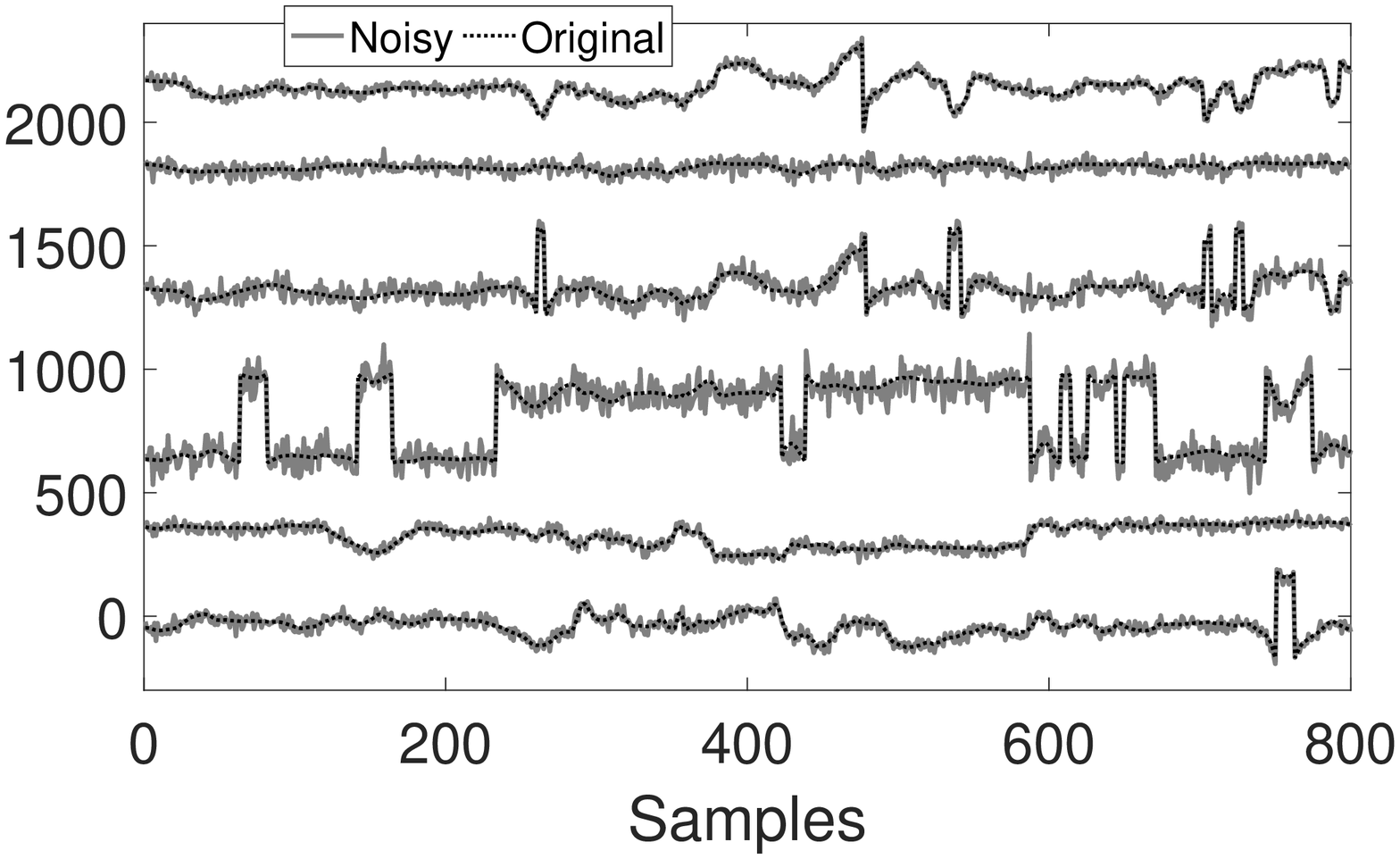}
		\caption{Noisy}
		%		\label{HvyDoppt}
	\end{subfigure}
	\hspace{-6mm}
	\begin{subfigure}{.4\textwidth}
		\centering
		\includegraphics[width=\linewidth]{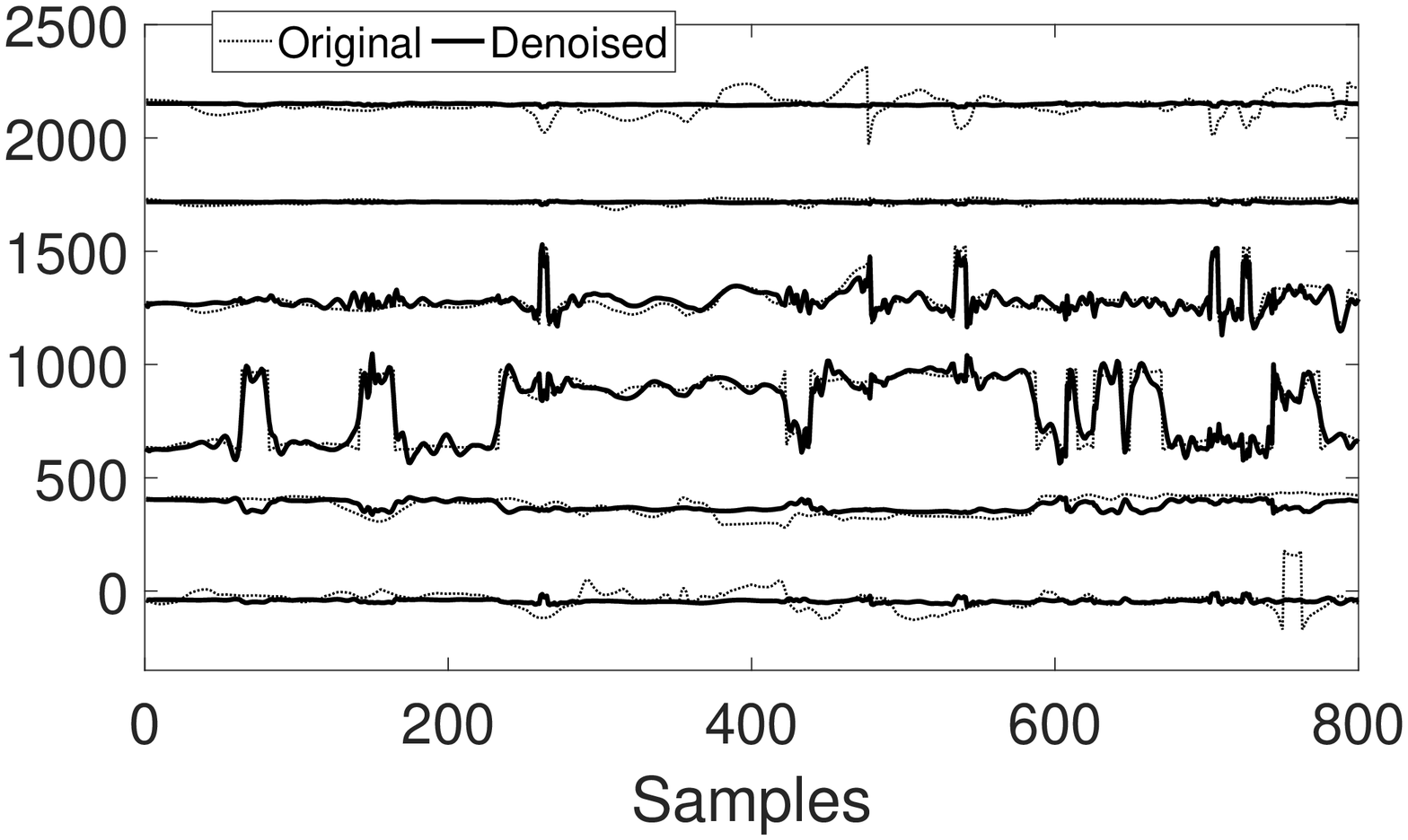}
		\caption{MWD}
		%		\label{WLS}
	\end{subfigure}
	%	\vspace{-2mm}
	
	\begin{subfigure}{.4\textwidth}
		\centering
		\includegraphics[width=\linewidth]{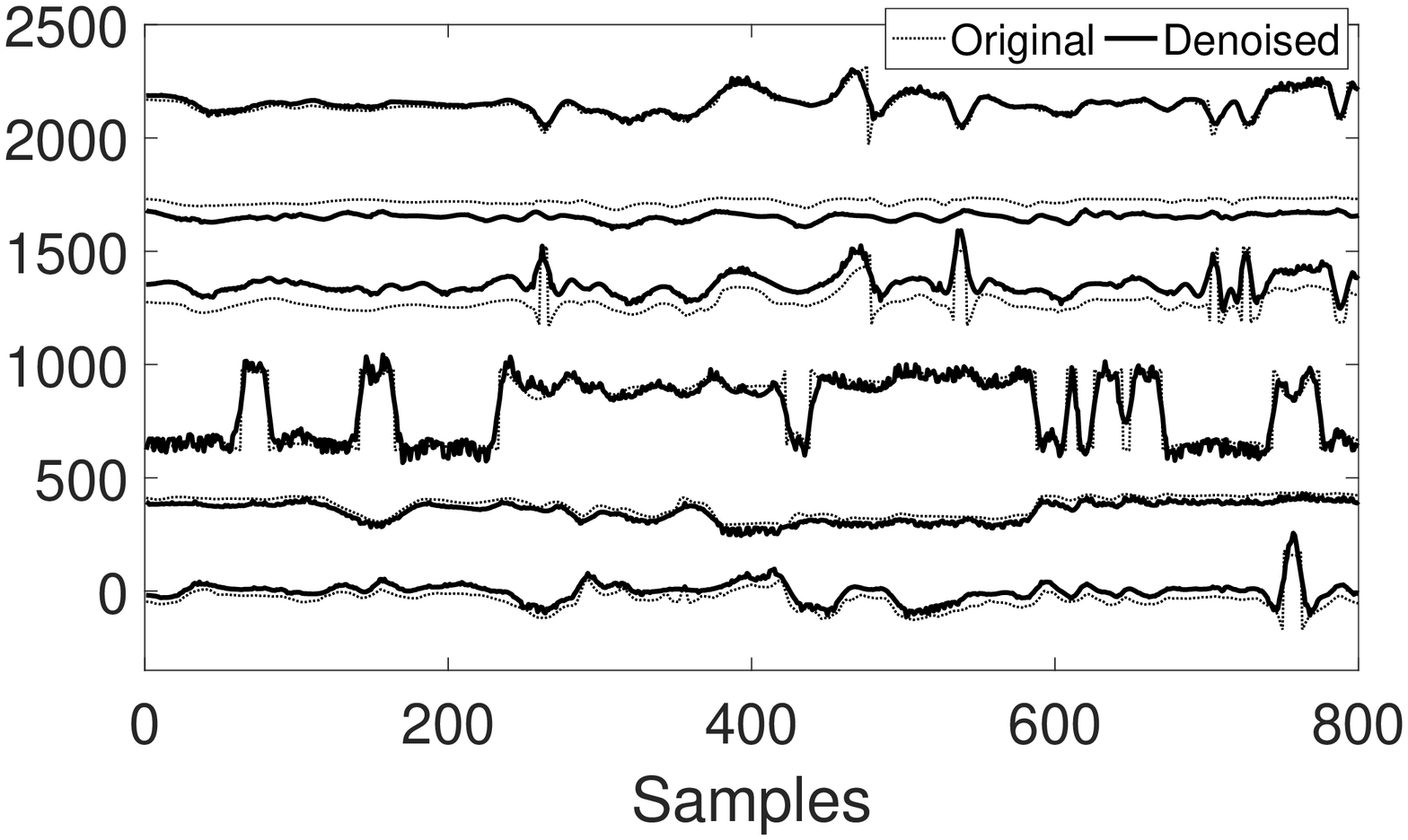}
		\caption{MWSD}
		%		\label{HvyDopp}
	\end{subfigure}
	\hspace{-6mm}
	\begin{subfigure}{.4\textwidth}
		\centering
		\includegraphics[width=\linewidth]{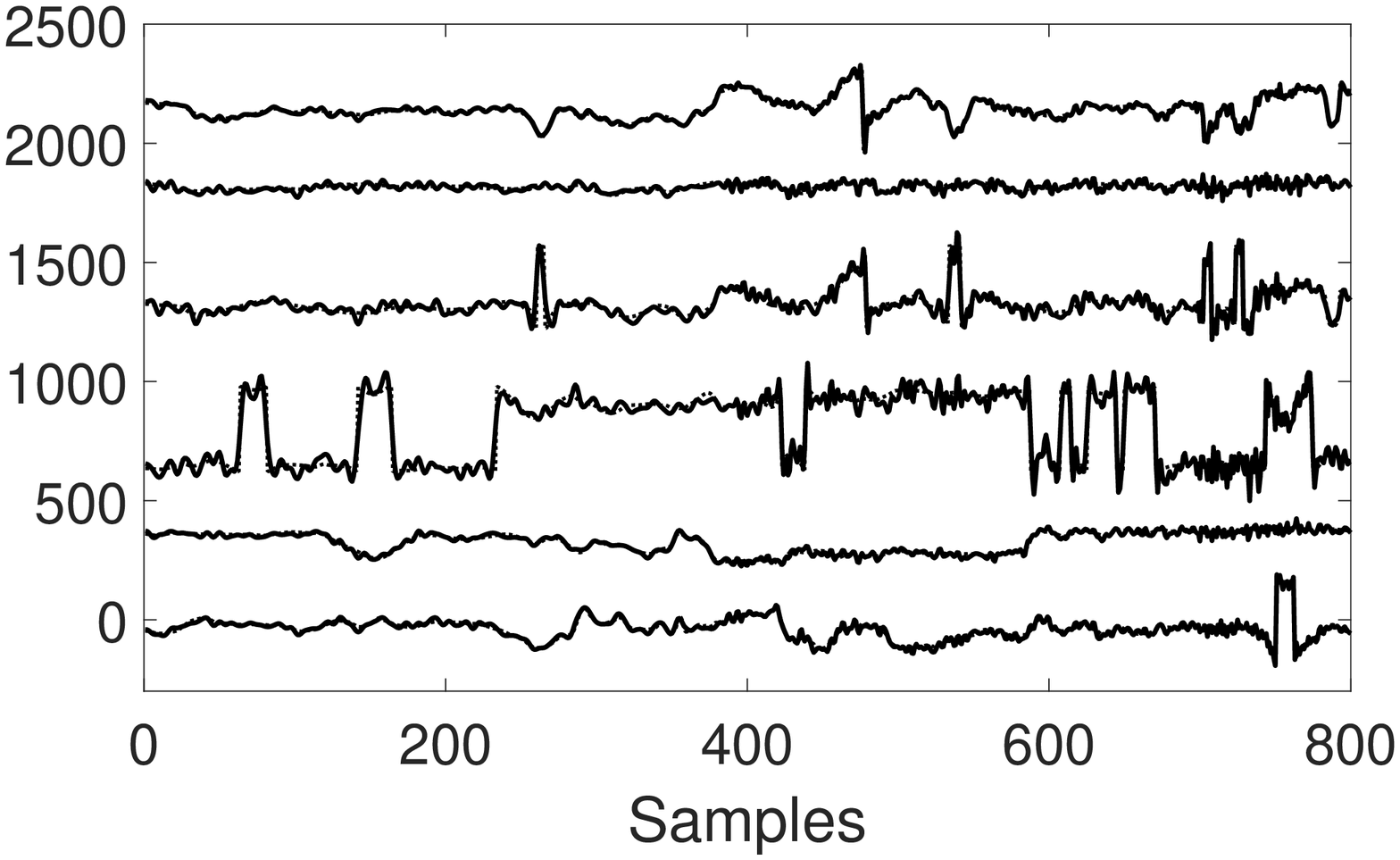}
		\caption{MGWD}
		%		\label{WLS}
	\end{subfigure}
	\caption{Noisy and denoised hexavariate 'Tai Chi' signals obtained from applying denoising methods on the the noisy 'Tai Chi' signal corresponding to input SNR $=10$ dB.}
	\label{Taichi}
	\vspace{-3mm}
\end{figure*}

To visually illustrate how well the proposed method recovered a true multivariate signal from its noisy observations, the denoised \textit{Sofar}, \textit{Heavy Sine \& Doppler} and \textit{Taichi} signals obtained from the proposed method are shown in Fig. \ref{ToyEx} - Fig. \ref{Taichi}. The discussion on the visual results of \textit{Sofar} signal in Fig. \ref{ToyEx} is already present in a preceding subsection demonstrating the efficacy of our method for correlated noise.

Fig. \ref{HvyDopp} displays the denoised trivariate `Heavy Doppler' signal obtained from \textit{MWD}, \textit{MMD} and \textit{MGWD} respectively in Fig. \ref{HvyDopp}  (b-d) where the noisy signal with input SNR $=10$ dB is plotted in the Fig. \ref{HvyDopp} (a). It can be seen from the Fig. \ref{HvyDopp} (d) that the proposed \textit{MGWD} method accurately recovered the signals in each channel except for some initial fluctuations in the \textit{Doppler} signal in the second channel.
The denoised signal by the \textit{MWD} method also yielded an overall good estimate of the original signal, however, the recovered signals showed few artifacts in the regions of high variations, see Fig. \ref{HvyDopp} (b). \textit{MMD}, on the contrary, yielded artifacts despite good signal recovery.

Fig. \ref{Taichi} shows denoising results of the comparative methods on the real world \textit{Taichi} signal at input SNR $=10$ dB. The noisy signal is shown in Fig. \ref{Taichi} (a) and the denoised signals from \textit{MWD}, \textit{MWSD} and \textit{MGWD} are respectively shown in Fig. \ref{Taichi} (b - d). It can be seen that, even with large number of input channels that exhibit complex structure and subtle variations; the proposed method recovered the true signal fairly accurately (see Fig. \ref{Taichi} (d)). The channels exhibiting high dynamics, i.e., channels $1,3,4,6$, were recovered accurately by our method whereas the comparative methods failed to do so because the \textit{MWSD} suffered from artifacts and phase distortions while \textit{MWD} was inconsistent across channels, see \ref{Taichi} (b \& c).

\subsection{Comparison with univariate denoising methods}
Fully multivariate denoising methods are still emerging and not yet fully established. As a result,  univariate denoising methods are often utilized separately and independently on multiple channels of a multivariate signal. By doing that, inter-channel dependencies within data channels from multivariate signal are not incorporated in the denoising process. Here, we are interested in showing that for given noisy multivariate observations exhibiting inter-channel dependencies, utilizing the proposed denoising method does offer improved performance when compared against the univariate denoising methods applied channel-wise on multivariate data. 

To demonstrate that, we evaluate our method against two popular univariate denoising methods \textit{BLFDR} \cite{lavrik2008BLFDR} and \textit{EMD-IT} \cite{kopsinis2009EMD-IT} that were applied on each channel separately for denoising multivariate signals. The input signals used in this experiment included the bivariate `Sofar', bivariate `EOG' and trivariate `Heavy Doppler' signals. The signals were corrupted with additive multivariate wGn having correlated balanced noise corresponding to input SNR $=-5, 0$ and $5$ dB. The quantitative results in terms of output SNR are shown in Table. \ref{tableunivariate} where it can be observed that the proposed method outperformed univariate methods for all cases. That was expected given the ability of the proposed method to exploit and effectively incorporate inter-channel correlation structure in the denoising process.   
\section{Conclusions and discussion}
In this article, we have presented a multivariate signal denoising algorithm that uses a novel multivariate Goodness of Fit (GoF) test based on empirical distribution function (EDF) statistic operating at multiple data scales obtained from discrete wavelet transform (DWT). The multivariate GoF test uses squared Mahalanobis distance (MD) measure to first transform DWT coefficients from multivariate noisy data residing in $\mathcal{R}^M$ to $\mathcal{R}_+$. Owing to the one-to-one correspondence between the empirical distributions of original multivariate data coefficients and its transformation via MD, our proposed multivariate GoF test utilizes the distribution of quadratically transformed multivariate additive white Gaussian noise (wGn) vectors as the \textit{reference CDF} which is compared to the \textit{test EDF} corresponding to the distribution of quadratically transformed multivariate noisy data coefficients. In this regard, a modified Anderson Darling (AD) test statistic $\tau$ is proposed to quantify the difference between the reference and test EDFs, which is then compared against a threshold value corresponding to a desired probability of false alarm. The detected noisy coefficients are discarded using a fully multivariate thresholding function that operates collectively on all channels of input data.  

The three important contributions in this work include: i) using Mahalnobis distance to define the multivariate GoF test that not only enables a unique definition of multivariate CDF (multivariate EDF) but also makes its computation and further processing convenient and efficient; ii) operating the multivariate GoF test based on Mahalanobis distance at multiple data scales obtained from DWT. This way, GoF framework which has been traditionally used for detection related applications is extended for denoising problems; iii) a fully multivariate thresholding function operating collectively on all data channels while fully incorporating inter-channel correlations in input noisy data.   
   \begin{table}[t]
   	\caption{Performance comparison of the proposed method against univariate denoising methods applied channel-wise to multivariate data exhibiting covariance among its channels. The comparison is given for `Sofar', `EOG' and `Heavy Doppler' datasets for input SNR values of $-5$, $0$, and $5$ dB. }
   	\large
   	%	\left
   	\scalebox{0.45}{
   		%		\resizebox{1\textwidth}{!}{
   		\setlength\extrarowheight{6pt}
   		\begin{tabular}{|c||cccc||cccc||cccc||}\thickhline
   			\textbf{Avg. Input SNR}
   			&\multicolumn{3}{c}{\textbf{-5}} & &\multicolumn{3}{c}{\textbf{0}} & &\multicolumn{3}{c}{\textbf{5}} &\\
   			\hline
   			\textbf{Channels}
   			&\textbf{C1} &\textbf{C2} & \textbf{C3} & \textbf{Avg} &\textbf{C1} &\textbf{C2} &\textbf{C3} & \textbf{Avg} &\textbf{C1} &\textbf{C2} &\textbf{C3} & \textbf{Avg} \\
   			\thickhline
   			%				\thickhline
   			\textbf{Test Signal}
   			&\multicolumn{11}{c}{\textbf{Sofar Signal}} &\\
   			\hline 
   			%				\textbf{Channels}
   			%				&\textbf{C1} &\textbf{C2} & & \textbf{Avg} &\textbf{C1} &\textbf{C2} & & \textbf{Avg} &\textbf{C1} &\textbf{C2} & & \textbf{Avg} \\
   			%				\textbf{Inp. SNR}
   			%				&\textbf{-5} &\textbf{-5} & & \textbf{-5} & \textbf{0} &\textbf{0} & & \textbf{0} & \textbf{5} & \textbf{5} & & \textbf{5} \\
   			%				\hline
   			\textbf{EMD-IT}
   			& 6.02 & 6.94 && 6.48 & 9.92 & 10.66 && 10.29 & 14.05 & 13.89 && 13.97 \\
   			\hline
   			\textbf{BLFDR}
   			& 0.78 & 0.72 && 0.75 & 6.97 & 6.62 && 6.79 & 12.52 & 13.18 && 12.85 \\
   			\hline
   			\textbf{MGWD}
   			%			&\textbf{9.12} & \textbf{9.18} && \textbf{9.15} &\textbf{13.80} & \textbf{13.43} && \textbf{13.62} &\textbf{15.89} & \textbf{16.02} && \textbf{15.95}\\
   			& \textbf{9.83} & \textbf{9.89} && \textbf{9.86} & \textbf{14.17} & \textbf{13.88} && \textbf{14.03} & \textbf{17.12} & \textbf{16.92} && \textbf{17.02} \\
   			\thickhline
   			\textbf{Test Signal}
   			&\multicolumn{11}{c}{\textbf{EOG Signal}} &\\
   			\hline 
   			%				\textbf{Channels}
   			%				&\textbf{C1} &\textbf{C2} & & \textbf{Avg} &\textbf{C1} &\textbf{C2} & & \textbf{Avg} &\textbf{C1} &\textbf{C2} & & \textbf{Avg} \\
   			%				\textbf{Inp. SNR}
   			%				&\textbf{-5} &\textbf{-5} & & \textbf{-5} & \textbf{0} &\textbf{0} & & \textbf{0} & \textbf{5} & \textbf{5} & & \textbf{5} \\
   			%				\hline
   			\textbf{EMD-IT}
   			& 1.66 & 2.93 && 2.30 & 5.38 &  6.54 && 5.96 & 9.17 & 10.20 && 9.68 \\
   			\hline
   			\textbf{BLFDR}
   			& 0.16 & 0.49 &&  0.33 & 4.83 & 5.74 && 5.29 & 9.21 & \textbf{10.23} && \textbf{9.72} \\
   			\hline
   			\textbf{MGWD}
   			%			&\textbf{3.02} & \textbf{3.51} && \textbf{3.26} &\textbf{5.91} & \textbf{6.83} && \textbf{6.37} &\textbf{9.24} & 10.07 && 9.65 \\
   			& \textbf{3.55} & \textbf{4.49} && \textbf{4.02} & \textbf{6.31} & \textbf{7.90} && \textbf{7.11} & \textbf{9.28} & \textbf{10.51} && \textbf{9.90} \\
   			\thickhline
   			\textbf{Test Signal}
   			&\multicolumn{11}{c}{\textbf{HeavyDoppler Signal}} &\\
   			\hline 
   			%				\textbf{Channels}
   			%				&\textbf{C1} &\textbf{C2} & \textbf{C3} & \textbf{Avg} &\textbf{C1} &\textbf{C2} &\textbf{C3} & \textbf{Avg} &\textbf{C1} &\textbf{C2} &\textbf{C3} & \textbf{Avg} \\
   			%				\textbf{Inp. SNR}
   			%				&\textbf{-5} &\textbf{-5} & \textbf{-5}& \textbf{-5} & \textbf{0} &\textbf{0} &\textbf{0} & \textbf{0} & \textbf{5} & \textbf{5} &\textbf{5} & \textbf{5} \\
   			%				\hline
   			\textbf{EMD-IT}
   			& 6.09 & 4.91 & 5.24 & 5.41 & 11.27 & 9.31 & 10.18 & 10.25 & 15.61 & \textbf{13.61} & 14.22 & 14.48 \\
   			\hline
   			\textbf{BLFDR}
   			& 0.61 & 0.84 & 0.50 & 0.65 & 7.08 & 6.58 &    6.81 & 6.83 & 13.85  & 12.53 & 13.41 & 13.26 \\
   			\hline
   			\textbf{MGWD}
   			%			&\textbf{9.70} &\textbf{7.13} &\textbf{9.45} &\textbf{8.76} & \textbf{14.82} &\textbf{9.14} &\textbf{13.67} & \textbf{12.54} & \textbf{16.52} & 11.31 & \textbf{15.64} & \textbf{14.49} \\
   			& \textbf{8.93} & \textbf{6.69} & \textbf{7.48} & \textbf{7.70} & \textbf{13.26} & \textbf{10.59} & \textbf{11.45} & \textbf{11.77} & \textbf{16.54} & \textbf{14.39} & \textbf{14.54} & \textbf{15.16} \\
   			\thickhline
   	\end{tabular}}
   	\label{tableunivariate}
   	\vspace{-4mm}
   \end{table}

Benefiting from the advantages mentioned above, the proposed method has been shown to perform exceedingly well on a wide range of multivariate synthetic and real world signals. Of particular importance has been the ability of the proposed method to yield accurate denoising results consistently across \textit{all} input data channels for both \textit{correlated and uncorrelated} noise cases. The existing multivariate denoising approaches, on the other hand, have been found to be highly inconsistent in this regard which could be attributed to their lack of fully incorporating inter-channel correlations within input data due to the use of \textit{channel-wise} thresholding operation. Moreover, we have also demonstrated the superiority of the proposed method by applying univariate denoising methods independently on multiple data channels.

In other experiments, we verified the robustness of our method with change in number of channels $M$ of input signal and $P_{fa}$. In addition, the experiments involving asynchronous signals were also conducted and it was observed that our \textit{MGWD} method consistently outperformed the state of the art. Moreover, we used the empirical GoF test in \cite{mcassey2013empiricalMGOF} within our denoising framework and compared its performance against that of the \textit{MGWD} to validate the superiority of our GoF test.

Moreover, proposed framework is flexible in terms of its extension to possibly remove multivariate non-Gaussian noise from input data. While the scope of this paper is limited to multivariate wGn noise, extension to other noise types is clearly possible. The only obstacle could be to obtain analytical relation of quadratic transformation of reference multivariate noise distribution, similar to \eqref{nullCDF}. 
This could be an interesting avenue for future research work.
\vspace{-1mm}
\section{Appendix}
\begin{theorem}
	\label{theorem2}
	\textit{Given the normally distributed multivariate data $\mathbf{x}\sim \mathcal{N}(\boldsymbol{\mu},\Sigma)$ with mean $\boldsymbol{\mu}$ and covariance matrix $\Sigma$ and its transformed version $\mathbf{y}=\mathcal{T}(\mathbf{x})$. Iff $\mathcal{T}(\cdot)$ is a linear transformation, i.e., $\mathcal{T}(\mathbf{x})=A\mathbf{x}+\mathbf{b}$ such that $A$ is the transformation matrix and $\mathbf{b}$ is a normally distributed multivariate data with zero mean, then the data $\mathcal{T}(\mathbf{x})$ must also be normally distributed \cite{wang2019marginal,ghosh1969only}, where the	mean $\boldsymbol{\mu}_{\mathcal{T}(\mathbf{x})}$ and covariance matrix $\Sigma_{\mathcal{T}(\mathbf{x})}$ of the transformed normally distributed data $\mathcal{T}(\mathbf{x})$ are given as follows}
		\vspace{-0.1mm}
%		If $\mathbf{A}^{-1}$ is the inverse of $\mathbf{A}$ then (\ref{MD}) can be rewritten as
%		{\color{blue}
		\begin{equation} 
%		\begin{split}
		\boldsymbol{\mu}_{\mathcal{T}(\mathbf{x})} =  E[A\mathbf{x}+\mathbf{b}]
		=E[A\mathbf{x}]+E[\mathbf{b}] =A \ E[\mathbf{x}]= A\mathbf{\mu}.
%		\Delta_X = \sqrt{(\mathbf{x}^t -\boldsymbol{\mu})^T\mathbf{A}^T(\mathbf{A}^T)^{-1}\Sigma^{-1}\mathbf{A}^{-1}\mathbf{A} (\mathbf{x}^t-\boldsymbol{\mu})},
		\label{MD} \nonumber
%\end{split}
		\end{equation}
		\vspace{-4mm}
			\begin{equation} 
%		\begin{split}
		\Sigma_{\mathcal{T}(\mathbf{x})}  =  \mathit{Cov}\left (A\mathbf{x}+\mathbf{b}\right )
		=\mathit{Cov}\left (A\mathbf{x}\right )+ \mathit{Cov}\left (\mathbf{b}\right ) =A \Sigma A^T + \Sigma_b.
		%		\Delta_X = \sqrt{(\mathbf{x}^t -\boldsymbol{\mu})^T\mathbf{A}^T(\mathbf{A}^T)^{-1}\Sigma^{-1}\mathbf{A}^{-1}\mathbf{A} (\mathbf{x}^t-\boldsymbol{\mu})},
		\label{MD} \nonumber
%		\end{split}
		\end{equation}
\end{theorem}
\vspace{-6mm}
%\vspace{-1mm} 
\bibliographystyle{IEEEtr}
\bibliography{Manuscript}
\end{document}